\documentclass[a4paper,twocolumn,11pt,unpublished,noarxiv,allowtoday]{quantumarticle}
\usepackage[utf8]{inputenc}
\usepackage[english]{babel}
\usepackage[T1]{fontenc}
\usepackage{amsmath}
\usepackage{hyperref}
\usepackage{mathtools, nccmath}
\usepackage{setspace}
\usepackage{outlines}
\usepackage{fancyvrb}
\usepackage{mathtools}
\usepackage{physics}
\usepackage{comment}
\usepackage{bbold}
\usepackage{cancel}
\usepackage{graphicx}
\usepackage[usenames, dvipsnames]{color}
\usepackage{soul}
\usepackage{tensor}
\usepackage{amsthm}
\usepackage[caption=false]{subfig}
\usepackage{algorithm}
\usepackage{algcompatible}

\usepackage{lipsum}
\usepackage{tkz-graph}
\usetikzlibrary{hobby,backgrounds,calc,trees}
\usepackage{xcolor}
\usepackage{soul}
\usetikzlibrary{positioning}
\usepackage[normalem]{ulem}
\usepackage{yquant}
\useyquantlanguage{groups}
\usepackage{makecell}
\usepackage{tikz,esvect,tikz-3dplot}
\usetikzlibrary{3d,calc,intersections}
\usepackage{tikz}
\usepackage{adjustbox}
\usepackage{zx-calculus}
\usetikzlibrary{zx-calculus}
\usepackage{tikz-cd}
\usepackage[official]{eurosym}

\usepackage{tensor}
\usepackage{accents}
\usetikzlibrary{angles,quotes}

\definecolor{zx_red}{RGB}{232, 165, 165}
\definecolor{zx_green}{RGB}{216, 248, 216}

\usepackage[numbers,sort&compress]{natbib}
\usepackage{iftex}
\usepackage{tikzit}
\usepackage{amsfonts,amssymb}
\usepackage{tablefootnote}

\tikzstyle{gate}=[shape=rectangle, text height=1.5ex, text depth=0.25ex, yshift=0.5mm, fill=white, draw=black, minimum height=5mm, yshift=-0.5mm, minimum width=5mm, font={\small}, tikzit category=circuit]
\tikzstyle{big gate}=[shape=rectangle, text height=1.5ex, text depth=0.25ex, yshift=0.5mm, fill=white, draw=black, minimum height=10mm, yshift=-0.5mm, minimum width=5mm, font={\small}, tikzit category=circuit]
\tikzstyle{Z dot}=[inner sep=0mm, minimum size=2mm, shape=circle, draw=black, fill={rgb,255: red,221; green,255; blue,221}, tikzit category=zx]
\tikzstyle{Z phase dot}=[minimum size=5mm, font={\footnotesize\boldmath}, shape=rectangle, rounded corners=2mm, inner sep=0.2mm, outer sep=-2mm, scale=0.8, tikzit shape=rectangle, draw=black, fill={rgb,255: red,221; green,255; blue,221}, tikzit draw=blue, tikzit category=zx]
\tikzstyle{X dot}=[Z dot, shape=circle, draw=black, fill={rgb,255: red,255; green,136; blue,136}, tikzit category=zx]
\tikzstyle{X phase dot}=[Z phase dot, tikzit shape=rectangle, tikzit draw=blue, fill={rgb,255: red,255; green,136; blue,136}, font={\footnotesize\boldmath}, tikzit category=zx]
\tikzstyle{hadamard}=[fill=yellow, draw=black, shape=rectangle, inner sep=0.6mm, minimum height=1.5mm, minimum width=1.5mm, tikzit category=zx]
\tikzstyle{paulibox}=[fill={rgb,255: red,221; green,221; blue,255}, draw=black, shape=rectangle, inner sep=0.6mm, minimum height=5mm, minimum width=5mm, font={\footnotesize}, text height=1.5ex, text depth=0.25ex, tikzit category=zx]
\tikzstyle{vertex}=[inner sep=0mm, minimum size=1mm, shape=circle, draw=black, fill=black, tikzit category=misc]
\tikzstyle{vertex set}=[inner sep=0mm, minimum size=1mm, shape=circle, draw=black, fill=white, font={\footnotesize\boldmath}, tikzit category=misc]
\tikzstyle{small black dot}=[fill=black, draw=black, shape=circle, inner sep=0pt, minimum width=1.2mm, tikzit category=circuit]
\tikzstyle{cnot ctrl}=[fill=black, draw=black, shape=circle, inner sep=0pt, minimum width=1.2mm, tikzit category=circuit]
\tikzstyle{cnot targ}=[fill=white, draw=white, shape=circle, tikzit category=circuit, label={center:$\oplus$}, inner sep=0pt, minimum width=2.1mm, tikzit fill={rgb,255: red,102; green,204; blue,255}, tikzit draw=black]
\tikzstyle{ket}=[fill=white, draw=black, shape=regular polygon, regular polygon sides=3, regular polygon rotate=-30, scale=0.7, inner sep=1pt, tikzit category=circuit, tikzit shape=rectangle, tikzit fill=green]
\tikzstyle{bra}=[fill=white, draw=black, shape=regular polygon, regular polygon sides=3, regular polygon rotate=30, scale=0.7, inner sep=1pt, tikzit category=circuit, tikzit shape=rectangle, tikzit fill=red]
\tikzstyle{scalar}=[shape=rectangle, text height=1.5ex, text depth=0.25ex, yshift=0.5mm, fill=white, draw=black, minimum height=5mm, yshift=-0.5mm, minimum width=5mm, font={\small}]
\tikzstyle{clabel}=[fill=white, draw=none, shape=rectangle, tikzit fill={rgb,255: red,56; green,255; blue,242}, font={\footnotesize}, inner sep=1pt, tikzit category=labels]
\tikzstyle{empty diagram}=[draw={gray!40!white}, dashed, shape=rectangle, minimum width=1cm, minimum height=1cm, tikzit category=misc]
\tikzstyle{amap}=[fill=white, draw=black, shape=NEbox, tikzit category=asymmetric, tikzit fill=yellow, tikzit shape=rectangle]
\tikzstyle{amap conj}=[fill=white, draw=black, shape=NWbox, tikzit category=asymmetric, tikzit fill=green, tikzit shape=rectangle]
\tikzstyle{amap adj}=[fill=white, draw=black, shape=SEbox, tikzit category=asymmetric, tikzit fill=red, tikzit shape=rectangle]
\tikzstyle{amap trans}=[fill=white, draw=black, shape=SWbox, tikzit category=asymmetric, tikzit fill=orange, tikzit shape=rectangle]
\tikzstyle{astate}=[fill=white, draw=black, shape=NEtriangle, tikzit category=asymmetric, tikzit shape=circle, tikzit fill=yellow]
\tikzstyle{astate conj}=[fill=white, draw=black, shape=NWtriangle, tikzit category=asymmetric, tikzit shape=circle, tikzit fill=green]
\tikzstyle{astate adj}=[fill=white, draw=black, shape=SEtriangle, tikzit category=asymmetric, tikzit shape=circle, tikzit fill=red]
\tikzstyle{astate trans}=[fill=white, draw=black, shape=SWtriangle, tikzit category=asymmetric, tikzit shape=circle, tikzit fill=orange]

\tikzstyle{hadamard edge}=[-, dashed, dash pattern=on 2pt off 0.5pt, thick, draw={rgb,255: red,68; green,136; blue,255}]
\tikzstyle{star edge}=[-, dashed, dash pattern=on 2pt off 0.5pt, thick, draw={rgb,255: red,255; green,136; blue,68}]
\tikzstyle{box edge}=[-, dashed, dash pattern=on 2pt off 0.5pt, thick, draw={rgb,255: red,203; green,192; blue,225}]
\tikzstyle{brace edge}=[-, tikzit draw=blue, decorate, decoration={brace,amplitude=1mm,raise=-1mm}]
\tikzstyle{diredge}=[->]
\tikzstyle{double edge}=[-, double, shorten <=-1mm, shorten >=-1mm, double distance=2pt]
\tikzstyle{gray edge}=[-, {gray!60!white}]
\tikzstyle{pointer edge}=[->, very thick, gray]
\tikzstyle{boldedge}=[-, line width=1.6pt, shorten <=-0.17mm, shorten >=-0.17mm]
\tikzstyle{bidir edge}=[<->, very thick, draw={rgb,255: red,191; green,191; blue,191}]
\tikzstyle{separator edge}=[-, dashed, dash pattern=on 2pt off 0.5pt, thick, draw={rgb,255: red,153; green,153; blue,153}]

\tikzstyle{xweb}=[-, line width=4.4pt, line cap=round, draw={rgb,255: red,219; green,51; blue,51}, opacity=0.50]
\tikzstyle{zweb}=[-, line width=4.4pt, line cap=round, draw={rgb,255: red,51; green,166; blue,64}, opacity=0.50]
\tikzstyle{yweb}=[-, line width=4.4pt, line cap=round, draw={rgb,255: red,58; green,80; blue,190}, opacity=0.50]
\tikzstyle{tile box}=[rectangle, minimum size=3.2mm, fill=violet!25, draw=violet!70!black, line width=0.6pt]
\tikzstyle{ghost box}=[rectangle, minimum size=3.2mm, fill=violet!10, draw=violet!35, line width=0.5pt]
\tikzstyle{event box}=[rectangle, minimum size=1.5mm, inner sep=0, fill=orange!75, draw=orange!55!black, line width=0.5pt]
\tikzstyle{qubit dot}=[circle, minimum size=1.3mm, inner sep=0, fill=black]
\tikzstyle{rim dot}=[circle, minimum size=1.5mm, inner sep=0, fill=white, draw=red!65!black, line width=0.6pt]
\tikzstyle{plus dot}=[circle, minimum size=1.5mm, inner sep=0, draw=black, line width=0.4pt, fill={rgb,255: red,221; green,255; blue,221}]
\tikzstyle{stub dot}=[circle, minimum size=1.1mm, inner sep=0, fill=white, draw=gray, line width=0.4pt]

\usepackage{contour}
\contourlength{0.3pt} 
\definecolor{MyMutedYellow}{HTML}{BAB330} 
\definecolor{MyMutedRed}{HTML}{8B0000}
\definecolor{MyMutedBlue}{HTML}{87afff}

\makeatletter
\DeclareRobustCommand{\rvdots}{%
  \vbox{
    \baselineskip4\p@\lineskiplimit\z@
    \kern-\p@
    \hbox{.}\hbox{.}\hbox{.}
  }}
\makeatother

\begin{document}

\title{Holographic codes seen through ZX-calculus}

\date{\today}

\author{Kwok Ho Wan}
\orcid{0000-0002-1762-1001}
\affiliation{Blackett Laboratory, Imperial College London, London SW7 2AZ, UK}
\affiliation{Mathematical Institute, University of Oxford, Woodstock Road, Oxford OX2 6GG, UK}
\email{((initials))1496((at))(($9.81$))mail.com}
\author{Henry C. W. Price}
\orcid{0000-0003-0756-0652}
\affiliation{Centre for Complexity Science, Imperial College London, London SW7 2AZ, UK}
\author{Qing Yao}
\orcid{0000-0002-5222-8977}
\affiliation{Centre for Complexity Science, Imperial College London, London SW7 2AZ, UK}
\affiliation{Mailman School of Public Health, Columbia University, New York, NY, USA.}
\author{Zhenghao Zhong}
\affiliation{Shanghai Institute for Mathematics and Interdisciplinary Sciences (SIMIS), Shanghai 200433, China}
\author{Ainhoa Zapirain}
\affiliation{Independent Researcher}

\begin{abstract}
We re-visit the pentagon holographic quantum error correcting code from a ZX-calculus perspective. By expressing the underlying tensors as ZX-diagrams, we study the stabiliser structure of the code via Pauli webs. In addition, we obtain a diagrammatic understanding of its logical operators, encoding isometries, R\'enyi entropy and toy models of black holes/wormholes. Then, motivated by the pentagon holographic code's ZX-diagram, we introduce a family of codes constructed from ZX-diagrams on its dual hyperbolic tessellations and study their logical error rates using belief propagation decoders. Finally, we show how to construct spacetime ZX-diagrams that realise a fault-tolerant quantum channel in which every internal edge is a protected fault location.
\end{abstract}

\maketitle

\section{Introduction}
Holographic tensor networks provide a concrete bridge between ideas from $\mathrm{AdS}/\mathrm{CFT}$ and quantum error correction: bulk degrees of freedom can be redundantly encoded into boundary degrees of freedom, and erasures on the boundary can be interpreted as correctable noise \cite{Almheiri:2014lwa,Pastawski:2015qua}. The pentagon holographic code \cite{Pastawski:2015qua} is an example of a holographic tensor network code built from perfect tensors arranged on a hyperbolic tessellation.

\begin{widetext}
\begin{equation}
    \label{eq:pent_l4_edited}
    \raisebox{-17ex}{\scalebox{0.4}{\includegraphics{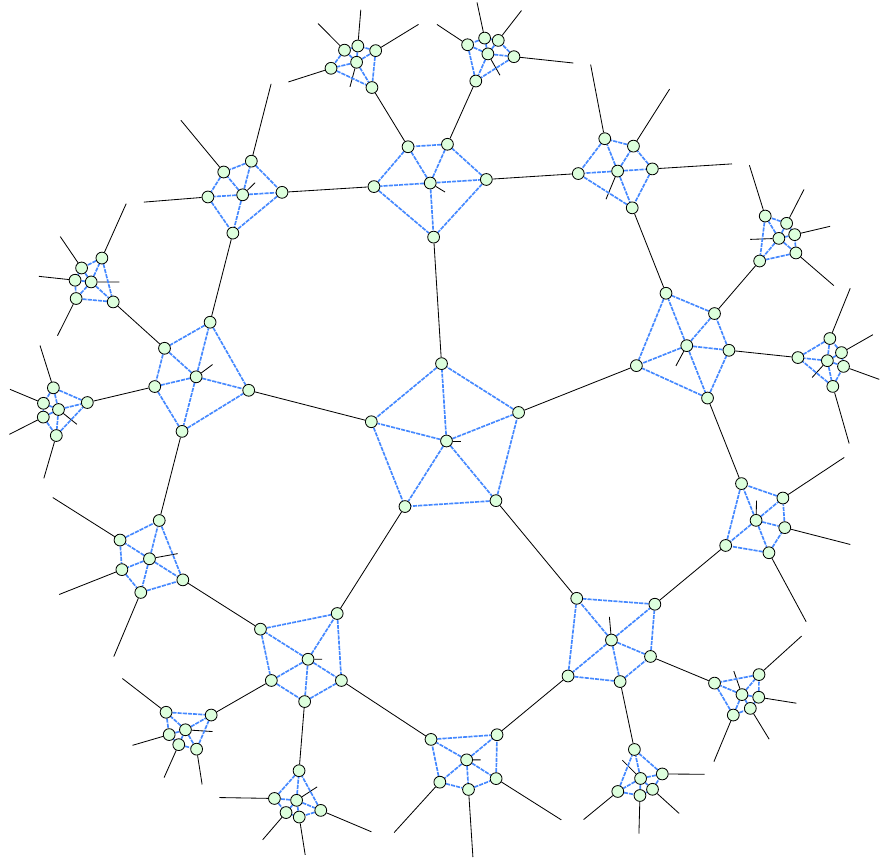}}} \hspace{1cm} \xleftrightarrow[\hspace{1cm}]{} \hspace{1cm} \raisebox{-17ex}{\scalebox{0.027}{\includegraphics{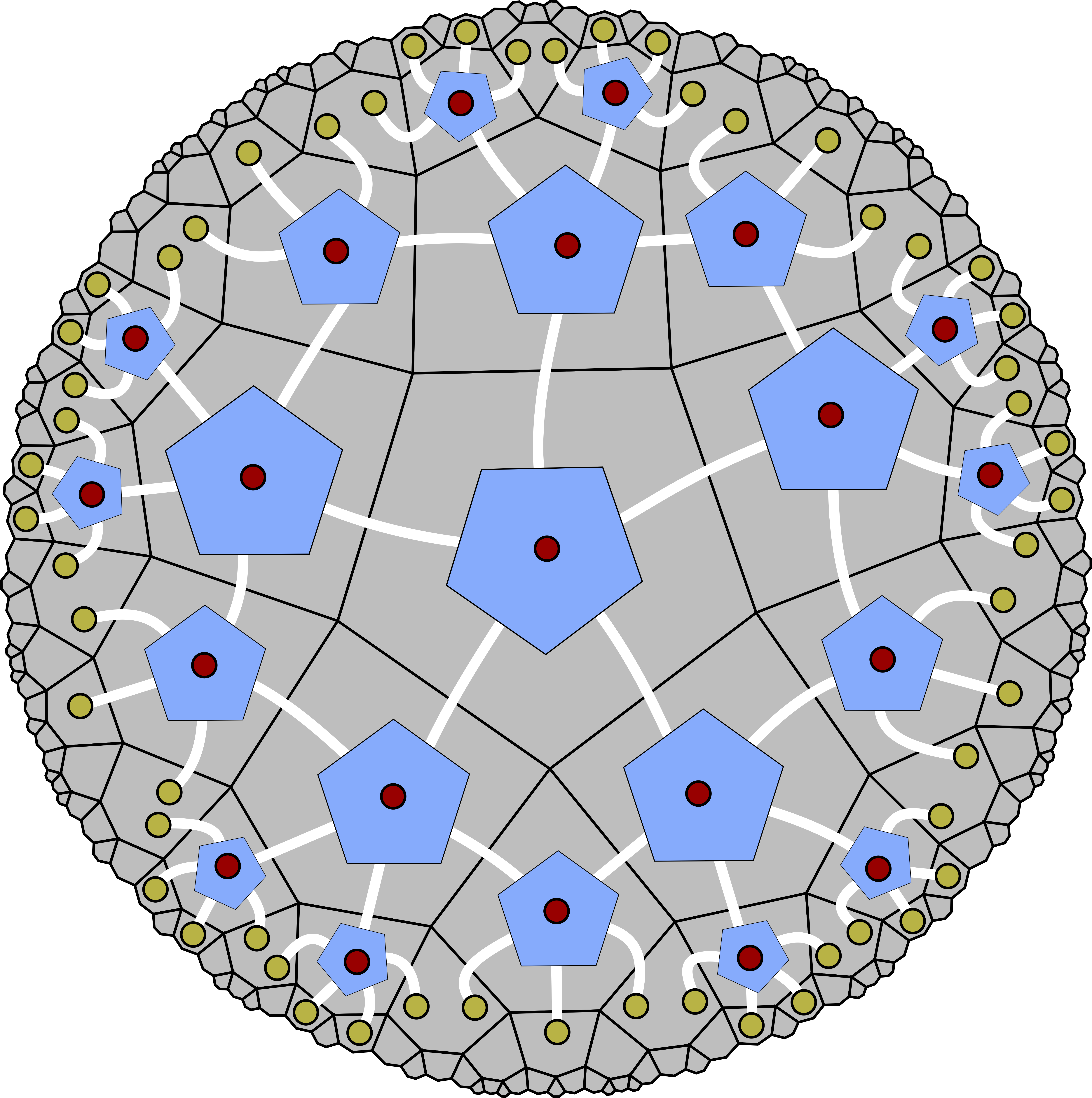}}} 
\end{equation}
\end{widetext}
Most analyses of such codes proceed directly in the tensor-network language, where geometry suggests entanglement structure and decoding is often phrased in terms of operator pushing and greedy algorithms \cite{Pastawski:2015qua,Steinberg_2025}. In this work we advocate a complementary viewpoint: for stabiliser codes, ZX-calculus provides a graphical language in which (i) tensor identities become diagrammatic equalities, and (ii) stabilisers and logical operators can be represented and manipulated as Pauli webs. Once a holographic encoder is expressed as a Clifford ZX-diagram, one can algorithmically extract stabiliser generators, parity check matrices, and boundary representatives of bulk logical operators using standard ZX tooling \cite{kissinger2020Pyzx,rüsch2025completenessfaultequivalenceclifford,Bombin_2024}.

The framework developed here is primarily tailored to stabiliser holographic codes. Our constructions and automated extractions rely on Clifford ZX-diagrams and its Pauli web. The ZX-diagram representation acts as a map to a graphical calculus for holographic codes, enabling direct diagrammatic access to stabilisers, logical correlators, and (in certain settings) entanglement computations.

In this manuscript, we re-visit the pentagon holographic code from \cite{Pastawski:2015qua} through the lens of ZX-calculus and Pauli webs, and we provide numerical simulations of isometry-defined codes based on generalisations of its ZX-diagram construction. We study logical performance under belief propagation decoders \cite{Roffe_LDPC_Python_tools_2022}, and we illustrate how generator choices (and simple gauge-fixing prescriptions) can dramatically affect decoder performance (in line with related observations from \cite{Steinberg_2025}). More broadly, we expect other hyperbolic tessellation holographic codes to admit ZX-calculus motivated decompositions into Clifford ZX-diagrams, allowing the same Pauli web and parity-check extraction techniques to be applied. Finally, we show how to foliate these constructions into spacetime ZX-diagrams realising a fault-tolerant quantum channel in which every internal edge is a protected fault location (section~\ref{sec:spacetimeZX}).

We assume basic familiarity with ZX-calculus and refer the reader to \cite{KissingerWetering2024Book} for an extended introduction. We begin with a short introduction to ZX-diagrams, and then present the pentagon holographic code \cite{Pastawski:2015qua}, whose tensor network admits a natural ZX-diagram realisation.

\section{A short introduction to ZX-diagrams}
ZX-diagrams \cite{KissingerWetering2024Book} provide a diagrammatic language for tensors, and in particular for tensors that represent quantum states and processes. They come equipped with a set of rewrite rules that leave the represented linear map/state invariant, allowing one to transform diagrams while preserving their meaning. Any ZX-diagram can be built from a small collection of building blocks. In this manuscript we will use only these three basic building blocks: the Hadamard tensor, the $Z$-spider, and the $X$-spider.

First, the $m$-legged $Z$-spider with phase $\alpha$ is the tensor \cite{Magdalena_de_la_Fuente_2025} with indices $i_1,i_2,\dots,i_m$ given by
\begin{equation}
    \label{eq:GHZ_spider}
    \scalebox{2}{\begin{tikzpicture}[yshift = 0.3cm,scale = 0.5]
    \begin{pgfonlayer}{nodelayer}
        \node [style=none,scale=.5] (0) at (1.00, 0.00) {$i_1$};
        \node [style=none,scale=.5] (1) at (1.00, -1.00) {$i_2$};
        \node [style=none,scale=.5] (2) at (1.00, -2.00) {$i_3$};
        \node [style=none,scale=.5] (3) at (2.00, -2.00) {$i_4$};
        \node [style=Z phase dot] (4) at (3.50, -0.00) {$\alpha$};
        \node [style=none,scale=.5] (20) at (6.00, 0.00) {\hspace{0.05cm}$i_{m}$};
        \node [style=none,scale=.5] (21) at (6.00, -1.00) {$i_{m-1}$};
        \node [style=none,scale=.5] (23) at (5.00, -2.00) {\hspace{0.2cm}$i_{m-2}$};
        \node [style=none,scale=.5] (24) at (3.50, -2.00) {$\dots$};
    \end{pgfonlayer}
    \begin{pgfonlayer}{edgelayer}
        \draw (0) to (4);
        \draw (1) to (4);
        \draw (2) to (4);
        \draw (3) to (4);
        \draw (4) to (20);
        \draw (4) to (21);
        \draw (4) to (23);
    \end{pgfonlayer}
\end{tikzpicture}}
    \;=\;
    e^{i\,\delta_{1,i_1}\alpha}\,\delta_{i_1,i_2,\dots,i_m}\,,
\end{equation}
where $\delta_{a,b,c,\dots}$ denotes the Kronecker delta, and each index ranges over $i_k\in\{0,1\}$. When viewed as an (unnormalised) quantum state, this corresponds to
$\ket{0}^{\otimes m}+e^{i\alpha}\ket{1}^{\otimes m}$.

Next, the Hadamard tensor is the single-qubit unitary which, in the computational basis $\{\ket{0},\ket{1}\}$, is
\begin{equation}
    \label{eq:h_box}
    \scalebox{2}{
\begin{tikzpicture}
    \begin{pgfonlayer}{nodelayer}
        \node [style=hadamard] (0) at (0.00, 0.00) {};
        \node [style=none] (1) at (-1.00, 0.00) {};
        \node [style=none] (2) at (1.00, 0.00) {};
    \end{pgfonlayer}
    \begin{pgfonlayer}{edgelayer}
        \draw (0) to (1);
        \draw (0) to (2);
    \end{pgfonlayer}
\end{tikzpicture}
}
    \;=\;
    \frac{1}{\sqrt{2}}
    \begin{pmatrix}
        1 & 1 \\
        1 & -1
    \end{pmatrix}
    \;=\; H\,.
\end{equation}
In ZX-diagrams it is often depicted as a blue dashed edge as a shorthand:
\begin{equation}
    \scalebox{2}{
\begin{tikzpicture}
    \begin{pgfonlayer}{nodelayer}
        \node [style=hadamard] (0) at (0.00, 0.00) {};
        \node [style=none] (1) at (-1.00, 0.00) {};
        \node [style=none] (2) at (1.00, 0.00) {};
    \end{pgfonlayer}
    \begin{pgfonlayer}{edgelayer}
        \draw (0) to (1);
        \draw (0) to (2);
    \end{pgfonlayer}
\end{tikzpicture}
}=\scalebox{2}{
\begin{tikzpicture}
    \begin{pgfonlayer}{nodelayer}
        \node [style=none] (1) at (-1.00, 0.00) {};
        \node [style=none] (2) at (1.00, 0.00) {};
    \end{pgfonlayer}
    \begin{pgfonlayer}{edgelayer}
        \draw [style=hadamard edge] (1) to (2);
    \end{pgfonlayer}
\end{tikzpicture}
} \ .
\end{equation}
Finally, the $X$-spider with phase $\alpha$ is defined by contracting each leg of a $Z$-spider by a Hadamard:
\begin{equation}
    \label{eq:X_spider_def}
    \scalebox{2}{\begin{tikzpicture}[yshift = 0.3cm,scale = 0.5]
    \begin{pgfonlayer}{nodelayer}
        \node [style=none,scale=.5] (0) at (1.00, 0.00) {$i_1$};
        \node [style=none,scale=.5] (1) at (1.00, -1.00) {$i_2$};
        \node [style=none,scale=.5] (2) at (1.00, -2.00) {$i_3$};
        \node [style=none,scale=.5] (3) at (2.00, -2.00) {$i_4$};
        \node [style=X phase dot] (4) at (3.50, -0.00) {$\alpha$};
        \node [style=none,scale=.5] (20) at (6.00, 0.00) {\hspace{0.05cm}$i_{m}$};
        \node [style=none,scale=.5] (21) at (6.00, -1.00) {$i_{m-1}$};
        \node [style=none,scale=.5] (23) at (5.00, -2.00) {\hspace{0.2cm}$i_{m-2}$};
        \node [style=none,scale=.5] (24) at (3.50, -2.00) {$\dots$};
    \end{pgfonlayer}
    \begin{pgfonlayer}{edgelayer}
        \draw (0) to (4);
        \draw (1) to (4);
        \draw (2) to (4);
        \draw (3) to (4);
        \draw (4) to (20);
        \draw (4) to (21);
        \draw (4) to (23);
    \end{pgfonlayer}
\end{tikzpicture}}
    \;=\;
    \scalebox{2}{\begin{tikzpicture}[yshift = 0.3cm,scale = 0.5]
    \begin{pgfonlayer}{nodelayer}
        \node [style=none,scale=.5] (0) at (1.00, 0.00) {$i_1$};
        \node [style=none,scale=.5] (1) at (1.00, -1.00) {$i_2$};
        \node [style=none,scale=.5] (2) at (1.00, -2.00) {$i_3$};
        \node [style=none,scale=.5] (3) at (2.00, -2.00) {$i_4$};
        \node [style=Z phase dot] (4) at (3.50, -0.00) {$\alpha$};
        \node [style=none,scale=.5] (20) at (6.00, 0.00) {\hspace{0.05cm}$i_{m}$};
        \node [style=none,scale=.5] (21) at (6.00, -1.00) {$i_{m-1}$};
        \node [style=none,scale=.5] (23) at (5.00, -2.00) {\hspace{0.2cm}$i_{m-2}$};
        \node [style=none,scale=.5] (24) at (3.50, -2.00) {$\dots$};
    \end{pgfonlayer}
    \begin{pgfonlayer}{edgelayer}
        \draw [style=hadamard edge] (0) to (4);
        \draw [style=hadamard edge] (1) to (4);
        \draw [style=hadamard edge] (2) to (4);
        \draw [style=hadamard edge] (3) to (4);
        \draw [style=hadamard edge] (4) to (20);
        \draw [style=hadamard edge] (4) to (21);
        \draw [style=hadamard edge] (4) to (23);
    \end{pgfonlayer}
\end{tikzpicture}} \,.
\end{equation}
Equivalently, the $m$-legged $X$-spider (indices $i_1,\dots,i_m$) can be written in components as
\begin{equation}
    \label{eq:X_spider_components}
    \sum_{j_1=0}^{1} H_{j_1,i_1}\cdots \sum_{j_m=0}^{1} H_{j_m,i_m}\;
    e^{i\,\delta_{1,j_1}\alpha}\,\delta_{j_1,j_2,\dots,j_m}\,,
\end{equation}
where $H_{j_k,i_k}$ are the matrix elements of the Hadamard tensor in equation \ref{eq:h_box}. More generally, composing these building blocks corresponds to contracting tensors along the wires (summing over the shared indices) associated with connected legs.

In the next section, we will introduce the pentagon holographic code and then explain its connection with ZX-diagrams. Compared to traditional tensor-network approaches, ZX-calculus provides a compositional, rewrite-based structure for tensors and, crucially, supports Pauli web annotations that make stabilisers and logical operators transparent (as we explain later).

\section{The pentagon holographic code}
The pentagon holographic code shown on the RHS of equation \ref{eq:pent_l4_edited} may be viewed as an encoding map which takes logical \textit{bulk} qubits (shown as \textcolor{MyMutedRed}{dark red} nodes) to physical \textit{boundary} qubits (shown as \textcolor{MyMutedYellow}{citronelle} coloured nodes). The bulk and boundary degrees of freedom are related via a network of six-qubit perfect tensors, depicted as \textcolor{MyMutedBlue}{blue} pentagons in \cite{Pastawski:2015qua} and on the right-hand side of equation \ref{eq:pent_l4_edited}. These \textcolor{MyMutedBlue}{blue} pentagons are connected together via white edges that are to be contracted to form the code \cite{Jahn_2021}. Geometrically, this tensor network corresponds to a (truncated) hyperbolic tiling with Schläfli symbol $\{p,q\}=\{5,4\}$, where each pentagon represents a perfect tensor\footnote{A perfect tensor defines an isometry for any bipartition of its legs into inputs and outputs, provided the input side contains at most half of the legs.
}.

Each six-qubit perfect tensor may be interpreted as an isometric encoding of a single bulk qubit into five qubits. Concretely, the corresponding state can be viewed as a logical Bell pair between one bulk qubit and the logical encoded qubit of the $[[5,1,3]]$ perfect code \cite{eczoo_stab_5_1_3}. Its stabiliser group is: 
\begin{equation}
\label{eq:perfect_tensor_stabs}
\begin{split}
\big\langle\;
& XZZXI \otimes I,\quad IXZZX \otimes I,\\
& XIXZZ \otimes I,\quad ZXIXZ \otimes I, \\
& \underbrace{XXXXX}_{\bar{X}_{[[5,1,3]]}} \otimes X,\quad \underbrace{ZZZZZ}_{\bar{Z}_{[[5,1,3]]}} \otimes Z
\;\big\rangle \ ,
\end{split}
\end{equation}
where the first four generators stabilise the five-qubit code, while the final two generators couple the bulk qubit to the logical $\bar{Z}$ and $\bar{X}$ operators of the code. In the logical basis, the six-qubit state takes the form:
\begin{equation}
\ket{\Psi}
\;\propto\;
\ket{\bar{0}}_{[[5,1,3]]}\otimes\ket{0}
\;+\;
\ket{\bar{1}}_{[[5,1,3]]}\otimes\ket{1},
\end{equation}
where $\ket{\bar{0}}_{[[5,1,3]]}$ and $\ket{\bar{1}}_{[[5,1,3]]}$ denote the logical codewords of the $[[5,1,3]]$ qubit perfect code. 

Gluing such tensors together by contracting pairs of legs yields a global stabiliser code whose logical qubits reside in the bulk and the remaining uncontracted qubits at the boundary are the physical qubits. More formally, $N_{\text{boundary}}$ boundary qubits live in the Hilbert space: $\mathcal{H}_{\partial} = (\mathbb{C}^2)^{\otimes N_{\text{boundary}}}$, similarly, $N_\text{bulk}$ bulk qubits reside in $\mathcal{H}_{\text{bulk}} = (\mathbb{C}^2)^{\otimes N_{\text{bulk}}}$. The isometric encoding map for this code is defined as:
\begin{equation}
\label{eq:V_iso}
V:\; \mathcal{H}_{\text{bulk}} \longrightarrow \mathcal{H}_{\partial},
\qquad
V^{\dagger} V ={I}_{\text{bulk}} \ ,
\end{equation}
where bulk and boundary states are related via: $\ket{A}_{\partial} = V\ket{B}_\text{bulk}$. One may construct a family of pentagon holographic codes by embedding the same local holographic tensor into progressively larger hyperbolic tessellations with an increasing number of layers: $n$. We built our tessellation with the python package $\mathtt{hypertiling}$ \cite{hypertiling}. In our notation, the pentagon holographic code from equation \ref{eq:pent_l4_edited} has $n=2$ layers, to be consistent with the original literature \cite{Pastawski:2015qua}. With the tensor network code defined, we can now express the tensor network as ZX-diagrams, providing a natural framework to visualise stabilisers, logical correlators and entanglement next.

\section{The ZX-calculus connection}
\label{sec:zx_connection}
Conveniently in ZX-calculus language, the 6-qubit state/encoding map\footnote{Also known as the pentagon holographic code with $n=0$ `layer'.} (up to a local Hadamard in the bulk qubit leg) can be written as a graph state using the convention from \cite{wu2024zxcalculusapproachconstructiongraph}:
\begin{equation}
    \label{eq:pent_l2}
    \scalebox{1.3}{\begin{tikzpicture}[scale=0.4]
    \begin{pgfonlayer}{nodelayer}
        \node [style=Z dot] (0) at (0.00, -0.00) {};
        \node [style=none] (1) at (3.50, -6.00) {};
        \node [style=none] (2) at (6.50, 3.00) {};
        \node [style=none] (3) at (0.00, 7.00) {};
        \node [style=none] (4) at (-6.50, 3.00) {};
        \node [style=none] (5) at (-3.50, -6.00) {};
        \node [style=Z dot] (6) at (2.50, -4.00) {};
        \node [style=Z dot] (7) at (4.50, 2.00) {};
        \node [style=Z dot] (8) at (0.00, 5.00) {};
        \node [style=Z dot] (9) at (-4.50, 2.00) {};
        \node [style=Z dot] (10) at (-2.50, -4.00) {};
        \node [style=none] (11) at (2.00, -1.00) {};
    \end{pgfonlayer}
    \begin{pgfonlayer}{edgelayer}
        \draw [style=hadamard edge] (0) to (6);
        \draw [style=hadamard edge] (0) to (7);
        \draw [style=hadamard edge] (0) to (8);
        \draw [style=hadamard edge] (0) to (9);
        \draw [style=hadamard edge] (0) to (10);
        \draw (0) to (11);
        \draw (1) to (6);
        \draw (2) to (7);
        \draw (3) to (8);
        \draw (4) to (9);
        \draw (5) to (10);
        \draw [style=hadamard edge] (6) to (10);
        \draw [style=hadamard edge] (6) to (7);
        \draw [style=hadamard edge] (7) to (8);
        \draw [style=hadamard edge] (8) to (9);
        \draw [style=hadamard edge] (9) to (10);
    \end{pgfonlayer}
\end{tikzpicture}} \ ,
\end{equation}
where the qubit extending from the centre of the pentagon represents the
logical (\textcolor{MyMutedRed}{bulk}) qubit, while the five qubits attached to the vertices of the pentagon correspond to the five physical (\textcolor{MyMutedYellow}{boundary}) qubits of the 6-qubit state in this 6-legged ZX-diagram \cite{wu2024zxcalculusapproachconstructiongraph,Pastawski:2015qua}. We suspect the decision to apply a local Hadamard to the central bulk qubit leg in \cite{wu2024zxcalculusapproachconstructiongraph} as opposed to using equation \ref{eq:perfect_tensor_stabs} is due to the ZX-diagram's (in equation \ref{eq:pent_l2}) interpretability as a graph state. We shall call this state the 6-qubit graph state from now onwards.

ZX-diagrams allow us to draw Pauli webs \cite{Bombin_2024} on top, providing a decorative visual representation of the stabilisers of the 6-qubit graph state for example (listed in the same order as in equation \ref{eq:perfect_tensor_stabs}, up to a bulk Hadamard).
\begin{equation}
\begin{split}
\Bigg\langle\;
& \scalebox{0.6}{\begin{tikzpicture}[scale=0.4]
    \begin{pgfonlayer}{nodelayer}
        \node [style=Z dot] (0) at (0.00, -0.00) {};
        \node [style=none] (1) at (3.50, -6.00) {};
        \node [style=none] (2) at (6.50, 3.00) {};
        \node [style=none] (3) at (0.00, 7.00) {};
        \node [style=none] (4) at (-6.50, 3.00) {};
        \node [style=none] (5) at (-3.50, -6.00) {};
        \node [style=Z dot] (6) at (2.50, -4.00) {};
        \node [style=Z dot] (7) at (4.50, 2.00) {};
        \node [style=Z dot] (8) at (0.00, 5.00) {};
        \node [style=Z dot] (9) at (-4.50, 2.00) {};
        \node [style=Z dot] (10) at (-2.50, -4.00) {};
        \node [style=none] (11) at (2.00, -1.00) {};
    \end{pgfonlayer}
    \begin{pgfonlayer}{edgelayer}
        \draw [style=hadamard edge] (0) to (6);
        \draw [style=hadamard edge] (0) to (7);
        \draw [style=hadamard edge] (0) to (8);
        \draw [style=hadamard edge] (0) to (9);
        \draw [style=hadamard edge] (0) to (10);
        \draw (0) to (11);
        \draw (1) to (6);
        \draw (2) to (7);
        \draw (3) to (8);
        \draw (4) to (9);
        \draw (5) to (10);
        \draw [style=hadamard edge] (6) to (10);
        \draw [style=hadamard edge] (6) to (7);
        \draw [style=hadamard edge] (7) to (8);
        \draw [style=hadamard edge] (8) to (9);
        \draw [style=hadamard edge] (9) to (10);
    \end{pgfonlayer}

    \begin{pgfonlayer}{edgelayer}
\draw[red, line width=2.5pt, opacity=0.4] (8) -- ($(8)!0.5!(3)$);
\draw[red, line width=2.5pt, opacity=0.4] (3) -- ($(3)!0.5!(8)$);
\draw[green, line width=2.5pt, opacity=0.4] (6) -- ($(6)!0.5!(1)$);
\draw[green, line width=2.5pt, opacity=0.4] (1) -- ($(1)!0.5!(6)$);
\draw[green, line width=2.5pt, opacity=0.4] (0) -- ($(0)!0.5!(8)$);
\draw[red, line width=2.5pt, opacity=0.4] (8) -- ($(8)!0.5!(0)$);
\draw[green, line width=2.5pt, opacity=0.4] (0) -- ($(0)!0.5!(10)$);
\draw[red, line width=2.5pt, opacity=0.4] (10) -- ($(10)!0.5!(0)$);
\draw[green, line width=2.5pt, opacity=0.4] (7) -- ($(7)!0.5!(8)$);
\draw[red, line width=2.5pt, opacity=0.4] (8) -- ($(8)!0.5!(7)$);
\draw[red, line width=2.5pt, opacity=0.4] (8) -- ($(8)!0.5!(9)$);
\draw[green, line width=2.5pt, opacity=0.4] (9) -- ($(9)!0.5!(8)$);
\draw[green, line width=2.5pt, opacity=0.4] (9) -- ($(9)!0.5!(10)$);
\draw[red, line width=2.5pt, opacity=0.4] (10) -- ($(10)!0.5!(9)$);
\draw[red, line width=2.5pt, opacity=0.4] (10) -- ($(10)!0.5!(5)$);
\draw[red, line width=2.5pt, opacity=0.4] (5) -- ($(5)!0.5!(10)$);
\draw[green, line width=2.5pt, opacity=0.4] (7) -- ($(7)!0.5!(2)$);
\draw[green, line width=2.5pt, opacity=0.4] (2) -- ($(2)!0.5!(7)$);
\draw[green, line width=2.5pt, opacity=0.4] (6) -- ($(6)!0.5!(10)$);
\draw[red, line width=2.5pt, opacity=0.4] (10) -- ($(10)!0.5!(6)$);
\end{pgfonlayer}
\end{tikzpicture}},\quad \scalebox{0.6}{\begin{tikzpicture}[scale=0.4]
    \begin{pgfonlayer}{nodelayer}
        \node [style=Z dot] (0) at (0.00, -0.00) {};
        \node [style=none] (1) at (3.50, -6.00) {};
        \node [style=none] (2) at (6.50, 3.00) {};
        \node [style=none] (3) at (0.00, 7.00) {};
        \node [style=none] (4) at (-6.50, 3.00) {};
        \node [style=none] (5) at (-3.50, -6.00) {};
        \node [style=Z dot] (6) at (2.50, -4.00) {};
        \node [style=Z dot] (7) at (4.50, 2.00) {};
        \node [style=Z dot] (8) at (0.00, 5.00) {};
        \node [style=Z dot] (9) at (-4.50, 2.00) {};
        \node [style=Z dot] (10) at (-2.50, -4.00) {};
        \node [style=none] (11) at (2.00, -1.00) {};
    \end{pgfonlayer}
    \begin{pgfonlayer}{edgelayer}
        \draw [style=hadamard edge] (0) to (6);
        \draw [style=hadamard edge] (0) to (7);
        \draw [style=hadamard edge] (0) to (8);
        \draw [style=hadamard edge] (0) to (9);
        \draw [style=hadamard edge] (0) to (10);
        \draw (0) to (11);
        \draw (1) to (6);
        \draw (2) to (7);
        \draw (3) to (8);
        \draw (4) to (9);
        \draw (5) to (10);
        \draw [style=hadamard edge] (6) to (10);
        \draw [style=hadamard edge] (6) to (7);
        \draw [style=hadamard edge] (7) to (8);
        \draw [style=hadamard edge] (8) to (9);
        \draw [style=hadamard edge] (9) to (10);
    \end{pgfonlayer}

    \begin{pgfonlayer}{edgelayer}
\draw[red, line width=2.5pt, opacity=0.4] (9) -- ($(9)!0.5!(4)$);
\draw[red, line width=2.5pt, opacity=0.4] (4) -- ($(4)!0.5!(9)$);
\draw[green, line width=2.5pt, opacity=0.4] (6) -- ($(6)!0.5!(1)$);
\draw[green, line width=2.5pt, opacity=0.4] (1) -- ($(1)!0.5!(6)$);
\draw[green, line width=2.5pt, opacity=0.4] (0) -- ($(0)!0.5!(7)$);
\draw[red, line width=2.5pt, opacity=0.4] (7) -- ($(7)!0.5!(0)$);
\draw[green, line width=2.5pt, opacity=0.4] (0) -- ($(0)!0.5!(9)$);
\draw[red, line width=2.5pt, opacity=0.4] (9) -- ($(9)!0.5!(0)$);
\draw[red, line width=2.5pt, opacity=0.4] (7) -- ($(7)!0.5!(8)$);
\draw[green, line width=2.5pt, opacity=0.4] (8) -- ($(8)!0.5!(7)$);
\draw[green, line width=2.5pt, opacity=0.4] (8) -- ($(8)!0.5!(9)$);
\draw[red, line width=2.5pt, opacity=0.4] (9) -- ($(9)!0.5!(8)$);
\draw[red, line width=2.5pt, opacity=0.4] (9) -- ($(9)!0.5!(10)$);
\draw[green, line width=2.5pt, opacity=0.4] (10) -- ($(10)!0.5!(9)$);
\draw[red, line width=2.5pt, opacity=0.4] (7) -- ($(7)!0.5!(2)$);
\draw[red, line width=2.5pt, opacity=0.4] (2) -- ($(2)!0.5!(7)$);
\draw[green, line width=2.5pt, opacity=0.4] (10) -- ($(10)!0.5!(5)$);
\draw[green, line width=2.5pt, opacity=0.4] (5) -- ($(5)!0.5!(10)$);
\draw[green, line width=2.5pt, opacity=0.4] (6) -- ($(6)!0.5!(7)$);
\draw[red, line width=2.5pt, opacity=0.4] (7) -- ($(7)!0.5!(6)$);
\end{pgfonlayer}
\end{tikzpicture}},\\
& \scalebox{0.6}{\begin{tikzpicture}[scale=0.4]
    \begin{pgfonlayer}{nodelayer}
        \node [style=Z dot] (0) at (0.00, -0.00) {};
        \node [style=none] (1) at (3.50, -6.00) {};
        \node [style=none] (2) at (6.50, 3.00) {};
        \node [style=none] (3) at (0.00, 7.00) {};
        \node [style=none] (4) at (-6.50, 3.00) {};
        \node [style=none] (5) at (-3.50, -6.00) {};
        \node [style=Z dot] (6) at (2.50, -4.00) {};
        \node [style=Z dot] (7) at (4.50, 2.00) {};
        \node [style=Z dot] (8) at (0.00, 5.00) {};
        \node [style=Z dot] (9) at (-4.50, 2.00) {};
        \node [style=Z dot] (10) at (-2.50, -4.00) {};
        \node [style=none] (11) at (2.00, -1.00) {};
    \end{pgfonlayer}
    \begin{pgfonlayer}{edgelayer}
        \draw [style=hadamard edge] (0) to (6);
        \draw [style=hadamard edge] (0) to (7);
        \draw [style=hadamard edge] (0) to (8);
        \draw [style=hadamard edge] (0) to (9);
        \draw [style=hadamard edge] (0) to (10);
        \draw (0) to (11);
        \draw (1) to (6);
        \draw (2) to (7);
        \draw (3) to (8);
        \draw (4) to (9);
        \draw (5) to (10);
        \draw [style=hadamard edge] (6) to (10);
        \draw [style=hadamard edge] (6) to (7);
        \draw [style=hadamard edge] (7) to (8);
        \draw [style=hadamard edge] (8) to (9);
        \draw [style=hadamard edge] (9) to (10);
    \end{pgfonlayer}

    \begin{pgfonlayer}{edgelayer}
\draw[green, line width=2.5pt, opacity=0.4] (8) -- ($(8)!0.5!(3)$);
\draw[green, line width=2.5pt, opacity=0.4] (3) -- ($(3)!0.5!(8)$);
\draw[red, line width=2.5pt, opacity=0.4] (9) -- ($(9)!0.5!(4)$);
\draw[red, line width=2.5pt, opacity=0.4] (4) -- ($(4)!0.5!(9)$);
\draw[red, line width=2.5pt, opacity=0.4] (6) -- ($(6)!0.5!(1)$);
\draw[red, line width=2.5pt, opacity=0.4] (1) -- ($(1)!0.5!(6)$);
\draw[green, line width=2.5pt, opacity=0.4] (0) -- ($(0)!0.5!(6)$);
\draw[red, line width=2.5pt, opacity=0.4] (6) -- ($(6)!0.5!(0)$);
\draw[green, line width=2.5pt, opacity=0.4] (0) -- ($(0)!0.5!(9)$);
\draw[red, line width=2.5pt, opacity=0.4] (9) -- ($(9)!0.5!(0)$);
\draw[green, line width=2.5pt, opacity=0.4] (8) -- ($(8)!0.5!(9)$);
\draw[red, line width=2.5pt, opacity=0.4] (9) -- ($(9)!0.5!(8)$);
\draw[red, line width=2.5pt, opacity=0.4] (9) -- ($(9)!0.5!(10)$);
\draw[green, line width=2.5pt, opacity=0.4] (10) -- ($(10)!0.5!(9)$);
\draw[red, line width=2.5pt, opacity=0.4] (6) -- ($(6)!0.5!(10)$);
\draw[green, line width=2.5pt, opacity=0.4] (10) -- ($(10)!0.5!(6)$);
\draw[red, line width=2.5pt, opacity=0.4] (6) -- ($(6)!0.5!(7)$);
\draw[green, line width=2.5pt, opacity=0.4] (7) -- ($(7)!0.5!(6)$);
\draw[green, line width=2.5pt, opacity=0.4] (7) -- ($(7)!0.5!(2)$);
\draw[green, line width=2.5pt, opacity=0.4] (2) -- ($(2)!0.5!(7)$);
\end{pgfonlayer}
    
\end{tikzpicture}},\quad \scalebox{0.6}{\begin{tikzpicture}[scale=0.4]
    \begin{pgfonlayer}{nodelayer}
        \node [style=Z dot] (0) at (0.00, -0.00) {};
        \node [style=none] (1) at (3.50, -6.00) {};
        \node [style=none] (2) at (6.50, 3.00) {};
        \node [style=none] (3) at (0.00, 7.00) {};
        \node [style=none] (4) at (-6.50, 3.00) {};
        \node [style=none] (5) at (-3.50, -6.00) {};
        \node [style=Z dot] (6) at (2.50, -4.00) {};
        \node [style=Z dot] (7) at (4.50, 2.00) {};
        \node [style=Z dot] (8) at (0.00, 5.00) {};
        \node [style=Z dot] (9) at (-4.50, 2.00) {};
        \node [style=Z dot] (10) at (-2.50, -4.00) {};
        \node [style=none] (11) at (2.00, -1.00) {};
    \end{pgfonlayer}
    \begin{pgfonlayer}{edgelayer}
        \draw [style=hadamard edge] (0) to (6);
        \draw [style=hadamard edge] (0) to (7);
        \draw [style=hadamard edge] (0) to (8);
        \draw [style=hadamard edge] (0) to (9);
        \draw [style=hadamard edge] (0) to (10);
        \draw (0) to (11);
        \draw (1) to (6);
        \draw (2) to (7);
        \draw (3) to (8);
        \draw (4) to (9);
        \draw (5) to (10);
        \draw [style=hadamard edge] (6) to (10);
        \draw [style=hadamard edge] (6) to (7);
        \draw [style=hadamard edge] (7) to (8);
        \draw [style=hadamard edge] (8) to (9);
        \draw [style=hadamard edge] (9) to (10);
    \end{pgfonlayer}

    \begin{pgfonlayer}{edgelayer}
\draw[green, line width=2.5pt, opacity=0.4] (8) -- ($(8)!0.5!(3)$);
\draw[green, line width=2.5pt, opacity=0.4] (3) -- ($(3)!0.5!(8)$);
\draw[green, line width=2.5pt, opacity=0.4] (9) -- ($(9)!0.5!(4)$);
\draw[green, line width=2.5pt, opacity=0.4] (4) -- ($(4)!0.5!(9)$);
\draw[green, line width=2.5pt, opacity=0.4] (0) -- ($(0)!0.5!(7)$);
\draw[red, line width=2.5pt, opacity=0.4] (7) -- ($(7)!0.5!(0)$);
\draw[red, line width=2.5pt, opacity=0.4] (7) -- ($(7)!0.5!(8)$);
\draw[green, line width=2.5pt, opacity=0.4] (8) -- ($(8)!0.5!(7)$);
\draw[red, line width=2.5pt, opacity=0.4] (7) -- ($(7)!0.5!(2)$);
\draw[red, line width=2.5pt, opacity=0.4] (2) -- ($(2)!0.5!(7)$);
\draw[green, line width=2.5pt, opacity=0.4] (6) -- ($(6)!0.5!(7)$);
\draw[red, line width=2.5pt, opacity=0.4] (7) -- ($(7)!0.5!(6)$);
\draw[green, line width=2.5pt, opacity=0.4] (0) -- ($(0)!0.5!(10)$);
\draw[red, line width=2.5pt, opacity=0.4] (10) -- ($(10)!0.5!(0)$);
\draw[green, line width=2.5pt, opacity=0.4] (9) -- ($(9)!0.5!(10)$);
\draw[red, line width=2.5pt, opacity=0.4] (10) -- ($(10)!0.5!(9)$);
\draw[red, line width=2.5pt, opacity=0.4] (10) -- ($(10)!0.5!(5)$);
\draw[red, line width=2.5pt, opacity=0.4] (5) -- ($(5)!0.5!(10)$);
\draw[green, line width=2.5pt, opacity=0.4] (6) -- ($(6)!0.5!(10)$);
\draw[red, line width=2.5pt, opacity=0.4] (10) -- ($(10)!0.5!(6)$);
\end{pgfonlayer}
\end{tikzpicture}}, \\
& \underbrace{\scalebox{0.6}{\begin{tikzpicture}[scale=0.4]
    \begin{pgfonlayer}{nodelayer}
        \node [style=Z dot] (0) at (0.00, -0.00) {};
        \node [style=none] (1) at (3.50, -6.00) {};
        \node [style=none] (2) at (6.50, 3.00) {};
        \node [style=none] (3) at (0.00, 7.00) {};
        \node [style=none] (4) at (-6.50, 3.00) {};
        \node [style=none] (5) at (-3.50, -6.00) {};
        \node [style=Z dot] (6) at (2.50, -4.00) {};
        \node [style=Z dot] (7) at (4.50, 2.00) {};
        \node [style=Z dot] (8) at (0.00, 5.00) {};
        \node [style=Z dot] (9) at (-4.50, 2.00) {};
        \node [style=Z dot] (10) at (-2.50, -4.00) {};
        \node [style=none] (11) at (2.00, -1.00) {};
    \end{pgfonlayer}
    \begin{pgfonlayer}{edgelayer}
        \draw [style=hadamard edge] (0) to (6);
        \draw [style=hadamard edge] (0) to (7);
        \draw [style=hadamard edge] (0) to (8);
        \draw [style=hadamard edge] (0) to (9);
        \draw [style=hadamard edge] (0) to (10);
        \draw (0) to (11);
        \draw (1) to (6);
        \draw (2) to (7);
        \draw (3) to (8);
        \draw (4) to (9);
        \draw (5) to (10);
        \draw [style=hadamard edge] (6) to (10);
        \draw [style=hadamard edge] (6) to (7);
        \draw [style=hadamard edge] (7) to (8);
        \draw [style=hadamard edge] (8) to (9);
        \draw [style=hadamard edge] (9) to (10);
    \end{pgfonlayer}

    \begin{pgfonlayer}{edgelayer}
\draw[red, line width=2.5pt, opacity=0.4] (6) -- ($(6)!0.5!(1)$);
\draw[red, line width=2.5pt, opacity=0.4] (1) -- ($(1)!0.5!(6)$);
\draw[red, line width=2.5pt, opacity=0.4] (7) -- ($(7)!0.5!(2)$);
\draw[red, line width=2.5pt, opacity=0.4] (2) -- ($(2)!0.5!(7)$);
\draw[red, line width=2.5pt, opacity=0.4] (8) -- ($(8)!0.5!(3)$);
\draw[red, line width=2.5pt, opacity=0.4] (3) -- ($(3)!0.5!(8)$);
\draw[red, line width=2.5pt, opacity=0.4] (9) -- ($(9)!0.5!(4)$);
\draw[red, line width=2.5pt, opacity=0.4] (4) -- ($(4)!0.5!(9)$);
\draw[red, line width=2.5pt, opacity=0.4] (10) -- ($(10)!0.5!(5)$);
\draw[red, line width=2.5pt, opacity=0.4] (5) -- ($(5)!0.5!(10)$);
\draw[green, line width=2.5pt, opacity=0.4] (0) -- ($(0)!0.5!(11)$);
\draw[green, line width=2.5pt, opacity=0.4] (11) -- ($(11)!0.5!(0)$);
\draw[green, line width=2.5pt, opacity=0.4] (0) -- ($(0)!0.5!(6)$);
\draw[red, line width=2.5pt, opacity=0.4] (6) -- ($(6)!0.5!(0)$);
\draw[green, line width=2.5pt, opacity=0.4] (0) -- ($(0)!0.5!(7)$);
\draw[red, line width=2.5pt, opacity=0.4] (7) -- ($(7)!0.5!(0)$);
\draw[green, line width=2.5pt, opacity=0.4] (0) -- ($(0)!0.5!(8)$);
\draw[red, line width=2.5pt, opacity=0.4] (8) -- ($(8)!0.5!(0)$);
\draw[green, line width=2.5pt, opacity=0.4] (0) -- ($(0)!0.5!(9)$);
\draw[red, line width=2.5pt, opacity=0.4] (9) -- ($(9)!0.5!(0)$);
\draw[green, line width=2.5pt, opacity=0.4] (0) -- ($(0)!0.5!(10)$);
\draw[red, line width=2.5pt, opacity=0.4] (10) -- ($(10)!0.5!(0)$);
\draw[blue, line width=2.5pt, opacity=0.4] (6) -- ($(6)!0.5!(10)$);
\draw[blue, line width=2.5pt, opacity=0.4] (10) -- ($(10)!0.5!(6)$);
\draw[blue, line width=2.5pt, opacity=0.4] (6) -- ($(6)!0.5!(7)$);
\draw[blue, line width=2.5pt, opacity=0.4] (7) -- ($(7)!0.5!(6)$);
\draw[blue, line width=2.5pt, opacity=0.4] (7) -- ($(7)!0.5!(8)$);
\draw[blue, line width=2.5pt, opacity=0.4] (8) -- ($(8)!0.5!(7)$);
\draw[blue, line width=2.5pt, opacity=0.4] (8) -- ($(8)!0.5!(9)$);
\draw[blue, line width=2.5pt, opacity=0.4] (9) -- ($(9)!0.5!(8)$);
\draw[blue, line width=2.5pt, opacity=0.4] (9) -- ($(9)!0.5!(10)$);
\draw[blue, line width=2.5pt, opacity=0.4] (10) -- ($(10)!0.5!(9)$);
\end{pgfonlayer}
    
\end{tikzpicture}}}_{XXXXX\otimes Z},\quad \underbrace{\scalebox{0.6}{\begin{tikzpicture}[scale=0.4]
    \begin{pgfonlayer}{nodelayer}
        \node [style=Z dot] (0) at (0.00, -0.00) {};
        \node [style=none] (1) at (3.50, -6.00) {};
        \node [style=none] (2) at (6.50, 3.00) {};
        \node [style=none] (3) at (0.00, 7.00) {};
        \node [style=none] (4) at (-6.50, 3.00) {};
        \node [style=none] (5) at (-3.50, -6.00) {};
        \node [style=Z dot] (6) at (2.50, -4.00) {};
        \node [style=Z dot] (7) at (4.50, 2.00) {};
        \node [style=Z dot] (8) at (0.00, 5.00) {};
        \node [style=Z dot] (9) at (-4.50, 2.00) {};
        \node [style=Z dot] (10) at (-2.50, -4.00) {};
        \node [style=none] (11) at (2.00, -1.00) {};
    \end{pgfonlayer}
    \begin{pgfonlayer}{edgelayer}
        \draw [style=hadamard edge] (0) to (6);
        \draw [style=hadamard edge] (0) to (7);
        \draw [style=hadamard edge] (0) to (8);
        \draw [style=hadamard edge] (0) to (9);
        \draw [style=hadamard edge] (0) to (10);
        \draw (0) to (11);
        \draw (1) to (6);
        \draw (2) to (7);
        \draw (3) to (8);
        \draw (4) to (9);
        \draw (5) to (10);
        \draw [style=hadamard edge] (6) to (10);
        \draw [style=hadamard edge] (6) to (7);
        \draw [style=hadamard edge] (7) to (8);
        \draw [style=hadamard edge] (8) to (9);
        \draw [style=hadamard edge] (9) to (10);
    \end{pgfonlayer}

    \begin{pgfonlayer}{edgelayer}
\draw[green, line width=2.5pt, opacity=0.4] (9) -- ($(9)!0.5!(4)$);
\draw[green, line width=2.5pt, opacity=0.4] (4) -- ($(4)!0.5!(9)$);
\draw[red, line width=2.5pt, opacity=0.4] (0) -- ($(0)!0.5!(6)$);
\draw[green, line width=2.5pt, opacity=0.4] (6) -- ($(6)!0.5!(0)$);
\draw[red, line width=2.5pt, opacity=0.4] (0) -- ($(0)!0.5!(9)$);
\draw[green, line width=2.5pt, opacity=0.4] (9) -- ($(9)!0.5!(0)$);
\draw[green, line width=2.5pt, opacity=0.4] (7) -- ($(7)!0.5!(2)$);
\draw[green, line width=2.5pt, opacity=0.4] (2) -- ($(2)!0.5!(7)$);
\draw[green, line width=2.5pt, opacity=0.4] (8) -- ($(8)!0.5!(3)$);
\draw[green, line width=2.5pt, opacity=0.4] (3) -- ($(3)!0.5!(8)$);
\draw[red, line width=2.5pt, opacity=0.4] (0) -- ($(0)!0.5!(11)$);
\draw[red, line width=2.5pt, opacity=0.4] (11) -- ($(11)!0.5!(0)$);
\draw[red, line width=2.5pt, opacity=0.4] (0) -- ($(0)!0.5!(7)$);
\draw[green, line width=2.5pt, opacity=0.4] (7) -- ($(7)!0.5!(0)$);
\draw[red, line width=2.5pt, opacity=0.4] (0) -- ($(0)!0.5!(8)$);
\draw[green, line width=2.5pt, opacity=0.4] (8) -- ($(8)!0.5!(0)$);
\draw[red, line width=2.5pt, opacity=0.4] (0) -- ($(0)!0.5!(10)$);
\draw[green, line width=2.5pt, opacity=0.4] (10) -- ($(10)!0.5!(0)$);
\draw[green, line width=2.5pt, opacity=0.4] (6) -- ($(6)!0.5!(1)$);
\draw[green, line width=2.5pt, opacity=0.4] (1) -- ($(1)!0.5!(6)$);
\draw[green, line width=2.5pt, opacity=0.4] (10) -- ($(10)!0.5!(5)$);
\draw[green, line width=2.5pt, opacity=0.4] (5) -- ($(5)!0.5!(10)$);
\end{pgfonlayer}
\end{tikzpicture}}}_{ZZZZZ\otimes X}
\;\Bigg\rangle \ .
\end{split}
\end{equation}
Where the {\color{red!40}R}{\color{green!40}G}{\color{blue!40}B} coloured Pauli webs represent Pauli-${\color{red!40}X}{\color{green!40}Z}{\color{blue!40}Y}$ operators respectively\footnote{\textbf{Not in the usual $XYZ$ = RGB.}}. Alternatively, you can view the ZX-diagram in equation \ref{eq:pent_l2} (since only-connectivity-matters) as a state with 6 qubits, where all the bulk and boundary qubit legs point to the right,
\begin{equation}
    \label{eq:6qb_bell_State}
    \scalebox{1.2}{\begin{tikzpicture}
    \begin{pgfonlayer}{nodelayer}
        \node [style=none] (0) at (4.00, -1.25) {};
        \node [style=Z dot] (1) at (1.75, 1.75) {};
        \node [style=none] (2) at (4.00, 1.75) {};
        \node [style=none] (3) at (4.00, -0.25) {};
        \node [style=Z dot] (4) at (2.00, 0.75) {};
        \node [style=none] (5) at (4.00, -2.25) {};
        \node [style=Z dot] (6) at (2.00, -0.25) {};
        \node [style=Z dot] (7) at (2.00, -1.25) {};
        \node [style=Z dot] (8) at (-1.50, 2.75) {};
        \node [style=none] (9) at (4.00, 0.75) {};
        \node [style=Z dot] (10) at (-1.50, -2.25) {};
        \node [style=none] (11) at (4.00, 2.75) {};
    \end{pgfonlayer}
    \begin{pgfonlayer}{edgelayer}
        \draw (0) to (7);
        \draw [style=hadamard edge] (1) to (8);
        \draw (1) to (2);
        \draw [style=hadamard edge] (1) to (4);
        \draw [style=hadamard edge] (1) to (10);
        \draw (3) to (6);
        \draw [style=hadamard edge] (4) to (6);
        \draw (4) to (9);
        \draw [style=hadamard edge] (4) to (10);
        \draw (5) to (10);
        \draw [style=hadamard edge] (6) to (7);
        \draw [style=hadamard edge] (6) to (10);
        \draw [style=hadamard edge] (7) to (8);
        \draw [style=hadamard edge] (7) to (10);
        \draw [style=hadamard edge] (8) to (10);
        \draw (8) to (11);
    \end{pgfonlayer}
\end{tikzpicture}} \ .
\end{equation}

\begin{figure}[!h]
    \centering
    \includegraphics[width=0.9\linewidth]{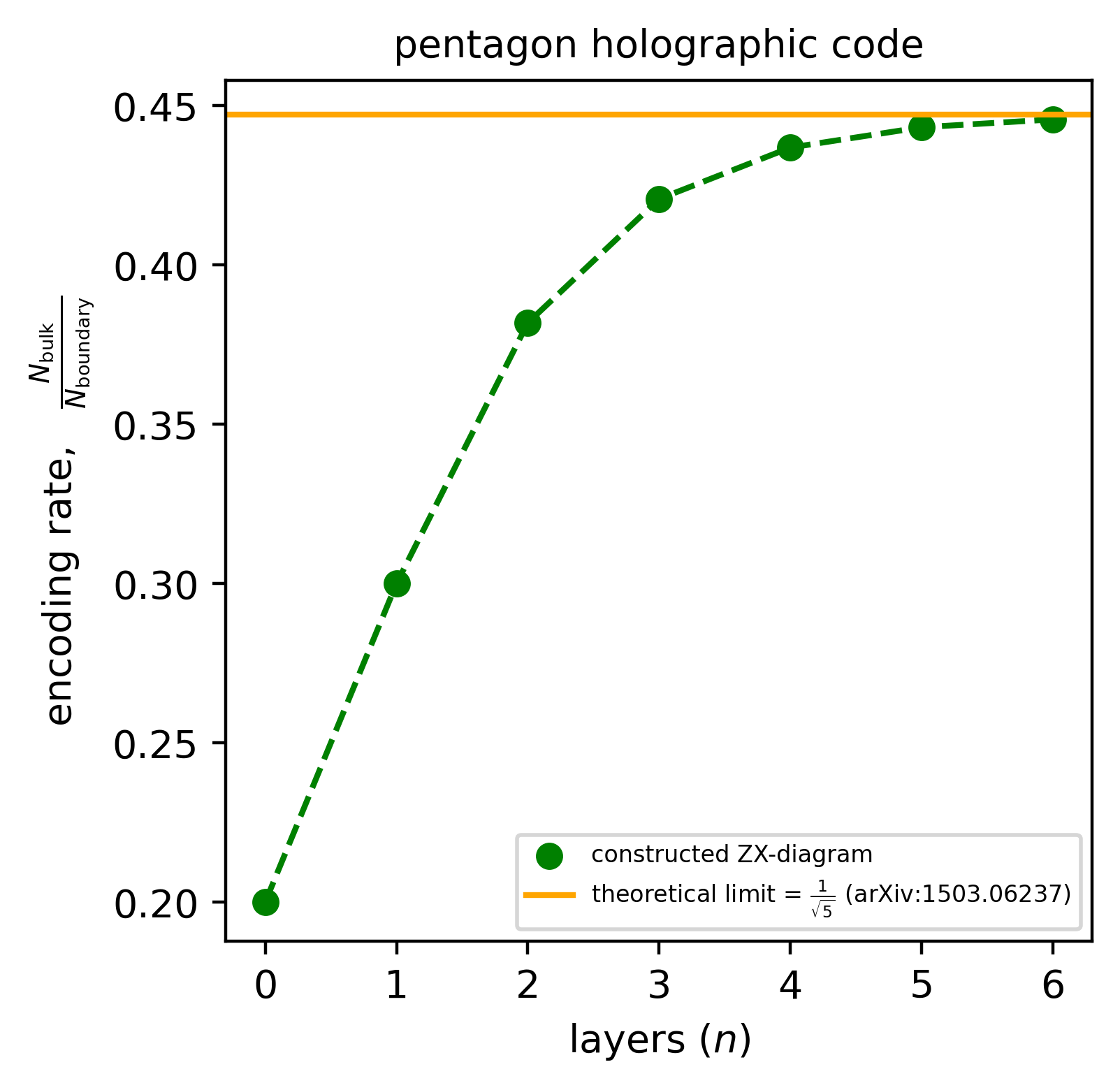}
    \caption{Encoding rate of the ZX-diagram equivalent to the pentagon holographic code.}
    \label{fig:encoding_rate_1_sqrt_5}
\end{figure}

Equipped with this knowledge, we can construct the pentagon holographic code completely as a ZX-diagram using 6-qubit graph state chunks, as shown in the LHS of equation \ref{eq:pent_l4_edited}. Firstly, as a sanity check, one can verify the ratio between the number of boundary and bulk qubits in our ZX-diagram, they should scale as:
\begin{equation}
    \frac{N_{\text{bulk}}}{N_{\text{boundary}}} \xrightarrow[]{n\rightarrow \infty} \frac{1}{\sqrt{5}} \ ,
\end{equation}
for increasing number of layers \cite{Pastawski:2015qua}, see figure \ref{fig:encoding_rate_1_sqrt_5}.
Explicitly, the total number of bulk and boundary legs for a tessellation of $n$ layers should be \cite{Pastawski:2015qua}:
\begin{equation}
    \begin{split}
        N_{\text{boundary}} & = 4f_n + 3g_n \\
        N_{\text{bulk}} & = 1+\sum_{k=1}^{n} (f_k + g_k)
    \end{split} \ ,
\end{equation}
for $n\geq 1$. The quantities $f_n$ and $g_n$ are computed by:
\begin{equation}
    \begin{split}
        \begin{pmatrix}
            f_n \\ g_n
        \end{pmatrix} = \begin{pmatrix}
            2 & 1 \\ 1 & 1
        \end{pmatrix}^{n-1}\begin{pmatrix}
            5 \\ 0
        \end{pmatrix} \ \text{ for $n\geq 1$} \ .
    \end{split}
\end{equation}
For $n=0$, $N_\text{bulk} = 1$ and $N_{\text{boundary}} = 5$.

\begin{figure}[!h]
    \centering
    \includegraphics[width=0.8\linewidth]{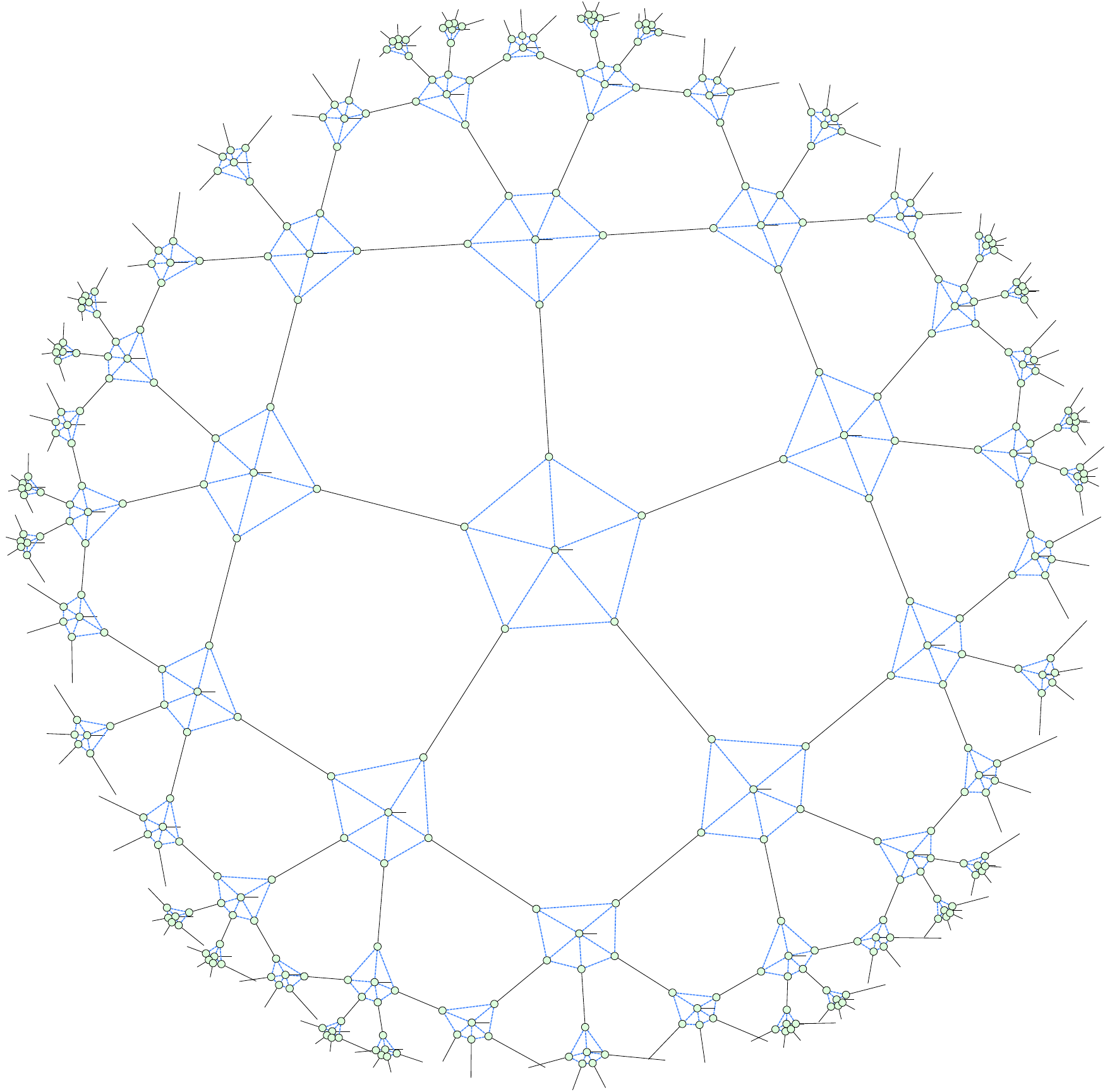}
    \caption{A pentagon holographic code ZX-diagram with $n=3$ layers.}
    \label{fig:pent_l5}
\end{figure}

Since we have written the pentagon holographic code as a ZX-diagram in $\mathtt{pyzx}$ \cite{kissinger2020Pyzx}, we can apply a whole host of tools to study it. We have added most of the ZX-diagram as .tikz files in the arXiv tex source, in case anyone wishes to look at them with $\mathtt{zxlive}$ \cite{zxcalc_zxlive}. For example, figure \ref{fig:pent_l5} shows a $n=3$ layers pentagon holographic code written as a ZX-diagram.

We can interpret the ZX-diagram for the pentagon holographic code as the encoding map $V$, mapping the bulk legs (as input(s)) to the boundary legs (as outputs). Conversely, if we switch the labelling of input and output, we have described the inverse map $V^{\dagger}$. Since this ZX-diagram is phase-less and only connectivity matters in ZX-diagram \cite{KissingerWetering2024Book}, we can call all the bulk and boundary legs as outputs (with no inputs). This describes Bell states between the bulk qubit leg and its logical encoded qubit on the boundary of the pentagon holographic code, on all the bulk qubits:
\begin{equation}
    \label{eq:bell_all}
    \ket{\Phi} \propto \bigotimes_{i\in \text{logical}} \big(\ket{\bar{0}}_{i}\ket{0}_{\text{bulk}_i} + \ket{\bar{1}}_{i}\ket{1}_{\text{bulk}_i}\big) \ .
\end{equation}
In other words, we can twist and bend the input and output legs as we wish.

\subsection{What can we do with the ZX-diagram?}
Now that we have the ZX-diagrams written down, what can we do with them? We can,
\begin{enumerate}
    \item draw Pauli webs to study its stabilisers diagrammatically,
    \item generate the pentagon holographic code's parity check matrix from its webs, run simulations for sanity check,
    \item compute entanglement entropy via ZX-calculus related simplifications,
    \item re-visit the black-hole and wormhole constructions from section 6 of \cite{Pastawski:2015qua},
\end{enumerate}
The above list will serve as the outline for the remainder of this manuscript on pentagon holographic codes, which we explore next.

\section{Stabiliser and logical operators of the pentagon holographic code}
Pauli webs are graphical annotations on ZX-diagrams that track the propagation of Pauli operators through (typically Clifford) ZX-calculus constructions, showcasing the stabiliser and logical operator structure. In $\mathtt{pyzx}$, Pauli webs (and related stabiliser information) can be generated automatically, see \cite{kissinger2020Pyzx, rüsch2025completenessfaultequivalenceclifford, tqec, mitosek2024algebraicinterpretationpauliflow, YehZXNormalForms} for details and alternative formalisms or generations. In the context of holographic tensor network codes, stabilisers and logical operators can also be obtained via operator-pushing routines such as $\mathtt{OperatorPush}$ in \cite{fan2025legohqecsoftwaretoolanalyzing}.

Let $V$ denote the encoding isometry represented by a holographic code ZX-diagram, mapping bulk (logical) degrees of freedom to boundary (physical) qubits from equation \ref{eq:V_iso}. A stabiliser of the code is a purely boundary supported Pauli operator $S_\partial$ that acts trivially on the encoded subspace, it satisfies:
\begin{equation}
    S_\partial\, V = V \ ,
\end{equation}
up to an overall phase. Graphically, such stabilisers appear as Pauli webs with support only on the boundary legs. For example, a stabiliser web of the code is
\begin{equation}
    \label{eq:sw_n_2_0}
    \scalebox{0.2}{\input{sw_n_2_0.tikz}} \ .
\end{equation}
Reading off the Pauli colouring ${\color{red!40}X}{\color{green!40}Z}{\color{blue!40}Y}$ on the boundary legs yields a Pauli string (e.g.\ $XXIZIXIZ\cdots$) and, upon converting to binary symplectic form, a row of the code's parity-check matrix \cite{gottesman1998heisenbergrepresentationquantumcomputers}.

Similarly, a Pauli web with support on the bulk leg corresponds to a pushed bulk Pauli operator: it gives a boundary representative $L_\partial$ of a logical operator $\bar L$ of the code, such that
\begin{equation}
    L_\partial\, V = V\, \bar L \ ,
\end{equation}
up to an overall phase. Boundary representatives of logical operators are defined up to multiplication by boundary stabilisers. For instance a logical web is:
\begin{equation}
    \label{eq:lw_n_2_0}
    \scalebox{0.2}{\input{lw_n_2_0.tikz}} \ .
\end{equation}
By reading the boundary colouring, we obtain a boundary Pauli string implementing $\bar{X}$ or $\bar{Z}$ for the corresponding bulk qubit (here, the central bulk qubit), up to multiplication(s) of stabiliser(s) of the code.

In our constructions we use modified variants of the operator-pushing approach of \cite{fan2025legohqecsoftwaretoolanalyzing} adapted to our ZX-diagram encoders to acquire stabilisers and logical operators of the holographic codes. With these stabiliser and logical representatives in hand, we next study recovery of the encoded information in the presence of erasures.

\subsection{Erasure decoding}
In \cite{Pastawski:2015qua}, the authors studied the logical recovery probability of the central qubit in the pentagon holographic code under an erasure error model with rate $p$ applied to each boundary qubit, using their greedy decoder; the results are shown in figure 26a of the arXiv version. Using the generated Pauli webs, we constructed a parity check matrix for the pentagon holographic code. We then applied the same erasure error model with probability $p$ on each boundary qubit and decoded the syndromes using a similar approach known as the peeling decoder (see \cite{Connolly_2024} for a review). This allowed us to reproduce an equivalent of figure 26a from \cite{Pastawski:2015qua}.
\begin{figure}[!h]
\centering
\includegraphics[width=0.95\linewidth]{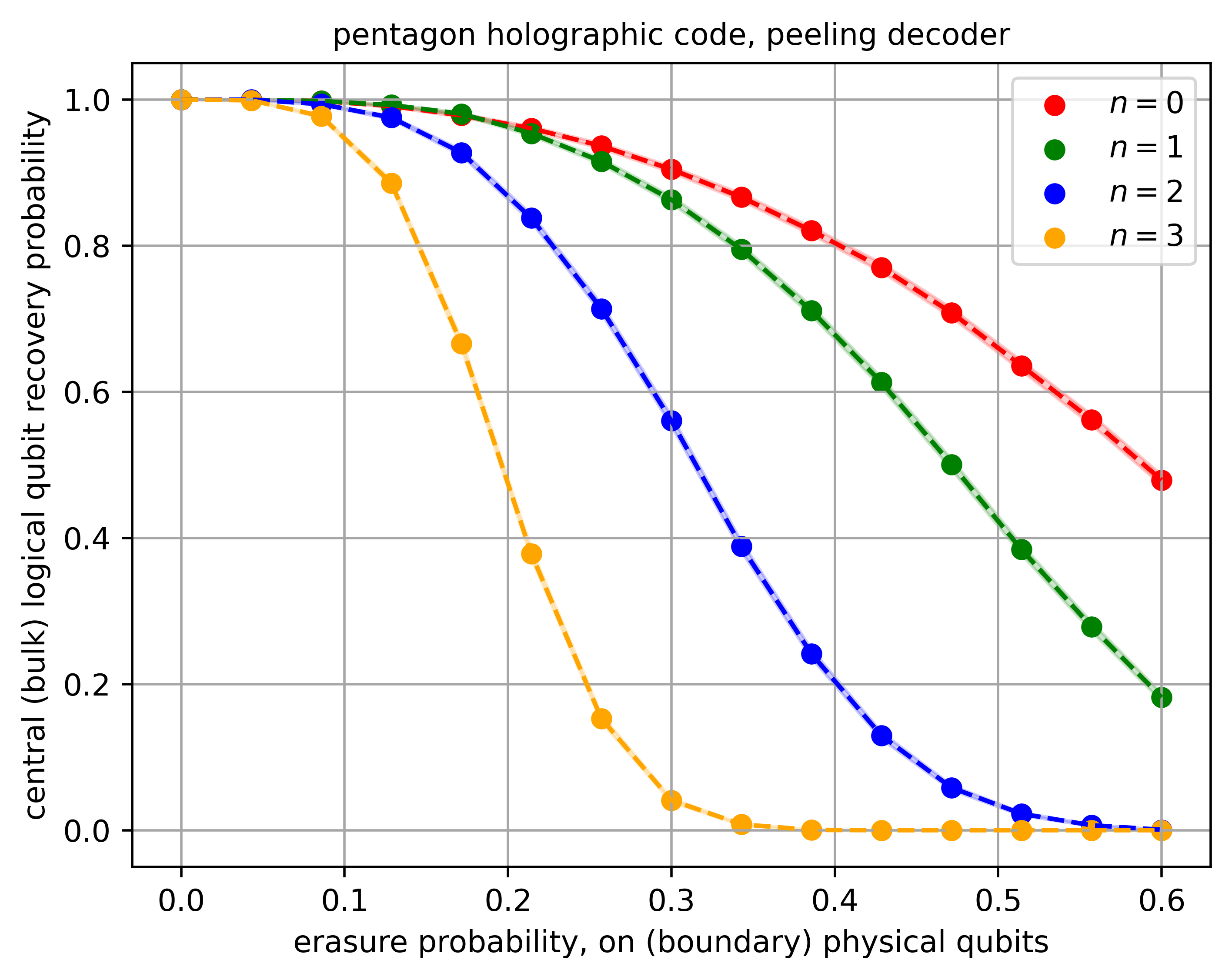}
\caption{Logical recovery probability of the central bulk qubit under an erasure error model, decoded with the peeling decoder.}
\label{fig:central_qb_recovery_peeling}
\end{figure}
The resulting curve shows a similar shape to figure 26a of \cite{Pastawski:2015qua}, providing reassurance. 

We now move on to studying the quantum correlations in these ZX-diagrams, which can be quantified using entanglement measures.

\section{Computing entanglement entropy, Ryu-Takayanagi formulae}
The holographic principle \cite{tHooft:1993dmi,Susskind:1994vu} relates a $d$-dimensional (bulk) quantum gravity theory to a $d-1$ dimensional (boundary) quantum field theory. The most concrete example of holography takes the form of the $\text{AdS}_d/\text{CFT}_{d-1}$ correspondence \cite{Maldacena:1997re,Gubser:1998bc,Witten:1998qj}. This correspondence allows us to study quantum gravity from the much more familiar setting of field theories. Following the development of the Ryu-Takayanagi (RT) formula \cite{Ryu:2006bv}, the correspondence takes a more geometric form where the entropy of a boundary subregion is related to certain minimal surfaces in the bulk spacetime. From this picture, the transition to tensor networks seems very natural where the minimal surfaces can be interpreted as the size of the minimal cut through the graph \cite{Hayden:2016cfa}. 

The RT formula also makes clear the connection between gravity and quantum information. Being a subregion duality, the RT formula implies that bulk degrees of freedom are redundantly encoded in the boundary theory. This is because the bulk state can be among overlapping regions of the entanglement wedge. This redundancy of bulk information stored in the boundary allows us to interpret it as a quantum error correcting code \cite{Almheiri:2014lwa} where states in the bulk nearer to the boundary are more susceptible to being erased than states deeper inside the bulk. A concrete realization of such a holographic quantum error-correcting code is provided by the pentagon holographic code \cite{Pastawski:2015qua}.

\subsection{Bipartite entanglement entropy via minimal cuts in the pentagon holographic codes}
In a pentagon holographic tensor network (see RHS of equation \ref{eq:pent_l4_edited}), each bulk qubit is represented as a perfect tensor connected to other tensors and the boundary via contracted indices. To compute the bipartite entanglement entropy between a boundary region $A$ and its complement $A^c$, one can follow a discrete analogue of the Ryu-Takayanagi prescription \cite{Pastawski:2015qua,Harlow_2017}:
\begin{enumerate}
    \item Select a boundary region $A$ and denote its complement by $A^c$.
    \item Consider a cut $c$ through the network that separates it into two disjoint chunks of tensors, $P$ and $Q$, such that the uncontracted legs of $P$ correspond exactly to $A$ and those of $Q$ to $A^c$. The cut $c$ intersects a set of edges (tensor legs) connecting $P$ and $Q$, and the number of such edges is called the \textit{length} of the cut, $|c|$.
    \item Identify the \textit{minimal cut} $\gamma_A$, i.e., the cut of shortest length whose boundary matches $A$. This cut realises a bipartition of the network into $P$ and $Q$, and the entanglement entropy of the bipartition is then:
    \begin{equation*}
        S(A:A^c) = |\gamma_A| \cdot \log_2 \chi \ ,
    \end{equation*}
    where $\chi$ is the Hilbert space dimension of each leg (for qubits, $\chi=2$).
\end{enumerate}

Intuitively, each edge crossing the minimal cut $\gamma_A$ carries one unit of bipartite entanglement (one `ebit'), so counting them gives a discrete realization of the Ryu-Takayanagi formula:
\begin{equation}
S(A:A^c) = \begin{pmatrix}
    \text{size of minimal cut} \\ \text{through the tensor network}
\end{pmatrix}.
\end{equation}
Counting the edges along $\gamma_A$ gives an exact value for the bipartite entanglement entropy of the corresponding boundary state (obtained by fixing/contracting the bulk legs), as in the pentagon holographic code.

\subsection{Why use ZX-calculus?}
A central problem of interest is the computation and visualisation of entanglement entropy between boundary regions. While tensor networks provide a natural geometric and numerical picture, ZX-calculus offers a more algebraically transparent graphical language for analysing such networks. A ZX-diagram encodes an explicit structure for the tensors, representing the pentagon holographic tensor network in a particular eigenbasis\footnote{All of ZX-calculus can be interpreted as a convenient representation of \textit{some} tensor networks \cite{Magdalena_de_la_Fuente_2025}.}. In addition, ZX rewrite rules \cite{KissingerWetering2024Book} (such as spider fusion, bi-algebra, and Hopf rule etc) allow large portions of the network to be systematically/intuitively simplified in calculations, something that is not easily accessible using purely geometric tensor-network manipulations. These allow entanglement structures to be analysed diagrammatically\footnote{See \cite{masmendoza2025graphicaldiagnostictopologicalorder} for a graphical treatment of topological codes and entanglement entropy via ZX.}. Let us work through two small examples now, computing entanglement entropy via ZX-calculus.

\subsection{Computing entanglement entropy between bulk-boundary as a scalar ZX-diagram}
In this subsection, we aim to compute the entanglement entropy of a section of the boundary qubits ($q
\in B$) with the rest of the \textbf{bulk and boundary qubits}. It is important to emphasise that the Ryu–Takayanagi formula computes entanglement between complementary regions of the boundary theory, not correlations between individual boundary and bulk degrees of freedom, as included in the following computation. 

Before computing the entanglement entropy of states in the pentagon holographic code, let us first review how to construct density operators and perform partial traces in ZX-calculus. A state $\ket{\psi}$ represented by a ZX-graph $g$ can be written as \cite{KissingerWetering2024Book,Dowling_2025}:
\begin{equation}
    \label{eq:zx_state}
    \ket{\psi} \propto \scalebox{1}{\begin{tikzpicture}[yshift=-6cm]
    \begin{pgfonlayer}{nodelayer}
        \node [style=Z dot] (0) at (4.25, 4.50) {};
        \node [style=none] (1) at (6.75, 6.00) {};
        \node [style=X dot] (3) at (4.25, 6.00) {};
        \node [style=none] (4) at (4.75, 4.25) {};
        \node [style=Z phase dot] (5) at (5.50, 5.75) {$b$};
        \node [style=X dot] (6) at (4.25, 7.00) {};
        \node [style=none] (13) at (6.75, 4.50) {};
        \node [style=none] (15) at (6.75, 7.00) {};
        \node [style=none] (17) at (6.25, 7.25) {};
    \end{pgfonlayer}
    \draw[fill=white, line width = .5pt] (4) rectangle (17);
    \node [style=none] (g) at (5.50, 5.75) {$g$};

    \begin{pgfonlayer}{edgelayer}
        \draw (0) to (13);
        \draw (1) to (3);
        \draw (6) to (15);
    \end{pgfonlayer}
\end{tikzpicture}} \ ,
\end{equation}
up to normalisation constant. The corresponding density operator, $\rho=\ketbra{\psi}{\psi}$, is obtained by taking the outer product of the state with its dual. In ZX-calculus, this amounts to ``combining'' the diagram with its conjugate ($g^{\dagger}$):
\begin{equation}
    \label{eq:zx_rho}
    \rho \propto \scalebox{1}{\begin{tikzpicture}[yshift = -6cm]
    \begin{pgfonlayer}{nodelayer}
        \node [style=none] (0) at (0.25, 4.50) {};
        \node [style=none] (5) at (0.25, 6.00) {};
        \node [style=none] (6) at (0.25, 7.00) {};
        \node [style=none] (7) at (0.750, 7.25) {};
        \node [style=none] (8) at (2.25, 4.25) {};
        \node [style=none] (9) at (4.75, 4.25) {};
        \node [style=none] (10) at (6.25, 7.25) {};
        \node [style=X dot] (11) at (2.75, 7.00) {};
        \node [style=X dot] (12) at (2.75, 6.00) {};
        \node [style=Z dot] (13) at (2.75, 4.50) {};
        \node [style=Z dot] (14) at (4.25, 4.50) {};
        \node [style=X dot] (15) at (4.25, 6.00) {};
        \node [style=none] (16) at (6.75, 7.00) {};
        \node [style=X dot] (17) at (4.25, 7.00) {};
        \node [style=none] (18) at (6.75, 6.00) {};
        \node [style=none] (19) at (6.75, 4.50) {};

    \end{pgfonlayer}
    \draw[fill=white, line width = .5pt] (7) rectangle (8);
    \draw[fill=white, line width = .5pt] (9) rectangle (10);
    \node [style=none] (g) at (1.50, 5.75) {$g^{\dagger}$};
    \node [style=none] (gdg) at (5.50, 5.75) {$g$};
    \begin{pgfonlayer}{edgelayer}
        \draw (0) to (13);
        \draw (5) to (12);
        \draw (6) to (11);
        \draw (14) to (19);
        \draw (15) to (18);
        \draw (16) to (17);
    \end{pgfonlayer}
\end{tikzpicture}} \ .
\end{equation}
For example, partially tracing out the top most qubit sub-system amounts to joining the top-most input and output legs together in equation \ref{eq:zx_rho}, namely:
\begin{equation}
    \label{eq:zx_rho_pt}
    \text{Tr}_1(\rho) = \scalebox{1}{\begin{tikzpicture}[yshift = -6cm]
    \begin{pgfonlayer}{nodelayer}
        \node [style=none] (0) at (0.25, 4.50) {};
        \node [style=none] (5) at (0.25, 6.00) {};
        \node [style=Z dot] (6) at (0.25, 7.00) {};
        \node [style=none] (7) at (0.750, 7.25) {};
        \node [style=none] (8) at (2.25, 4.25) {};
        \node [style=none] (9) at (4.75, 4.25) {};
        \node [style=none] (10) at (6.25, 7.25) {};
        \node [style=X dot] (11) at (2.75, 7.00) {};
        \node [style=X dot] (12) at (2.75, 6.00) {};
        \node [style=Z dot] (13) at (2.75, 4.50) {};
        \node [style=Z dot] (14) at (4.25, 4.50) {};
        \node [style=X dot] (15) at (4.25, 6.00) {};
        \node [style=Z dot] (16) at (6.75, 7.00) {};
        \node [style=X dot] (17) at (4.25, 7.00) {};
        \node [style=none] (18) at (6.75, 6.00) {};
        \node [style=none] (19) at (6.75, 4.50) {};

    \end{pgfonlayer}
    \draw[fill=white, line width = .5pt] (7) rectangle (8);
    \draw[fill=white, line width = .5pt] (9) rectangle (10);
    \node [style=none] (g) at (1.50, 5.75) {$g^{\dagger}$};
    \node [style=none] (gdg) at (5.50, 5.75) {$g$};
    \begin{pgfonlayer}{edgelayer}
        \draw (0) to (13);
        \draw (5) to (12);
        \draw (6) to (11);
        \draw (14) to (19);
        \draw (15) to (18);
        \draw (16) to (17);
        \draw (6) -- ($(6)+(0,1)$) -- ($(16)+(0,1)$) -- (16);
    \end{pgfonlayer}
\end{tikzpicture}} \ .
\end{equation}
With this in mind, we can compute the entanglement entropy of the holographic code using ZX-calculus. If we were to interpret an entire ZX-diagram (e.g. a $n=2$ pentagon holographic code isometry here) as a state $\ket{\Phi}$ with all the free (bulk and boundary) legs, we can effectively write this state as:
\begin{equation}
    \label{eq:phi_n_2}
    \ket{\Phi} =
    \raisebox{-17ex}{\scalebox{0.4}{\includegraphics{nontikz_figures/pent_l4_edited.pdf}}} 
\end{equation}
We can compute the reduced density matrix: $\rho_B=\text{Tr}_{B'}(\ketbra{\Phi}{\Phi})$ pertaining to regions $B$\footnote{$B'$ is any qubits/free legs not in $B$, bulk and boundary.}, labelled with qubits with dotted magenta circles below:
\begin{equation}
    \raisebox{-17ex}{\scalebox{0.4}{\includegraphics{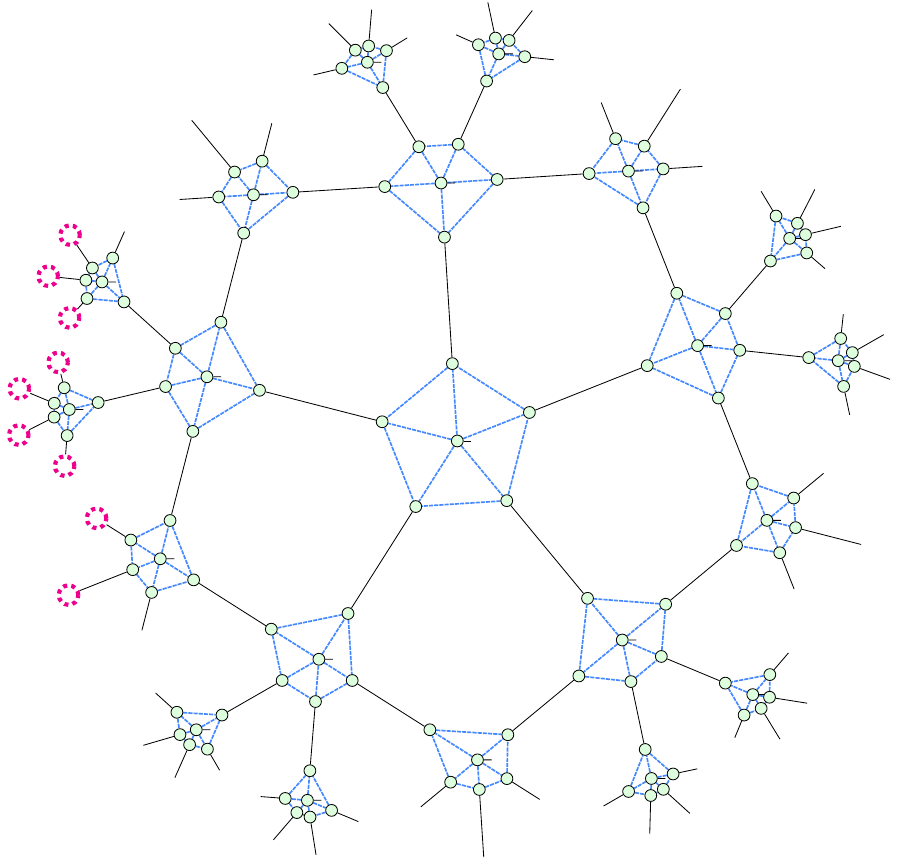}}} \ ,
\end{equation}
in ZX-calculus, the reduced density matrix can be formed by contracting all the other legs not in $B$, with another `bra' $\bra{\Phi}$ copy of itself \cite{KissingerWetering2024Book,Dowling_2025,masmendoza2025graphicaldiagnostictopologicalorder}.
\begin{equation}
    \rho_B = C \cdot \raisebox{-6ex}{\scalebox{0.18}{\includegraphics{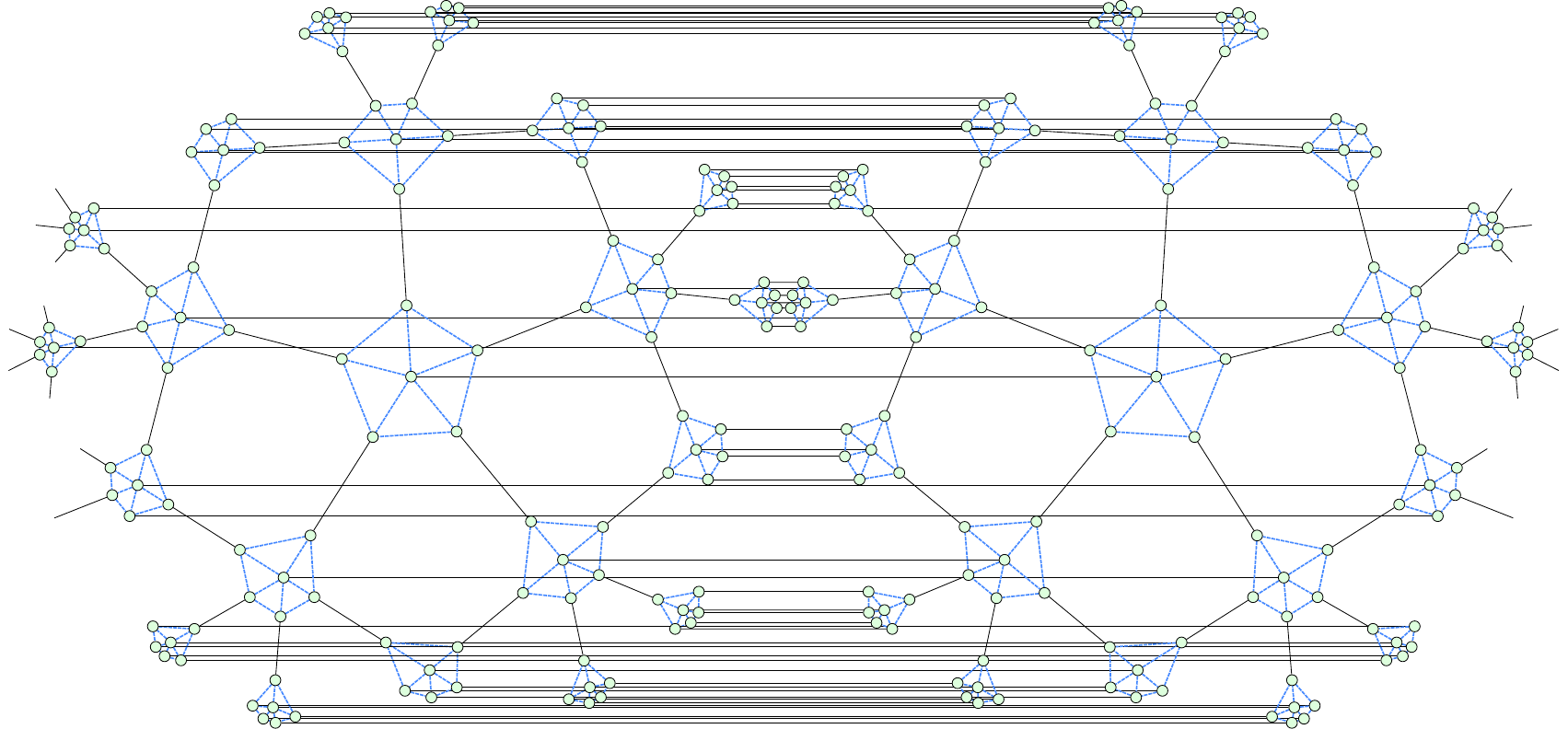}}} \ ,
\end{equation}
where the normalisation constant $C^{-1}$ is the ZX-diagram with all legs fully contracted and
\begin{equation}
    \frac{1}{C}= \raisebox{-6ex}{\scalebox{0.3}{\includegraphics{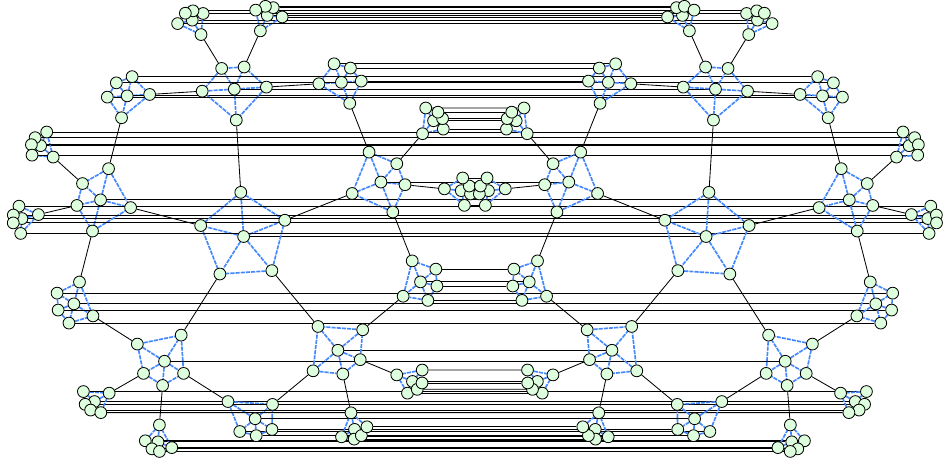}}} \ ,
\end{equation}
In this case, $C = \sqrt{2}^{218}$, where the exponent counts the total number of contracted legs (equivalently, Bell-pair normalisation factors) in the fully closed ZX-diagram defining $C$. Note that scalar ZX-diagrams have the following values:
\begin{equation}
    \begin{split}
       \scalebox{1}{
\begin{tikzpicture}
    \begin{pgfonlayer}{nodelayer}
        \node [style=Z phase dot] (0) at (0.00, -0.00) {$\beta$};
    \end{pgfonlayer}
    \begin{pgfonlayer}{edgelayer}
    \end{pgfonlayer}
\end{tikzpicture}
} & =  1+e^{i\beta}\\
       \scalebox{1}{
\begin{tikzpicture}
    \begin{pgfonlayer}{nodelayer}
        \node [style=Z phase dot] (1) at (-1.00, -0.00) {$\beta$};
        \node [style=X phase dot] (2) at (1.00, -0.00) {$\gamma$};
    \end{pgfonlayer}
    \begin{pgfonlayer}{edgelayer}
        \draw (1) to (2);
    \end{pgfonlayer}
\end{tikzpicture}
} & = \frac{1}{\sqrt{2}}\big( 1+ e^{i\beta}+ e^{i\gamma}- e^{i(\beta+\gamma)})
    \end{split} \ .
\end{equation} 
We can then perform the standard ZX-contraction and reduction of the Clifford ZX-diagram corresponding to $\rho_B$, we arrive at:
\begin{equation}
    \label{eq:rho_b}
    \rho_B = \sqrt{2}^{-2}\scalebox{0.8}{\begin{tikzpicture}[yshift = +4cm]
    \begin{pgfonlayer}{nodelayer}
        \node [style=none] (1) at (0.00, -5.00) {};
        \node [style=none] (2) at (8.00, -4.00) {};
        \node [style=none] (3) at (8.00, -8.00) {};
        \node [style=Z dot] (4) at (7.00, -4.00) {};
        \node [style=none] (5) at (0.00, -3.00) {};
        \node [style=none] (6) at (0.00, -1.00) {};
        \node [style=none] (7) at (8.00, -1.00) {};
        \node [style=none] (8) at (8.00, -5.00) {};
        \node [style=Z dot] (9) at (1.75, -3.00) {};
        \node [style=none] (10) at (0.00, -7.00) {};
        \node [style=none] (11) at (0.00, -8.00) {};
        \node [style=none] (12) at (0.00, -4.00) {};
        \node [style=none] (13) at (8.00, -6.00) {};
        \node [style=Z dot] (14) at (1.75, -6.00) {};
        \node [style=X phase dot] (15) at (7.00, -5.00) {$\pi$};
        \node [style=none] (16) at (0.00, -2.00) {};
        \node [style=none] (17) at (8.00, -3.00) {};
        \node [style=none] (18) at (8.00, -0.00) {};
        \node [style=none] (19) at (8.00, -7.00) {};
        \node [style=none] (20) at (8.00, -2.00) {};
        \node [style=X dot] (21) at (1.00, -5.00) {};
        \node [style=X dot] (22) at (1.00, -4.00) {};
        \node [style=X dot] (23) at (6.50, -6.00) {};
        \node [style=none] (24) at (0.00, -0.00) {};
        \node [style=none] (25) at (0.00, -6.00) {};
        \node [style=Z phase dot] (26) at (6.50, -3.00) {$\pi$};
    \end{pgfonlayer}
    \begin{pgfonlayer}{edgelayer}
        \draw (1) to (21);
        \draw (2) to (4);
        \draw (3) to (11);
        \draw (4) to (15);
        \draw (4) to (21);
        \draw (4) to (22);
        \draw (5) to (9);
        \draw (6) to (7);
        \draw (8) to (15);
        \draw (9) to (23);
        \draw [style=hadamard edge] (9) to (26);
        \draw [style=hadamard edge] (9) to (14);
        \draw (9) to (15);
        \draw (9) to (22);
        \draw (10) to (19);
        \draw (12) to (22);
        \draw (13) to (23);
        \draw (14) to (23);
        \draw [style=hadamard edge] (14) to (26);
        \draw (14) to (25);
        \draw (14) to (15);
        \draw (14) to (21);
        \draw [style=hadamard edge] (15) to (21);
        \draw [style=hadamard edge] (15) to (22);
        \draw (16) to (20);
        \draw (17) to (26);
        \draw (18) to (24);
        \draw (21) to (26);
        \draw [style=hadamard edge] (21) to (22);
        \draw (22) to (26);
        \draw (23) to (26);
    \end{pgfonlayer}
\end{tikzpicture}} \ .
\end{equation}
For a sanity check, let's compute $\text{Tr}(\rho_B)$ which should equate to $1$, tracing out the remaining legs of $\rho_B$ from equation \ref{eq:rho_b}, and performing further spider simplifications, we have:
\begin{equation}
    \text{Tr}(\rho_B) = \sqrt{2}^{-1}\underbrace{\scalebox{1}{\begin{tikzpicture}
    \begin{pgfonlayer}{nodelayer}
        \node [style=X phase dot] (1) at (4.00, -0.00) {$\pi$};
        \node [style=Z dot] (2) at (3.00, 1.00) {};
        \node [style=Z dot] (3) at (3.00, 2.00) {};
        \node [style=X dot] (4) at (2.00, -0.00) {};
        \node [style=Z dot] (5) at (3.00, -2.00) {};
        \node [style=Z dot] (6) at (3.00, -1.00) {};
        \node [style=Z dot] (7) at (3.00, 3.00) {};
    \end{pgfonlayer}
    \begin{pgfonlayer}{edgelayer}
        \draw [style=hadamard edge] (1) to (4);
        \draw (1) to (2);
        \draw (1) to (3);
        \draw (1) to (5);
        \draw (1) to (6);
        \draw (1) to (7);
        \draw (2) to (4);
        \draw (3) to (4);
        \draw (4) to (5);
        \draw (4) to (6);
        \draw (4) to (7);
    \end{pgfonlayer}
\end{tikzpicture}}}_{=\sqrt{2}} = 1 \ ,
\end{equation}
as expected. 

The R\'enyi entropy of $\rho_B$ from equation \ref{eq:rho_b} can be computed via,
\begin{equation}
    S_{\alpha}^{(R)}(B) = \frac{1}{1-\alpha}\log_2 \text{Tr}(\rho_B^{\alpha}) \ .
\end{equation}
For an integer $\alpha \geq 2$, this can be achieved by fusing the ZX-diagram with itself repeatedly and taking a further trace \cite{Dowling_2025}. For example with $\alpha=2$:
\begin{equation}
    \label{eq:rho_b_sq}
    \rho_B^2 = \sqrt{2}^{-4}\scalebox{.8}{\begin{tikzpicture}[yshift=+4cm]
    \begin{pgfonlayer}{nodelayer}
        \node [style=X dot] (0) at (1.00, -4.00) {};
        \node [style=none] (1) at (0.00, 0.00) {};
        \node [style=none] (2) at (0.00, -1.00) {};
        \node [style=none] (3) at (0.00, -4.00) {};
        \node [style=none] (4) at (0.00, -5.00) {};
        \node [style=none] (5) at (0.00, -8.00) {};
        \node [style=X phase dot] (6) at (5.00, -5.00) {$\pi$};
        \node [style=X dot] (7) at (1.00, -5.00) {};
        \node [style=Z dot] (8) at (2.00, -3.00) {};
        \node [style=Z dot] (9) at (2.00, -6.00) {};
        \node [style=none] (10) at (0.00, -2.00) {};
        \node [style=none] (11) at (0.00, -3.00) {};
        \node [style=none] (12) at (0.00, -6.00) {};
        \node [style=none] (13) at (0.00, -7.00) {};
        \node [style=X dot] (14) at (4.00, -6.00) {};
        \node [style=Z phase dot] (15) at (4.00, -3.00) {$\pi$};
        \node [style=Z dot] (16) at (5.00, -4.00) {};
        \node [style=X dot] (17) at (6.00, -4.00) {};
        \node [style=X phase dot] (18) at (10.00, -5.00) {$\pi$};
        \node [style=X dot] (19) at (6.00, -5.00) {};
        \node [style=Z dot] (20) at (7.00, -3.00) {};
        \node [style=Z dot] (21) at (7.00, -6.00) {};
        \node [style=none] (22) at (11.00, 0.00) {};
        \node [style=none] (23) at (11.00, -1.00) {};
        \node [style=none] (24) at (11.00, -4.00) {};
        \node [style=none] (25) at (11.00, -5.00) {};
        \node [style=none] (26) at (11.00, -8.00) {};
        \node [style=X dot] (27) at (9.00, -6.00) {};
        \node [style=Z phase dot] (28) at (9.00, -3.00) {$\pi$};
        \node [style=Z dot] (29) at (10.00, -4.00) {};
        \node [style=none] (30) at (11.00, -2.00) {};
        \node [style=none] (31) at (11.00, -3.00) {};
        \node [style=none] (32) at (11.00, -6.00) {};
        \node [style=none] (33) at (11.00, -7.00) {};
    \end{pgfonlayer}
    \begin{pgfonlayer}{edgelayer}
        \draw [style=hadamard edge] (0) to (6);
        \draw (0) to (15);
        \draw (0) to (16);
        \draw (0) to (3);
        \draw [style=hadamard edge] (0) to (7);
        \draw (0) to (8);
        \draw (1) to (22);
        \draw (2) to (23);
        \draw (4) to (7);
        \draw (5) to (26);
        \draw (6) to (9);
        \draw (6) to (8);
        \draw [style=hadamard edge] (6) to (7);
        \draw (6) to (16);
        \draw (6) to (19);
        \draw (7) to (15);
        \draw (7) to (16);
        \draw (7) to (9);
        \draw (8) to (14);
        \draw [style=hadamard edge] (8) to (15);
        \draw (8) to (11);
        \draw [style=hadamard edge] (8) to (9);
        \draw (9) to (14);
        \draw [style=hadamard edge] (9) to (15);
        \draw (9) to (12);
        \draw (10) to (30);
        \draw (13) to (33);
        \draw (14) to (15);
        \draw (14) to (21);
        \draw (15) to (20);
        \draw (16) to (17);
        \draw [style=hadamard edge] (17) to (18);
        \draw (17) to (28);
        \draw (17) to (29);
        \draw [style=hadamard edge] (17) to (19);
        \draw (17) to (20);
        \draw (18) to (21);
        \draw (18) to (20);
        \draw [style=hadamard edge] (18) to (19);
        \draw (18) to (25);
        \draw (18) to (29);
        \draw (19) to (28);
        \draw (19) to (29);
        \draw (19) to (21);
        \draw (20) to (27);
        \draw [style=hadamard edge] (20) to (28);
        \draw [style=hadamard edge] (20) to (21);
        \draw (21) to (27);
        \draw [style=hadamard edge] (21) to (28);
        \draw (24) to (29);
        \draw (27) to (32);
        \draw (27) to (28);
        \draw (28) to (31);
    \end{pgfonlayer}
\end{tikzpicture}} \ .
\end{equation}
Taking the trace and further ZX-diagram simplifications leads to:
\begin{equation}
    \label{eq:rho_b_sq_simp}
    \text{Tr}(\rho_B^2)= \sqrt{2}^{4}\underbrace{\scalebox{.8}{\begin{tikzpicture}
    \begin{pgfonlayer}{nodelayer}
        \node [style=Z dot] (0) at (4.50, 4.75) {};
        \node [style=Z dot] (1) at (1.00, 3.50) {};
        \node [style=Z phase dot] (2) at (1.00, -0.75) {$\pi$};
        \node [style=Z dot] (3) at (1.00, -1.75) {};
        \node [style=X dot] (4) at (2.00, 2.50) {};
        \node [style=X phase dot] (5) at (4.00, 1.50) {$\pi$};
        \node [style=X dot] (6) at (2.00, 1.50) {};
        \node [style=X dot] (7) at (4.00, -0.75) {};
        \node [style=Z phase dot] (8) at (5.25, -3.25) {$\pi$};
        \node [style=Z dot] (9) at (4.00, 2.50) {};
        \node [style=Z dot] (10) at (5.00, 3.50) {};
        \node [style=Z dot] (11) at (3.00, -2.00) {};
        \node [style=X dot] (12) at (7.00, 2.50) {};
        \node [style=X phase dot] (13) at (9.00, 1.50) {$\pi$};
        \node [style=X dot] (14) at (7.00, 1.50) {};
        \node [style=X dot] (15) at (3.00, -3.25) {};
    \end{pgfonlayer}
    \begin{pgfonlayer}{edgelayer}
        \draw (0) to (4);
        \draw (0) to (12);
        \draw (0) to (13);
        \draw (0) to (14);
        \draw (1) to (6);
        \draw (1) to (13);
        \draw (2) to (4);
        \draw (2) to (5);
        \draw (2) to (7);
        \draw (2) to (12);
        \draw (2) to (14);
        \draw (2) to (15);
        \draw [style=hadamard edge] (2) to (3);
        \draw [style=hadamard edge] (2) to (11);
        \draw (3) to (15);
        \draw (3) to (5);
        \draw (3) to (6);
        \draw (3) to (7);
        \draw [style=hadamard edge] (3) to (8);
        \draw [style=hadamard edge] (4) to (5);
        \draw (4) to (8);
        \draw (4) to (9);
        \draw [style=hadamard edge] (4) to (6);
        \draw [style=hadamard edge] (5) to (6);
        \draw (5) to (9);
        \draw (5) to (10);
        \draw (6) to (8);
        \draw (6) to (9);
        \draw (7) to (8);
        \draw (7) to (11);
        \draw (8) to (12);
        \draw (8) to (13);
        \draw (8) to (15);
        \draw [style=hadamard edge] (8) to (11);
        \draw (9) to (12);
        \draw (10) to (14);
        \draw (11) to (13);
        \draw (11) to (14);
        \draw (11) to (15);
        \draw [style=hadamard edge] (12) to (13);
        \draw [style=hadamard edge] (12) to (14);
        \draw [style=hadamard edge] (13) to (14);
    \end{pgfonlayer}
\end{tikzpicture}}}_{2^{-9}} \ .
\end{equation}
Hence $S_{2}^{(R)}(B) = \frac{1}{1-2} \text{log}_2(2^{-7}) = 7$, which coincides with the numerically calculated value\footnote{By exporting the ZX-diagram in $\mathtt{pyzx}$ to a matrix and computing the entropy with the standard matrix.}. The computed value $S_2^{(R)}(B)$ includes contributions from bulk legs entangled with $B$, and can therefore exceed the number of edges crossed by a naive minimal cut through the network. We will now re-perform a similar calculation applied to a boundary state of the pentagon holographic code in the next subsection.

\subsection{Computing boundary entanglement entropy via ZX-diagrams}
The RT formula relates the entanglement of the boundary states. In order to obtain a boundary state in our ZX-diagram, we can encode a reference state (say $\ket{+}$) on all the bulk legs, effectively project all bulk legs onto a $\scalebox{1}{
\begin{tikzpicture}
    \begin{pgfonlayer}{nodelayer}
        \node [style=Z dot] (1) at (-1.00, -0.00) {};
        \node [style=none] (2) at (0.00, -0.00) {};
    \end{pgfonlayer}
    \begin{pgfonlayer}{edgelayer}
        \draw (1) to (2);
    \end{pgfonlayer}
\end{tikzpicture}
}$ spider, effectively applying $\tensor{^{N_{\text{bulk}}\otimes}\bra{+}}{}$ to both sides of equation \ref{eq:phi_n_2}, let's call this state $\ket{\partial} = \tensor{^{N_{\text{bulk}}\otimes} \langle +}{} |\Phi\rangle$:
\begin{equation}
     \ket{\partial}=\raisebox{-17ex}{\scalebox{0.38}{\includegraphics{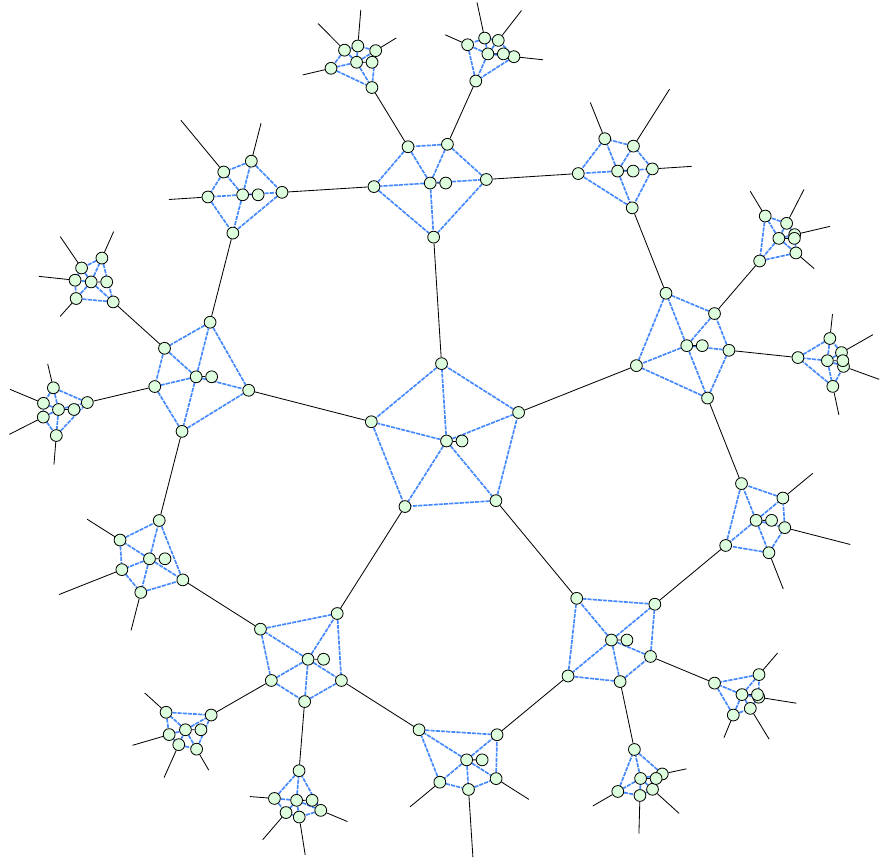}}} \ .
\end{equation}
Similar to previous calculation, we assign boundary qubits labelled with dotted magenta circled to live $\in A$ and we wish to compute the entanglement entropy between $A$ and its complement $A^c$ on the boundary:
\begin{equation}
    \label{eq:ec_orange}
     \raisebox{-17ex}{\scalebox{0.4}{\includegraphics{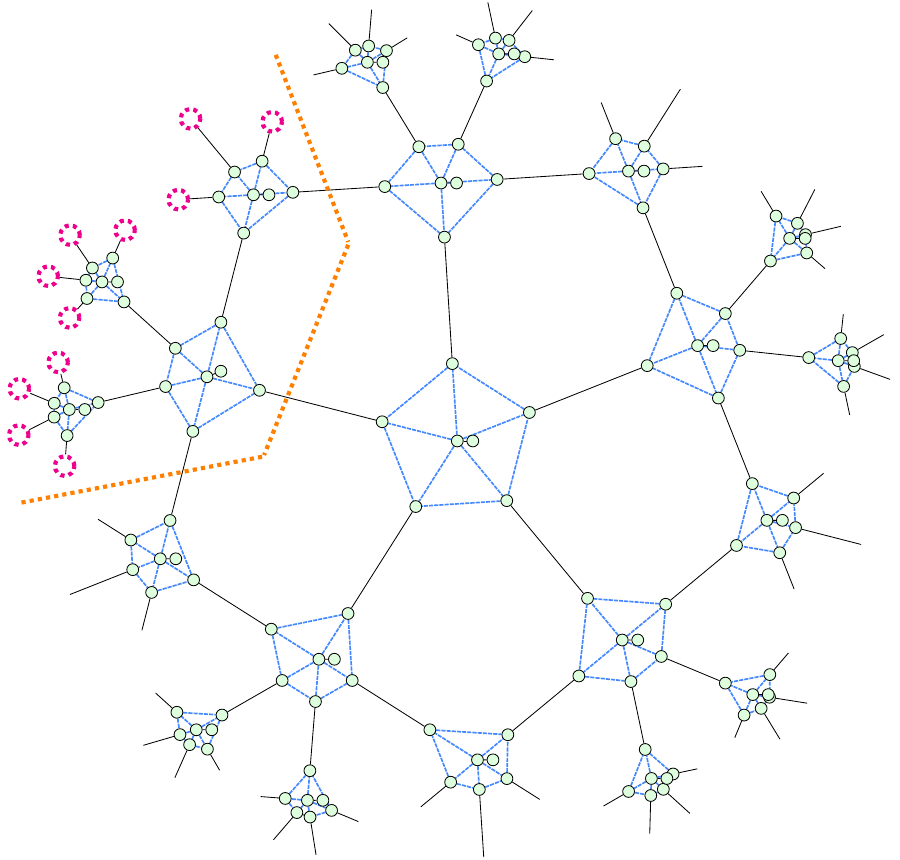}}} \ ,
\end{equation}
the orange dotted lines indicate the minimal cut as dictated by the RT formula and separates $A$ from $A^c$. To compute $S_2^{(R)}(A)$, we first compute the reduced density matrix $\rho_A = \text{Tr}_{A^c}(\ketbra{\partial}{\partial})$:
\begin{equation}
    \rho_A = C' \cdot \raisebox{-8ex}{\scalebox{0.2}{\includegraphics{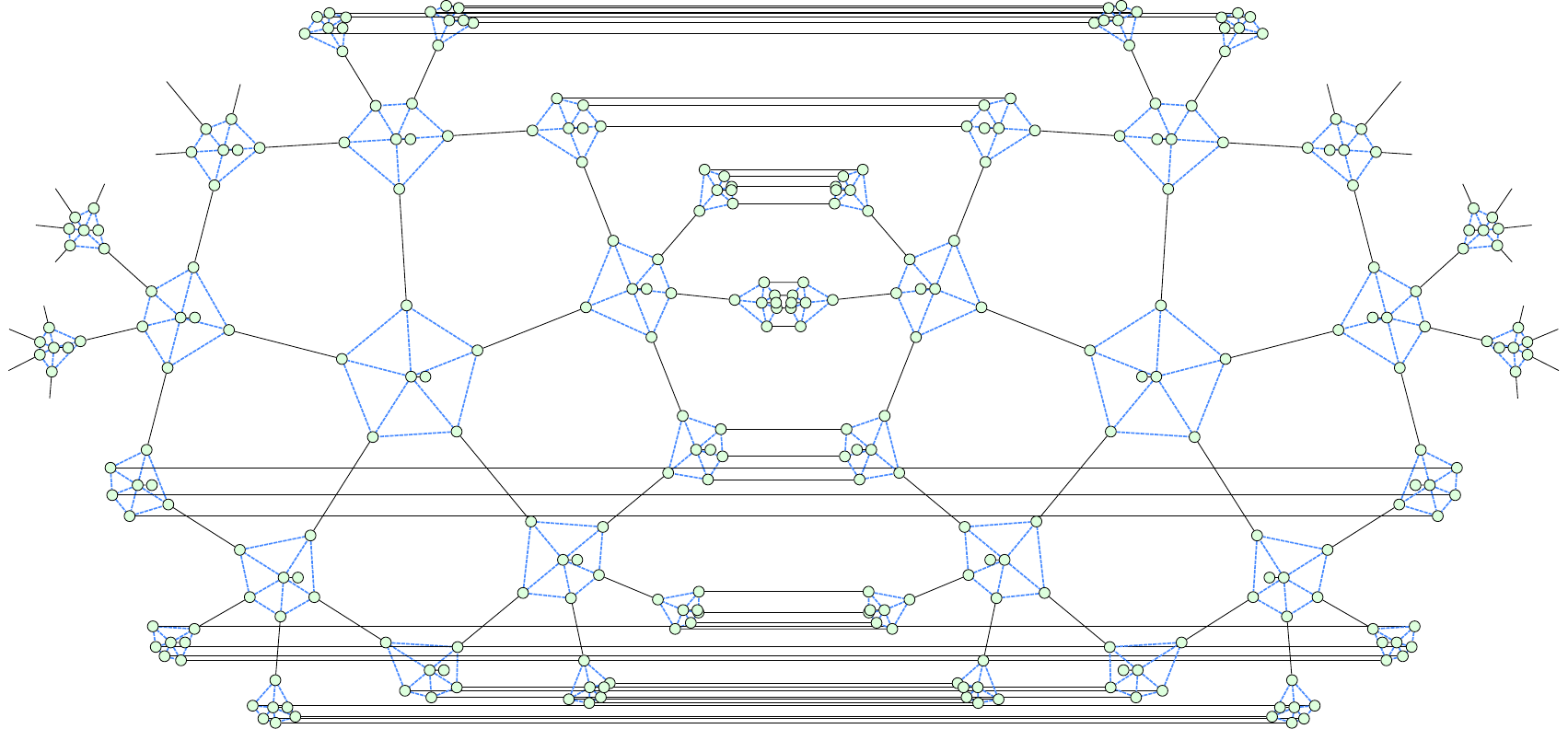}}} \ ,
\end{equation}
we can simplify this ZX-diagram and then contract it again with itself to acquire $\text{Tr}(\rho_A^2)$:
\begin{equation}
    \label{eq:tr_rho_a_sq}
    \text{Tr}(\rho_A^2)= 2^{39}\underbrace{\scalebox{.8}{\begin{tikzpicture}[yshift=+5cm]
    \begin{pgfonlayer}{nodelayer}
        \node [style=Z dot] (11) at (1.00, 0.00) {};
        \node [style=X phase dot] (12) at (1.00, -1.00) {$\pi$};
        \node [style=X phase dot] (13) at (1.00, -2.00) {$\pi$};
        \node [style=X dot] (14) at (1.00, -3.00) {};
        \node [style=Z dot] (15) at (1.00, -4.00) {};
        \node [style=X phase dot] (16) at (1.00, -5.00) {$\pi$};
        \node [style=Z dot] (17) at (1.00, -6.00) {};
        \node [style=Z dot] (18) at (1.00, -7.00) {};
        \node [style=Z dot] (19) at (1.00, -8.00) {};
        \node [style=X phase dot] (20) at (1.00, -9.00) {$\pi$};
        \node [style=Z phase dot] (21) at (1.00, -10.00) {$\pi$};
        \node [style=X dot] (22) at (2.00, 0.00) {};
        \node [style=X dot] (23) at (2.00, -6.00) {};
        \node [style=Z phase dot] (24) at (2.00, -8.00) {$\pi$};
        \node [style=X dot] (25) at (4.00, -8.00) {};
        \node [style=Z phase dot] (33) at (4.00, -10.00) {$\pi$};
        \node [style=X phase dot] (34) at (4.00, 0.00) {$\pi$};
        \node [style=Z dot] (35) at (4.00, -5.00) {};
        \node [style=Z phase dot] (37) at (4.00, -9.00) {$\pi$};
        \node [style=Z phase dot] (38) at (4.00, -3.00) {$\pi$};
        \node [style=X phase dot] (39) at (4.00, -7.00) {$\pi$};
        \node [style=Z dot] (40) at (4.00, -2.00) {};
        \node [style=X phase dot] (41) at (4.00, -1.00) {$\pi$};
        \node [style=X phase dot] (42) at (4.00, -4.00) {$\pi$};
        \node [style=Z dot] (43) at (4.00, -6.00) {};
        \node [style=X dot] (45) at (5.00, -6.00) {};
        \node [style=Z phase dot] (47) at (7.00, -8.00) {$\pi$};
        \node [style=Z dot] (49) at (5.00, -2.00) {};
        \node [style=Z dot] (50) at (5.00, -3.00) {};
        \node [style=X phase dot] (52) at (5.00, -5.00) {$\pi$};
        \node [style=Z phase dot] (53) at (5.00, -7.00) {$\pi$};
        \node [style=Z phase dot] (54) at (5.00, -10.00) {$\pi$};
        \node [style=X phase dot] (55) at (7.00, -10.00) {$\pi$};
        \node [style=X phase dot] (56) at (7.00, 0.00) {$\pi$};
        \node [style=Z dot] (57) at (7.00, -5.00) {};
        \node [style=Z dot] (63) at (5.00, -9.00) {};
        \node [style=Z phase dot] (64) at (7.00, -9.00) {$\pi$};
        \node [style=Z phase dot] (65) at (7.00, -3.00) {$\pi$};
        \node [style=Z phase dot] (66) at (7.00, -7.00) {$\pi$};
        \node [style=Z dot] (67) at (7.00, -2.00) {};
    \end{pgfonlayer}
    \begin{pgfonlayer}{edgelayer}
        \draw (11) to (22);
        \draw (11) to (56);
        \draw [style=hadamard edge] (12) to (23);
        \draw (12) to (38);
        \draw (12) to (43);
        \draw (12) to (35);
        \draw (12) to (65);
        \draw (12) to (18);
        \draw (12) to (19);
        \draw [style=hadamard edge] (13) to (23);
        \draw (13) to (57);
        \draw (13) to (18);
        \draw (14) to (24);
        \draw (14) to (47);
        \draw (15) to (22);
        \draw [style=hadamard edge] (15) to (24);
        \draw (15) to (25);
        \draw (15) to (34);
        \draw [style=hadamard edge] (15) to (37);
        \draw (15) to (41);
        \draw [style=hadamard edge] (15) to (64);
        \draw (15) to (16);
        \draw (16) to (47);
        \draw (16) to (54);
        \draw (16) to (57);
        \draw (16) to (63);
        \draw (16) to (64);
        \draw (16) to (65);
        \draw (16) to (67);
        \draw [style=hadamard edge] (16) to (22);
        \draw (16) to (24);
        \draw (16) to (40);
        \draw (16) to (37);
        \draw (16) to (38);
        \draw (16) to (43);
        \draw (16) to (35);
        \draw (16) to (19);
        \draw (16) to (21);
        \draw [style=hadamard edge] (16) to (34);
        \draw (17) to (22);
        \draw [style=hadamard edge] (17) to (40);
        \draw (17) to (34);
        \draw [style=hadamard edge] (17) to (67);
        \draw [style=hadamard edge] (18) to (65);
        \draw (18) to (23);
        \draw (19) to (23);
        \draw (19) to (45);
        \draw (19) to (56);
        \draw (19) to (42);
        \draw (19) to (34);
        \draw (19) to (41);
        \draw (20) to (24);
        \draw (20) to (66);
        \draw (20) to (21);
        \draw (21) to (55);
        \draw (21) to (22);
        \draw [style=hadamard edge] (21) to (24);
        \draw (21) to (25);
        \draw [style=hadamard edge] (21) to (37);
        \draw (21) to (41);
        \draw (21) to (34);
        \draw [style=hadamard edge] (22) to (25);
        \draw [style=hadamard edge] (22) to (34);
        \draw (22) to (37);
        \draw (22) to (38);
        \draw [style=hadamard edge] (22) to (41);
        \draw (22) to (43);
        \draw (22) to (35);
        \draw (22) to (24);
        \draw (23) to (38);
        \draw (23) to (43);
        \draw (23) to (35);
        \draw (24) to (25);
        \draw [style=hadamard edge] (24) to (37);
        \draw (24) to (41);
        \draw (24) to (34);
        \draw (25) to (33);
        \draw (25) to (47);
        \draw (25) to (64);
        \draw [style=hadamard edge] (25) to (56);
        \draw (25) to (54);
        \draw (25) to (63);
        \draw (25) to (53);
        \draw (33) to (39);
        \draw [style=hadamard edge] (33) to (54);
        \draw (34) to (38);
        \draw (34) to (43);
        \draw (34) to (35);
        \draw (34) to (37);
        \draw (34) to (47);
        \draw [style=hadamard edge] (34) to (56);
        \draw (34) to (64);
        \draw (34) to (65);
        \draw (34) to (57);
        \draw (34) to (49);
        \draw (34) to (54);
        \draw (34) to (63);
        \draw [style=hadamard edge] (34) to (41);
        \draw [style=hadamard edge] (35) to (38);
        \draw (35) to (41);
        \draw (35) to (52);
        \draw (37) to (39);
        \draw (37) to (41);
        \draw [style=hadamard edge] (37) to (63);
        \draw (38) to (42);
        \draw (38) to (41);
        \draw [style=hadamard edge] (38) to (50);
        \draw (39) to (53);
        \draw (40) to (41);
        \draw [style=hadamard edge] (40) to (49);
        \draw (41) to (43);
        \draw (41) to (47);
        \draw (41) to (67);
        \draw (41) to (64);
        \draw (41) to (65);
        \draw (41) to (57);
        \draw (41) to (54);
        \draw (41) to (63);
        \draw (42) to (43);
        \draw [style=hadamard edge] (42) to (45);
        \draw (42) to (50);
        \draw (42) to (65);
        \draw (42) to (57);
        \draw (43) to (45);
        \draw (45) to (65);
        \draw (45) to (57);
        \draw [style=hadamard edge] (45) to (52);
        \draw (45) to (50);
        \draw [style=hadamard edge] (47) to (63);
        \draw [style=hadamard edge] (47) to (54);
        \draw (47) to (55);
        \draw (47) to (56);
        \draw [style=hadamard edge] (49) to (67);
        \draw (49) to (56);
        \draw (50) to (52);
        \draw [style=hadamard edge] (53) to (54);
        \draw [style=hadamard edge] (54) to (64);
        \draw (54) to (56);
        \draw (55) to (66);
        \draw (56) to (63);
        \draw (56) to (65);
        \draw (56) to (57);
        \draw (56) to (64);
        \draw [style=hadamard edge] (57) to (65);
        \draw [style=hadamard edge] (63) to (64);
        \draw [style=hadamard edge] (64) to (66);
    \end{pgfonlayer}
\end{tikzpicture}}}_{2^{-42}} \ .
\end{equation}
The structure of this scalar ZX-diagram offers no insights theoretically and miraculously reduced to $2^{-42}$, which implies: $\text{Tr}(\rho_A^2) = 2^{-3} \Rightarrow S_2^{(R)} = -\text{log}_2(2^{-3}) = 3$. Which coincides with the number of minimal cuts, shown in orange (equation \ref{eq:ec_orange}). In summary, we have demonstrated two examples to computing entanglement entropy using ZX-diagrams. Under what conditions does the computed $S_2^{(R)}$ coincide exactly with the Ryu–Takayanagi minimal-cut prediction?

Next, we turn to a brief discussion of black holes and wormholes, concluding our study of the pentagon holographic code.

\section{Black holes and wormholes}
In this section, we shall make some short remarks on toy models of black hole and wormholes constructed with the pentagon holographic code. A toy model of a black hole can be constructed via removing the central tensor and leaving the $5$ uncontracted tensor legs as additional free bulk legs \cite{Pastawski:2015qua}. These uncontracted legs can be interpreted as the horizon degrees of freedom. This construction can be represented as the following ZX-diagram:
\begin{equation}
    \label{eq:black_hole_n_2}
    \raisebox{-17ex}{\scalebox{0.4}{\includegraphics{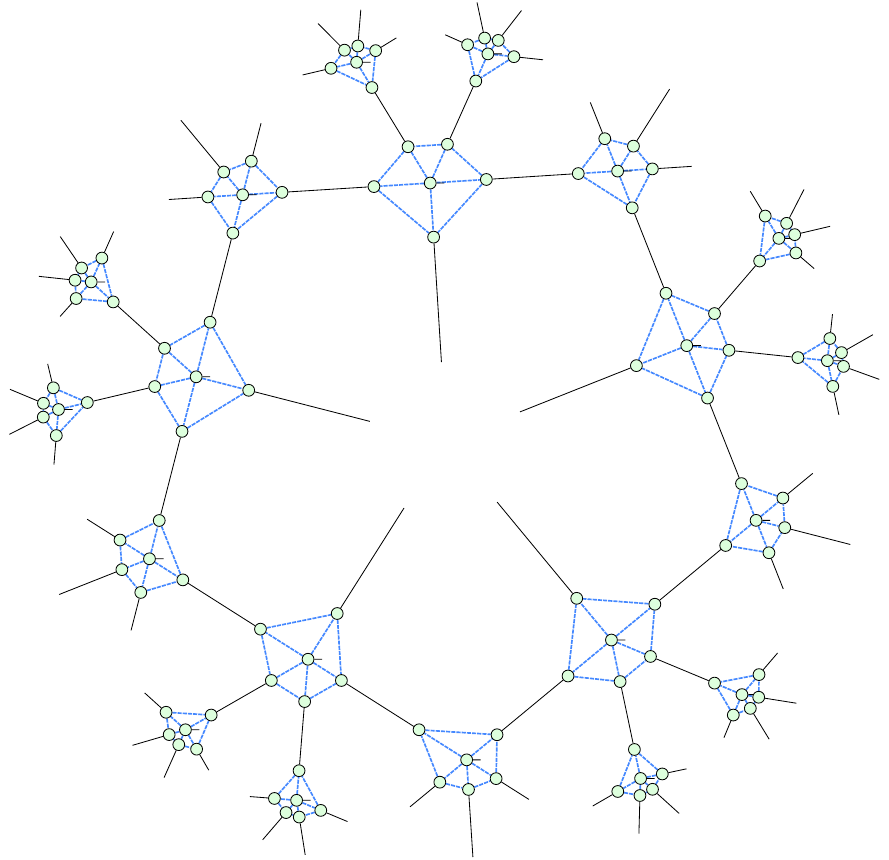}}} \ .
\end{equation}
Firstly, we can derive the reduced density operator of the 5 central horizon legs:
\begin{equation}
    \label{eq:bh_rho_hori}
    \rho_{{h}} \propto
    \raisebox{-8ex}{\scalebox{0.2}{\includegraphics{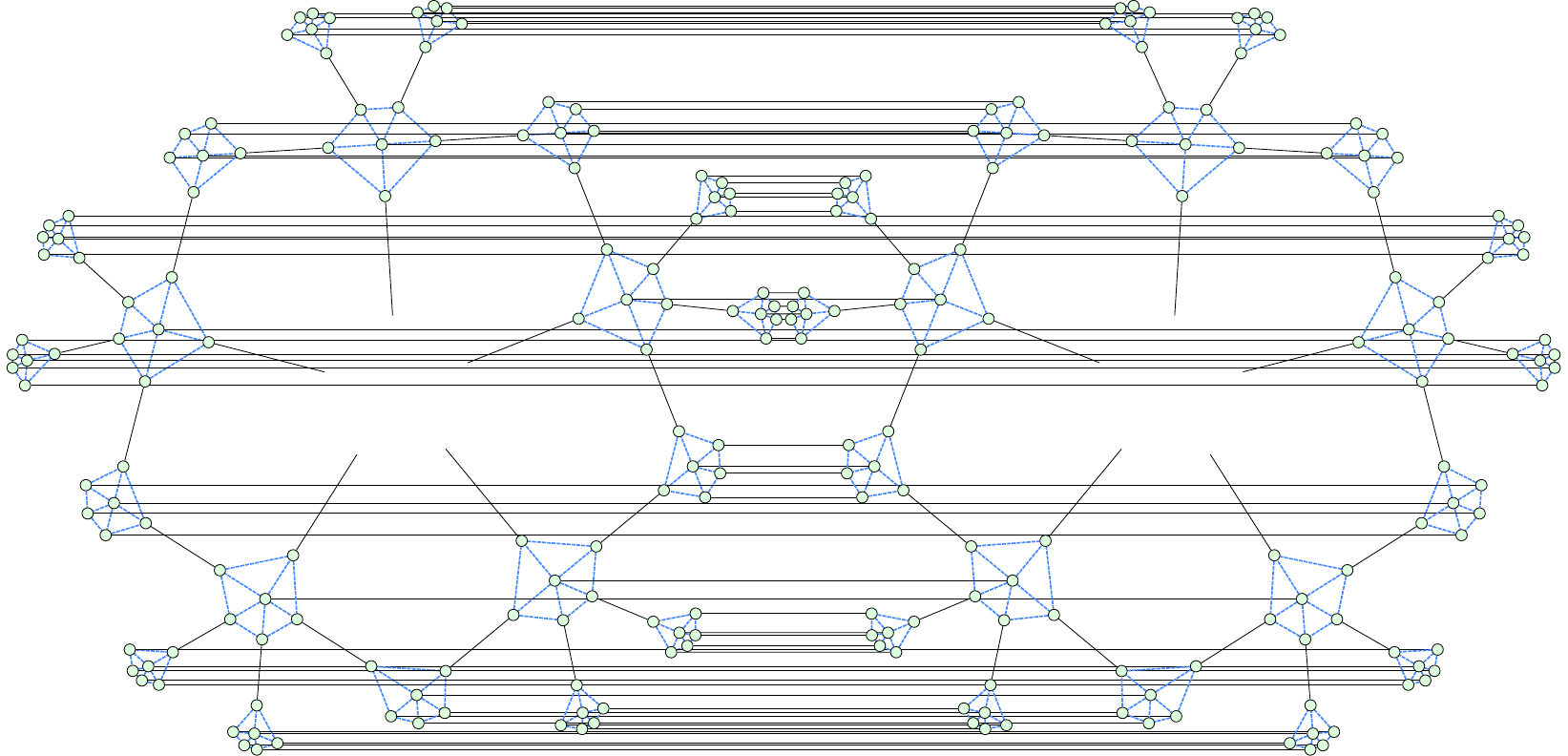}}} \ .
\end{equation}
This ZX-diagram reduces to $\rho_h=\frac{I^{\otimes 5}}{2^{5}}$ exactly, the maximally mixed state, via ZX-diagram simplification techniques. Any operator acting on these legs cannot be reconstructed from the boundary alone. These 5 legs on the horizon represent maximally scrambled degrees of freedom, hence a toy version of a black hole, as expected.

\begin{figure}[!h]
    \centering
    \input{figures/wh3d.tikz}
    \caption{Wormhole from \cite{Pastawski:2015qua} and one of its Pauli webs, spanning both sides of the wormhole.}
    \label{fig:wormhole_with_webs}
\end{figure}

Furthermore, a toy model of a wormhole can be constructed by taking two copies of these black holes ZX-diagrams and fusing its newly acquired additional bulk legs (near the event horizon), leading to the following ZX-diagram in figure \ref{fig:wormhole_with_webs}, where we can see Pauli webs spanning both sides of the wormhole. Dynamics would be needed to produce interesting effects, one suggestion would be to foliate it in a non-trivial way \cite{Bolt_2016,nickerson2018measurementbasedfaulttolerance}.  See section~\ref{sec:spacetimeZX} for the spacetime code construction.

Building on the ZX-diagram constructions of the pentagon holographic code, we now study the construction and decoding of ZX-diagram–inspired holographic codes on the dual $\{4,5\}$ hyperbolic tessellation.

\section{Constructions and decoding of ZX-diagrams inspired holographic codes}
Motivated by the ZX-diagram formulation of the pentagon holographic code on the hyperbolic tessellation with Schläfli symbol $\{p,q\}=\{5,4\}$, we introduce a ZX-diagram inspired realisation of a holographic code related to the hyperinvariant tensor network code of Evenbly \cite{Steinberg_2023}. This construction is naturally associated with the dual tessellation $\{p,q\}=\{4,5\}$ \cite{jahn2023holographic}. In contrast to \cite{Steinberg_2023,Steinberg_2025}, where a $[[4,1,2]]$ encoding tensor is placed at each vertex, we replace every tensor by the following ZX-diagram:
\begin{equation}
    \label{eq:r4}
    \scalebox{1}{\begin{tikzpicture}[scale=0.525]
    \begin{pgfonlayer}{nodelayer}
        \node [style=none] (0) at (-1.00, 3.00) {};
        \node [style=Z dot] (1) at (3.00, -1.00) {};
        \node [style=Z dot] (2) at (0.00, 2.00) {};
        \node [style=Z dot] (3) at (6.00, 2.00) {};
        \node [style=Z dot] (4) at (0.00, -4.00) {};
        \node [style=none] (5) at (7.00, 3.00) {};
        \node [style=none] (6) at (4.00, -1.00) {};
        \node [style=none] (7) at (-1.00, -5.00) {};
        \node [style=none] (8) at (7.00, -5.00) {};
        \node [style=Z dot] (9) at (6.00, -4.00) {};
    \end{pgfonlayer}
    \begin{pgfonlayer}{edgelayer}
        \draw (0) to (2);
        \draw [style=hadamard edge] (1) to (3);
        \draw [style=hadamard edge] (1) to (2);
        \draw [style=hadamard edge] (1) to (4);
        \draw [style=hadamard edge] (1) to (9);
        \draw (1) to (6);
        \draw [style=hadamard edge] (2) to (4);
        \draw [style=hadamard edge] (2) to (3);
        \draw (3) to (5);
        \draw [style=hadamard edge] (3) to (9);
        \draw (4) to (7);
        \draw [style=hadamard edge] (4) to (9);
        \draw (8) to (9);
    \end{pgfonlayer}
\end{tikzpicture}} \ .
\end{equation}
The central uncontracted leg in equation \ref{eq:r4} represents a bulk qubit, while the four corner legs are to be contracted with neighbouring tensors. 

\begin{widetext}
\begin{equation}
\label{eq:zxholo_codes}
    \begin{array}{ccccc}
        n=0 & n=1 & n=2 & n=3 & \\
        \scalebox{1.4}{\includegraphics{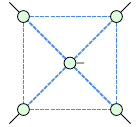}} &
        \scalebox{0.22}{\includegraphics{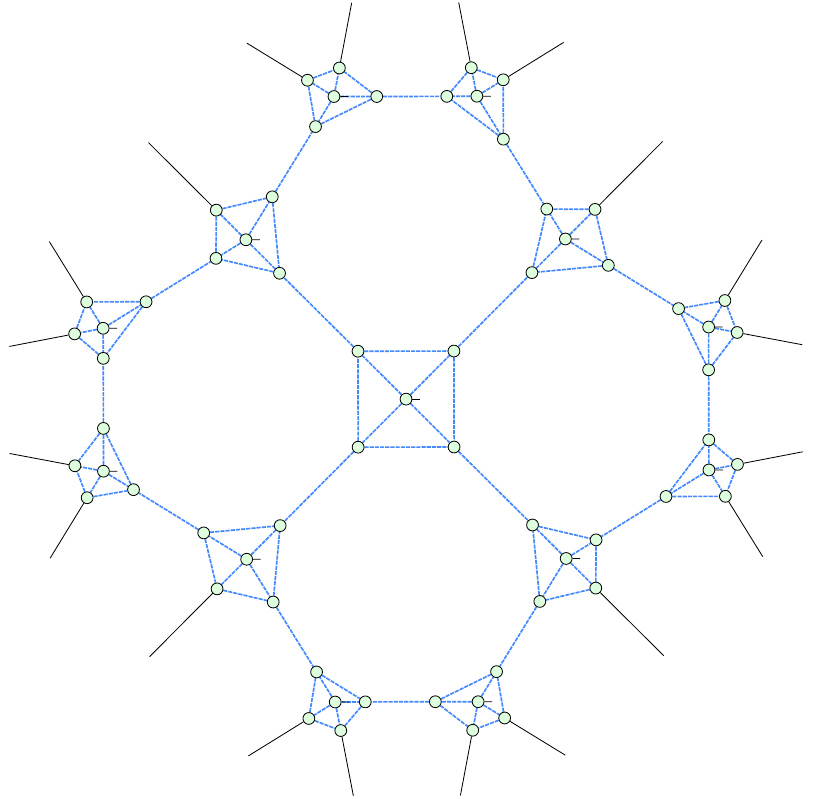}} &
        \scalebox{0.1}{\includegraphics{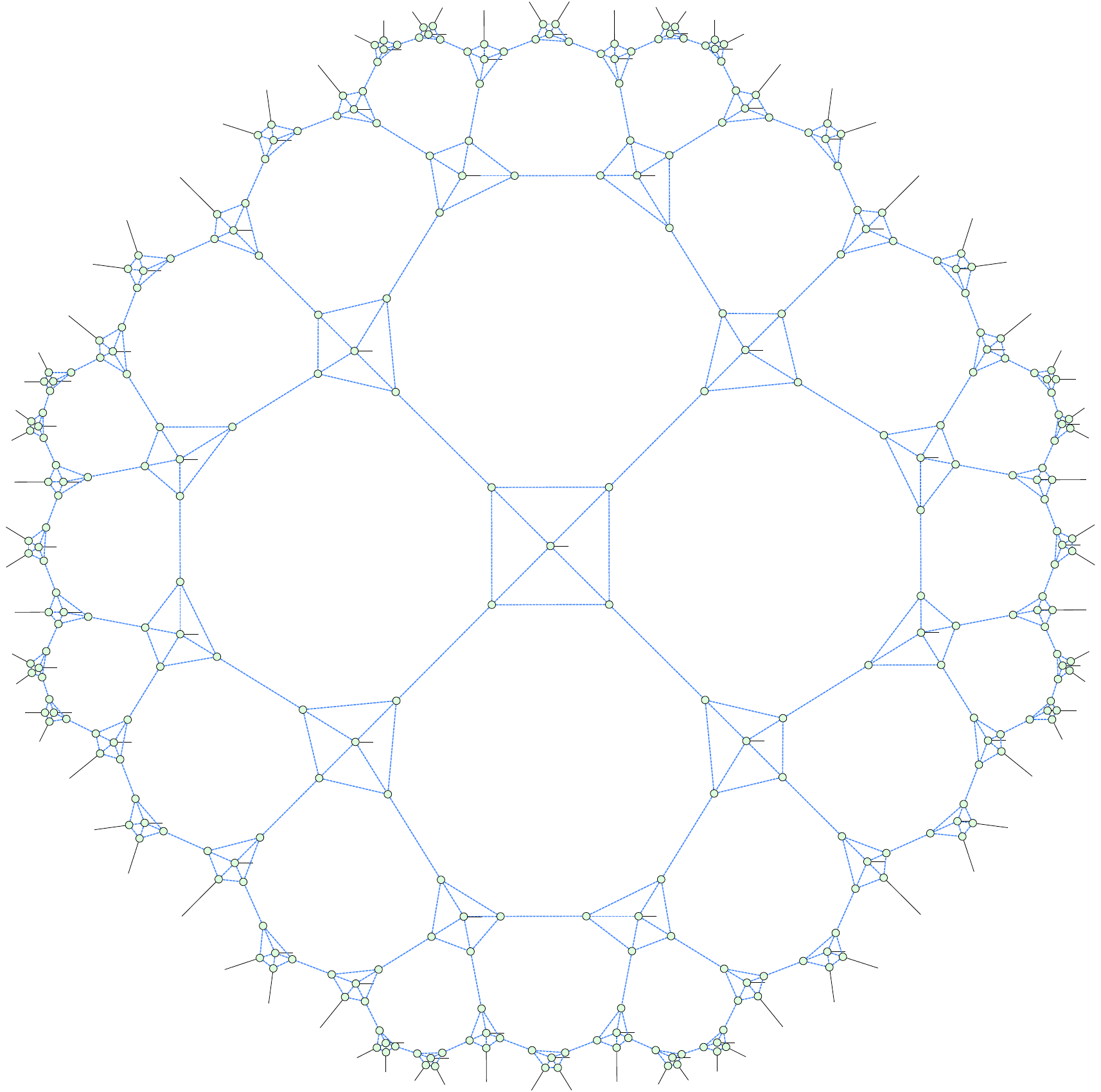}} &
        \scalebox{0.1}{\includegraphics{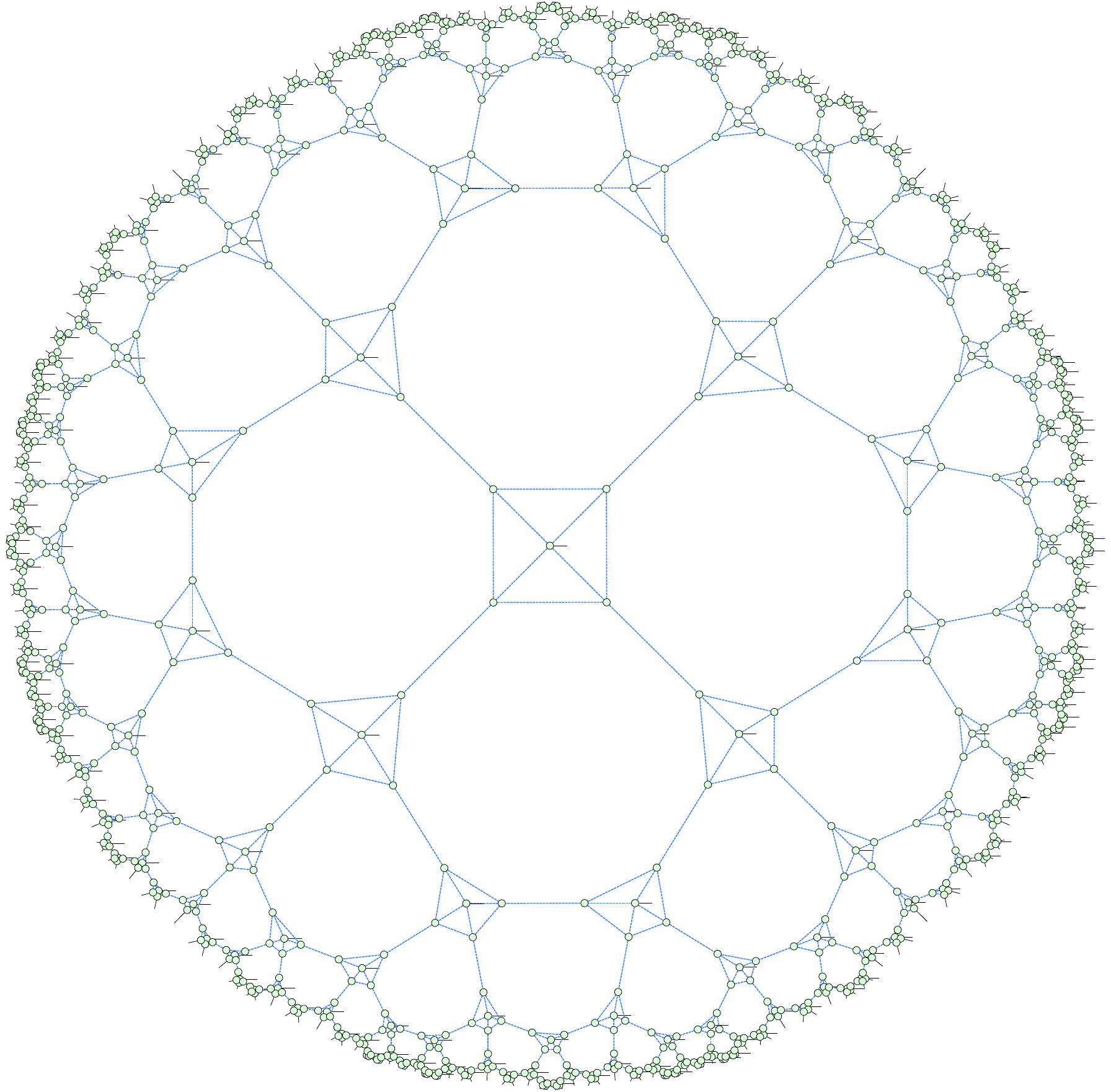}} &
        \hspace{.5cm} \raisebox{10ex}{\dots}
    \end{array}
\end{equation}
\end{widetext}

The resultant five qubit state is stabilised by
\begin{equation}
\begin{split}
\Bigg\langle\;
& \overbrace{\scalebox{0.8}{\begin{tikzpicture}[scale=0.525]
    \begin{pgfonlayer}{nodelayer}
        \node [style=none] (0) at (-1.00, 4.00) {};
        \node [style=Z dot] (1) at (3.00, -0.00) {};
        \node [style=Z dot] (2) at (0.00, 3.00) {};
        \node [style=Z dot] (3) at (6.00, 3.00) {};
        \node [style=Z dot] (4) at (0.00, -3.00) {};
        \node [style=none] (5) at (7.00, 4.00) {};
        \node [style=none] (6) at (4.00, -0.00) {};
        \node [style=none] (7) at (-1.00, -4.00) {};
        \node [style=none] (8) at (7.00, -4.00) {};
        \node [style=Z dot] (9) at (6.00, -3.00) {};
    \end{pgfonlayer}
    \begin{pgfonlayer}{edgelayer}
        \draw (0) to (2);
        \draw [style=hadamard edge] (1) to (3);
        \draw [style=hadamard edge] (1) to (2);
        \draw [style=hadamard edge] (1) to (4);
        \draw [style=hadamard edge] (1) to (9);
        \draw (1) to (6);
        \draw [style=hadamard edge] (2) to (4);
        \draw [style=hadamard edge] (2) to (3);
        \draw (3) to (5);
        \draw [style=hadamard edge] (3) to (9);
        \draw (4) to (7);
        \draw [style=hadamard edge] (4) to (9);
        \draw (8) to (9);
    \end{pgfonlayer}

\begin{pgfonlayer}{edgelayer}
\draw[blue, line width=2.5pt, opacity=0.4] (4) -- ($(4)!0.5!(7)$);
\draw[blue, line width=2.5pt, opacity=0.4] (7) -- ($(7)!0.5!(4)$);
\draw[green, line width=2.5pt, opacity=0.4] (2) -- ($(2)!0.5!(0)$);
\draw[green, line width=2.5pt, opacity=0.4] (0) -- ($(0)!0.5!(2)$);
\draw[blue, line width=2.5pt, opacity=0.4] (9) -- ($(9)!0.5!(8)$);
\draw[blue, line width=2.5pt, opacity=0.4] (8) -- ($(8)!0.5!(9)$);
\draw[green, line width=2.5pt, opacity=0.4] (1) -- ($(1)!0.5!(4)$);
\draw[red, line width=2.5pt, opacity=0.4] (4) -- ($(4)!0.5!(1)$);
\draw[green, line width=2.5pt, opacity=0.4] (2) -- ($(2)!0.5!(4)$);
\draw[red, line width=2.5pt, opacity=0.4] (4) -- ($(4)!0.5!(2)$);
\draw[blue, line width=2.5pt, opacity=0.4] (4) -- ($(4)!0.5!(9)$);
\draw[blue, line width=2.5pt, opacity=0.4] (9) -- ($(9)!0.5!(4)$);
\draw[green, line width=2.5pt, opacity=0.4] (3) -- ($(3)!0.5!(5)$);
\draw[green, line width=2.5pt, opacity=0.4] (5) -- ($(5)!0.5!(3)$);
\draw[green, line width=2.5pt, opacity=0.4] (1) -- ($(1)!0.5!(9)$);
\draw[red, line width=2.5pt, opacity=0.4] (9) -- ($(9)!0.5!(1)$);
\draw[green, line width=2.5pt, opacity=0.4] (3) -- ($(3)!0.5!(9)$);
\draw[red, line width=2.5pt, opacity=0.4] (9) -- ($(9)!0.5!(3)$);
\end{pgfonlayer}
\end{tikzpicture}}}^{ZZYY\otimes I},\quad
  \overbrace{\scalebox{0.8}{\begin{tikzpicture}[scale=0.525]
    \begin{pgfonlayer}{nodelayer}
        \node [style=none] (0) at (-1.00, 4.00) {};
        \node [style=Z dot] (1) at (3.00, -0.00) {};
        \node [style=Z dot] (2) at (0.00, 3.00) {};
        \node [style=Z dot] (3) at (6.00, 3.00) {};
        \node [style=Z dot] (4) at (0.00, -3.00) {};
        \node [style=none] (5) at (7.00, 4.00) {};
        \node [style=none] (6) at (4.00, -0.00) {};
        \node [style=none] (7) at (-1.00, -4.00) {};
        \node [style=none] (8) at (7.00, -4.00) {};
        \node [style=Z dot] (9) at (6.00, -3.00) {};
    \end{pgfonlayer}
    \begin{pgfonlayer}{edgelayer}
        \draw (0) to (2);
        \draw [style=hadamard edge] (1) to (3);
        \draw [style=hadamard edge] (1) to (2);
        \draw [style=hadamard edge] (1) to (4);
        \draw [style=hadamard edge] (1) to (9);
        \draw (1) to (6);
        \draw [style=hadamard edge] (2) to (4);
        \draw [style=hadamard edge] (2) to (3);
        \draw (3) to (5);
        \draw [style=hadamard edge] (3) to (9);
        \draw (4) to (7);
        \draw [style=hadamard edge] (4) to (9);
        \draw (8) to (9);
    \end{pgfonlayer}

\begin{pgfonlayer}{edgelayer}
\draw[red, line width=2.5pt, opacity=0.4] (2) -- ($(2)!0.5!(0)$);
\draw[red, line width=2.5pt, opacity=0.4] (0) -- ($(0)!0.5!(2)$);
\draw[red, line width=2.5pt, opacity=0.4] (9) -- ($(9)!0.5!(8)$);
\draw[red, line width=2.5pt, opacity=0.4] (8) -- ($(8)!0.5!(9)$);
\draw[green, line width=2.5pt, opacity=0.4] (1) -- ($(1)!0.5!(2)$);
\draw[red, line width=2.5pt, opacity=0.4] (2) -- ($(2)!0.5!(1)$);
\draw[green, line width=2.5pt, opacity=0.4] (1) -- ($(1)!0.5!(9)$);
\draw[red, line width=2.5pt, opacity=0.4] (9) -- ($(9)!0.5!(1)$);
\draw[red, line width=2.5pt, opacity=0.4] (2) -- ($(2)!0.5!(4)$);
\draw[green, line width=2.5pt, opacity=0.4] (4) -- ($(4)!0.5!(2)$);
\draw[red, line width=2.5pt, opacity=0.4] (2) -- ($(2)!0.5!(3)$);
\draw[green, line width=2.5pt, opacity=0.4] (3) -- ($(3)!0.5!(2)$);
\draw[green, line width=2.5pt, opacity=0.4] (3) -- ($(3)!0.5!(9)$);
\draw[red, line width=2.5pt, opacity=0.4] (9) -- ($(9)!0.5!(3)$);
\draw[green, line width=2.5pt, opacity=0.4] (4) -- ($(4)!0.5!(9)$);
\draw[red, line width=2.5pt, opacity=0.4] (9) -- ($(9)!0.5!(4)$);
\end{pgfonlayer}
\end{tikzpicture}}}^{XIXI\otimes I},\quad
  \overbrace{\scalebox{0.8}{\begin{tikzpicture}[scale=0.525]
    \begin{pgfonlayer}{nodelayer}
        \node [style=none] (0) at (-1.00, 4.00) {};
        \node [style=Z dot] (1) at (3.00, -0.00) {};
        \node [style=Z dot] (2) at (0.00, 3.00) {};
        \node [style=Z dot] (3) at (6.00, 3.00) {};
        \node [style=Z dot] (4) at (0.00, -3.00) {};
        \node [style=none] (5) at (7.00, 4.00) {};
        \node [style=none] (6) at (4.00, -0.00) {};
        \node [style=none] (7) at (-1.00, -4.00) {};
        \node [style=none] (8) at (7.00, -4.00) {};
        \node [style=Z dot] (9) at (6.00, -3.00) {};
    \end{pgfonlayer}
    \begin{pgfonlayer}{edgelayer}
        \draw (0) to (2);
        \draw [style=hadamard edge] (1) to (3);
        \draw [style=hadamard edge] (1) to (2);
        \draw [style=hadamard edge] (1) to (4);
        \draw [style=hadamard edge] (1) to (9);
        \draw (1) to (6);
        \draw [style=hadamard edge] (2) to (4);
        \draw [style=hadamard edge] (2) to (3);
        \draw (3) to (5);
        \draw [style=hadamard edge] (3) to (9);
        \draw (4) to (7);
        \draw [style=hadamard edge] (4) to (9);
        \draw (8) to (9);
    \end{pgfonlayer}

\begin{pgfonlayer}{edgelayer}
\draw[red, line width=2.5pt, opacity=0.4] (3) -- ($(3)!0.5!(5)$);
\draw[red, line width=2.5pt, opacity=0.4] (5) -- ($(5)!0.5!(3)$);
\draw[red, line width=2.5pt, opacity=0.4] (4) -- ($(4)!0.5!(7)$);
\draw[red, line width=2.5pt, opacity=0.4] (7) -- ($(7)!0.5!(4)$);
\draw[green, line width=2.5pt, opacity=0.4] (1) -- ($(1)!0.5!(3)$);
\draw[red, line width=2.5pt, opacity=0.4] (3) -- ($(3)!0.5!(1)$);
\draw[green, line width=2.5pt, opacity=0.4] (1) -- ($(1)!0.5!(4)$);
\draw[red, line width=2.5pt, opacity=0.4] (4) -- ($(4)!0.5!(1)$);
\draw[green, line width=2.5pt, opacity=0.4] (2) -- ($(2)!0.5!(4)$);
\draw[red, line width=2.5pt, opacity=0.4] (4) -- ($(4)!0.5!(2)$);
\draw[green, line width=2.5pt, opacity=0.4] (2) -- ($(2)!0.5!(3)$);
\draw[red, line width=2.5pt, opacity=0.4] (3) -- ($(3)!0.5!(2)$);
\draw[red, line width=2.5pt, opacity=0.4] (3) -- ($(3)!0.5!(9)$);
\draw[green, line width=2.5pt, opacity=0.4] (9) -- ($(9)!0.5!(3)$);
\draw[red, line width=2.5pt, opacity=0.4] (4) -- ($(4)!0.5!(9)$);
\draw[green, line width=2.5pt, opacity=0.4] (9) -- ($(9)!0.5!(4)$);
\end{pgfonlayer}
\end{tikzpicture}}}^{IXIX\otimes I}, \\
& \underbrace{\scalebox{0.8}{\begin{tikzpicture}[scale=0.525]
    \begin{pgfonlayer}{nodelayer}
        \node [style=none] (0) at (-1.00, 4.00) {};
        \node [style=Z dot] (1) at (3.00, -0.00) {};
        \node [style=Z dot] (2) at (0.00, 3.00) {};
        \node [style=Z dot] (3) at (6.00, 3.00) {};
        \node [style=Z dot] (4) at (0.00, -3.00) {};
        \node [style=none] (5) at (7.00, 4.00) {};
        \node [style=none] (6) at (4.00, -0.00) {};
        \node [style=none] (7) at (-1.00, -4.00) {};
        \node [style=none] (8) at (7.00, -4.00) {};
        \node [style=Z dot] (9) at (6.00, -3.00) {};
    \end{pgfonlayer}
    \begin{pgfonlayer}{edgelayer}
        \draw (0) to (2);
        \draw [style=hadamard edge] (1) to (3);
        \draw [style=hadamard edge] (1) to (2);
        \draw [style=hadamard edge] (1) to (4);
        \draw [style=hadamard edge] (1) to (9);
        \draw (1) to (6);
        \draw [style=hadamard edge] (2) to (4);
        \draw [style=hadamard edge] (2) to (3);
        \draw (3) to (5);
        \draw [style=hadamard edge] (3) to (9);
        \draw (4) to (7);
        \draw [style=hadamard edge] (4) to (9);
        \draw (8) to (9);
    \end{pgfonlayer}

\begin{pgfonlayer}{edgelayer}
\draw[red, line width=2.5pt, opacity=0.4] (4) -- ($(4)!0.5!(7)$);
\draw[red, line width=2.5pt, opacity=0.4] (7) -- ($(7)!0.5!(4)$);
\draw[green, line width=2.5pt, opacity=0.4] (2) -- ($(2)!0.5!(0)$);
\draw[green, line width=2.5pt, opacity=0.4] (0) -- ($(0)!0.5!(2)$);
\draw[green, line width=2.5pt, opacity=0.4] (1) -- ($(1)!0.5!(6)$);
\draw[green, line width=2.5pt, opacity=0.4] (6) -- ($(6)!0.5!(1)$);
\draw[green, line width=2.5pt, opacity=0.4] (9) -- ($(9)!0.5!(8)$);
\draw[green, line width=2.5pt, opacity=0.4] (8) -- ($(8)!0.5!(9)$);
\draw[green, line width=2.5pt, opacity=0.4] (1) -- ($(1)!0.5!(4)$);
\draw[red, line width=2.5pt, opacity=0.4] (4) -- ($(4)!0.5!(1)$);
\draw[green, line width=2.5pt, opacity=0.4] (2) -- ($(2)!0.5!(4)$);
\draw[red, line width=2.5pt, opacity=0.4] (4) -- ($(4)!0.5!(2)$);
\draw[red, line width=2.5pt, opacity=0.4] (4) -- ($(4)!0.5!(9)$);
\draw[green, line width=2.5pt, opacity=0.4] (9) -- ($(9)!0.5!(4)$);
\end{pgfonlayer}
\end{tikzpicture}}}_{ZIZX\otimes Z},\quad
  \underbrace{\scalebox{0.8}{\begin{tikzpicture}[scale=0.525]
    \begin{pgfonlayer}{nodelayer}
        \node [style=none] (0) at (-1.00, 4.00) {};
        \node [style=Z dot] (1) at (3.00, -0.00) {};
        \node [style=Z dot] (2) at (0.00, 3.00) {};
        \node [style=Z dot] (3) at (6.00, 3.00) {};
        \node [style=Z dot] (4) at (0.00, -3.00) {};
        \node [style=none] (5) at (7.00, 4.00) {};
        \node [style=none] (6) at (4.00, -0.00) {};
        \node [style=none] (7) at (-1.00, -4.00) {};
        \node [style=none] (8) at (7.00, -4.00) {};
        \node [style=Z dot] (9) at (6.00, -3.00) {};
    \end{pgfonlayer}
    \begin{pgfonlayer}{edgelayer}
        \draw (0) to (2);
        \draw [style=hadamard edge] (1) to (3);
        \draw [style=hadamard edge] (1) to (2);
        \draw [style=hadamard edge] (1) to (4);
        \draw [style=hadamard edge] (1) to (9);
        \draw (1) to (6);
        \draw [style=hadamard edge] (2) to (4);
        \draw [style=hadamard edge] (2) to (3);
        \draw (3) to (5);
        \draw [style=hadamard edge] (3) to (9);
        \draw (4) to (7);
        \draw [style=hadamard edge] (4) to (9);
        \draw (8) to (9);
    \end{pgfonlayer}

\begin{pgfonlayer}{edgelayer}
\draw[red, line width=2.5pt, opacity=0.4] (1) -- ($(1)!0.5!(6)$);
\draw[red, line width=2.5pt, opacity=0.4] (6) -- ($(6)!0.5!(1)$);
\draw[red, line width=2.5pt, opacity=0.4] (4) -- ($(4)!0.5!(7)$);
\draw[red, line width=2.5pt, opacity=0.4] (7) -- ($(7)!0.5!(4)$);
\draw[red, line width=2.5pt, opacity=0.4] (9) -- ($(9)!0.5!(8)$);
\draw[red, line width=2.5pt, opacity=0.4] (8) -- ($(8)!0.5!(9)$);
\draw[red, line width=2.5pt, opacity=0.4] (1) -- ($(1)!0.5!(3)$);
\draw[green, line width=2.5pt, opacity=0.4] (3) -- ($(3)!0.5!(1)$);
\draw[red, line width=2.5pt, opacity=0.4] (1) -- ($(1)!0.5!(2)$);
\draw[green, line width=2.5pt, opacity=0.4] (2) -- ($(2)!0.5!(1)$);
\draw[blue, line width=2.5pt, opacity=0.4] (1) -- ($(1)!0.5!(4)$);
\draw[blue, line width=2.5pt, opacity=0.4] (4) -- ($(4)!0.5!(1)$);
\draw[blue, line width=2.5pt, opacity=0.4] (1) -- ($(1)!0.5!(9)$);
\draw[blue, line width=2.5pt, opacity=0.4] (9) -- ($(9)!0.5!(1)$);
\draw[green, line width=2.5pt, opacity=0.4] (2) -- ($(2)!0.5!(4)$);
\draw[red, line width=2.5pt, opacity=0.4] (4) -- ($(4)!0.5!(2)$);
\draw[green, line width=2.5pt, opacity=0.4] (3) -- ($(3)!0.5!(9)$);
\draw[red, line width=2.5pt, opacity=0.4] (9) -- ($(9)!0.5!(3)$);
\draw[blue, line width=2.5pt, opacity=0.4] (4) -- ($(4)!0.5!(9)$);
\draw[blue, line width=2.5pt, opacity=0.4] (9) -- ($(9)!0.5!(4)$);
\end{pgfonlayer}
\end{tikzpicture}}}_{IIXX\otimes X}
\;\Bigg\rangle \ .
\end{split}
\end{equation}
These stabilisers closely mirror those of the $[[4,1,2]]$ encoder, $\langle ZZZZ\otimes I,\; XIXI\otimes I,\; IXIX\otimes I,\; XXII\otimes X,\; ZIZI\otimes Z \rangle$, with the central bulk degree of freedom playing the role of the logical qubit. 

Following \cite{jahn2023holographic,Steinberg_2025}, Hadamard edges are retained between all contracted tensor legs. Iterating this construction over the layers of a hyperbolic tessellation generates a family of ZX-diagrams, from which we extract quantum codes by computing Pauli webs supported exclusively on the boundary qubits, as shown in equation \ref{eq:zxholo_codes}. We refer to this family as the $\{4,5\}$ ZX-holographic code. Next, we shall study the logical error rate of this family of codes over the erasure and depolarising error model under the code capacity setting.

\subsection{Erasure decoding}
\begin{figure}[!h]
    \centering
    \includegraphics[width=1\linewidth]{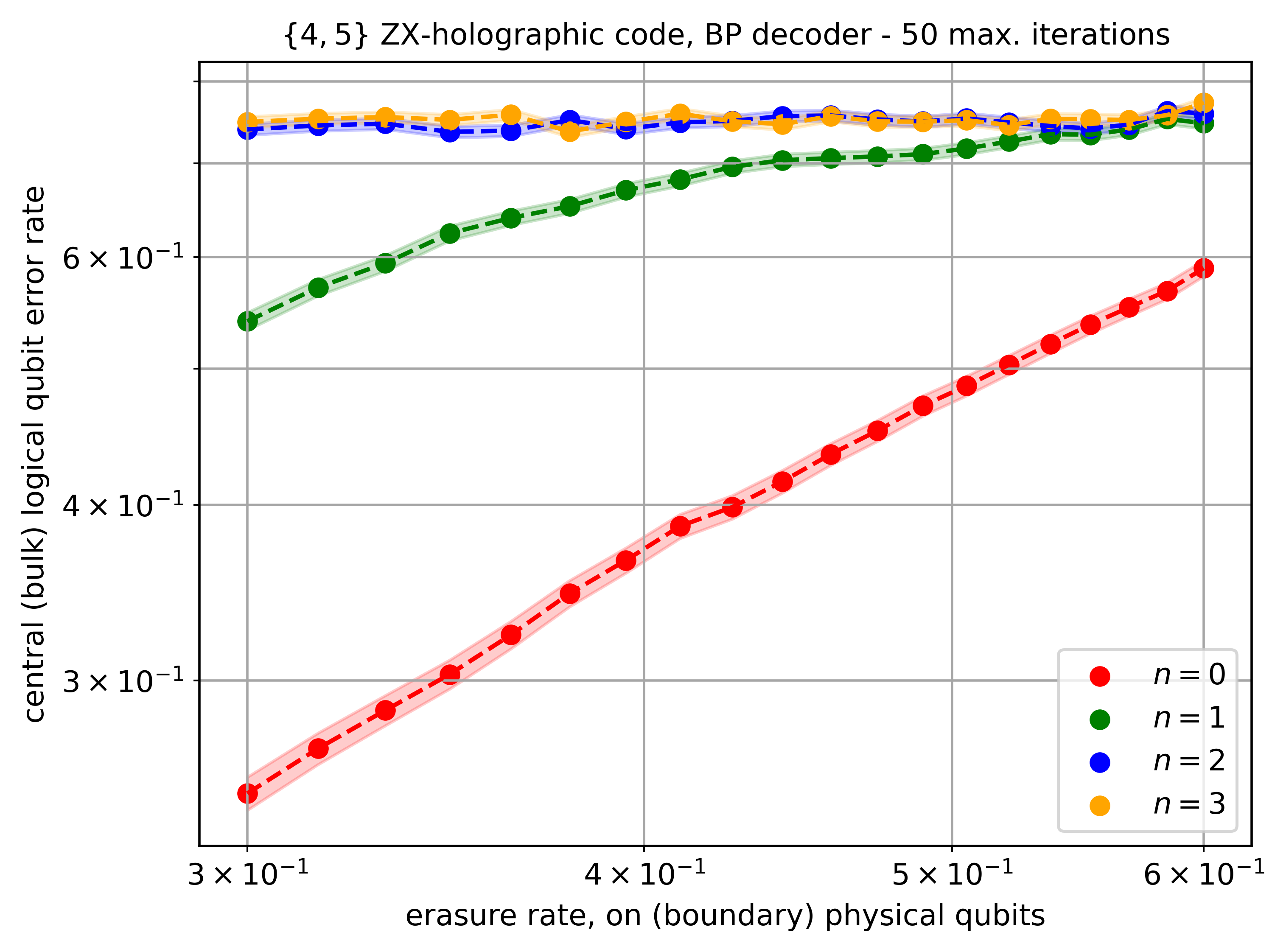}
    \caption{Logical error rate of the central bulk qubit for the $\{4,5\}$ ZX-holographic code without gauge fixing, decoded using belief propagation, under erasure errors at the boundary.}
    \label{fig:54zx_holo_no_gauge}
\end{figure}

Firstly, we consider an erasure noise model acting on the boundary qubits, where each qubit is erased independently with probability $p$ at known locations. Decoding is performed using the $\mathtt{product\_sum}$ belief-propagation (BP) decoder implemented in \texttt{ldpc} \cite{roffe_decoding_2020} using a parity check matrix generated. As in \cite{Steinberg_2025}, we focus on the logical error rate of the central bulk qubit. Without gauge fixing, we observe no erasure threshold as the number of layers $n$ increases (see figure \ref{fig:54zx_holo_no_gauge}).

Following a gauge-fixing prescription, we can project all bulk qubits except the central one onto fixed eigenstates of either $X$ or $Z$, thereby converting them into logical gauge qubits. In the corresponding ZX-diagram, this is equivalent to fusing all but one bulk leg with either $\scalebox{1}{
\begin{tikzpicture}
    \begin{pgfonlayer}{nodelayer}
        \node [style=X dot] (1) at (0.00, -0.00) {};
    \end{pgfonlayer}
\end{tikzpicture}
}$ or $\scalebox{1}{
\begin{tikzpicture}
    \begin{pgfonlayer}{nodelayer}
        \node [style=Z dot] (1) at (0.00, -0.00) {};
    \end{pgfonlayer}
\end{tikzpicture}
}$ at the yellow nodes ($\scalebox{1}{
\begin{tikzpicture}
    \begin{pgfonlayer}{nodelayer}
        \node [style=Z dot,yellow,draw=black] (1) at (0.00, -0.00) {};
    \end{pgfonlayer}
\end{tikzpicture}
}$) in equation \ref{eq:zx_holo_0} (for example taking $n=1$):
\begin{equation}
    \label{eq:zx_holo_0}
    \raisebox{-17ex}{\scalebox{0.4}{\includegraphics{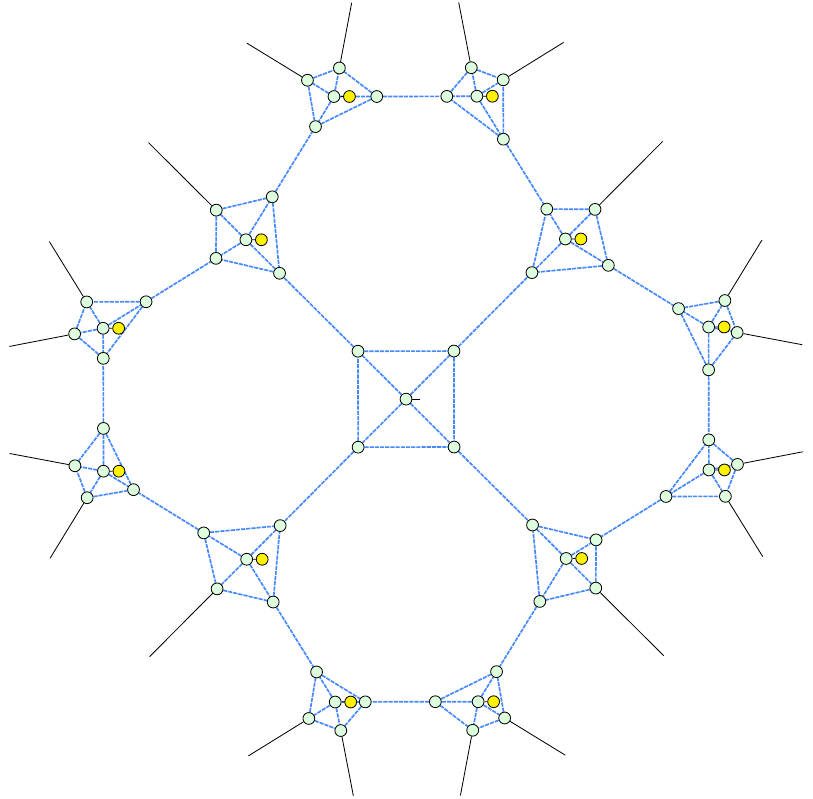}}} \ .
\end{equation}

\begin{figure}[!h]
    \centering
    \includegraphics[width=0.95\linewidth]{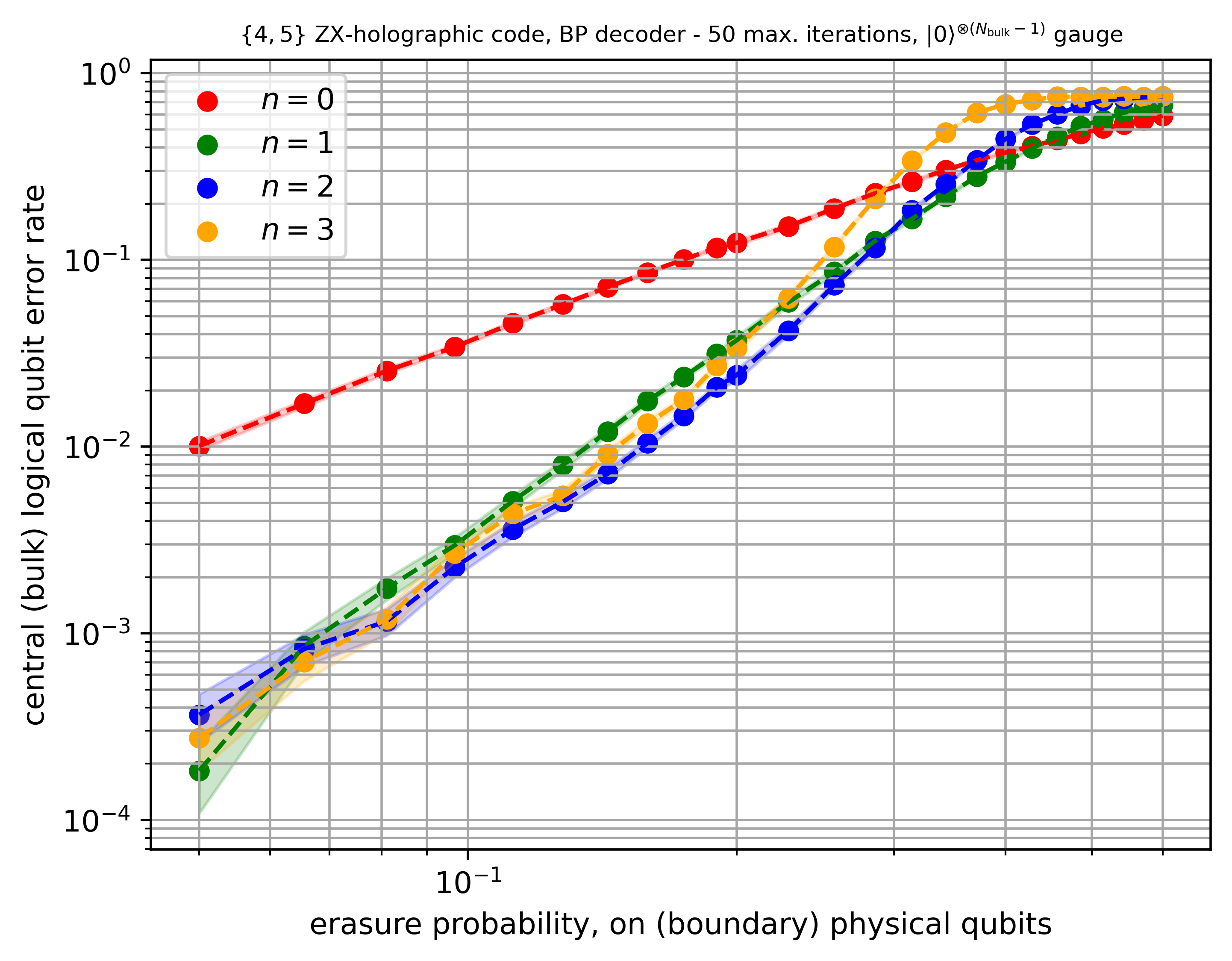}
    \caption{Logical error rate of the central bulk qubit for $\ket{0}^{\otimes (N_{\mathrm{bulk}}-1)}$ gauge fixing, decoded using BP under erasure errors at the boundary.}
    \label{fig:54zx_holo_0_gauge_BP}
\end{figure}

Fixing the gauge qubits to $\ket{0}^{\otimes (N_{\mathrm{bulk}}-1)}$ (corresponding to $Z$-eigenstates projection) yields no threshold and no improvement in logical performance with increasing $n$ (figure \ref{fig:54zx_holo_0_gauge_BP}). In contrast, projecting the gauge qubits onto $\ket{+}^{\otimes (N_{\mathrm{bulk}}-1)}$ (corresponding to $X$-eigenstate projection) produces clear sub-threshold scaling under BP decoding.

\begin{figure}[!h]
    \centering
    \includegraphics[width=0.95\linewidth]{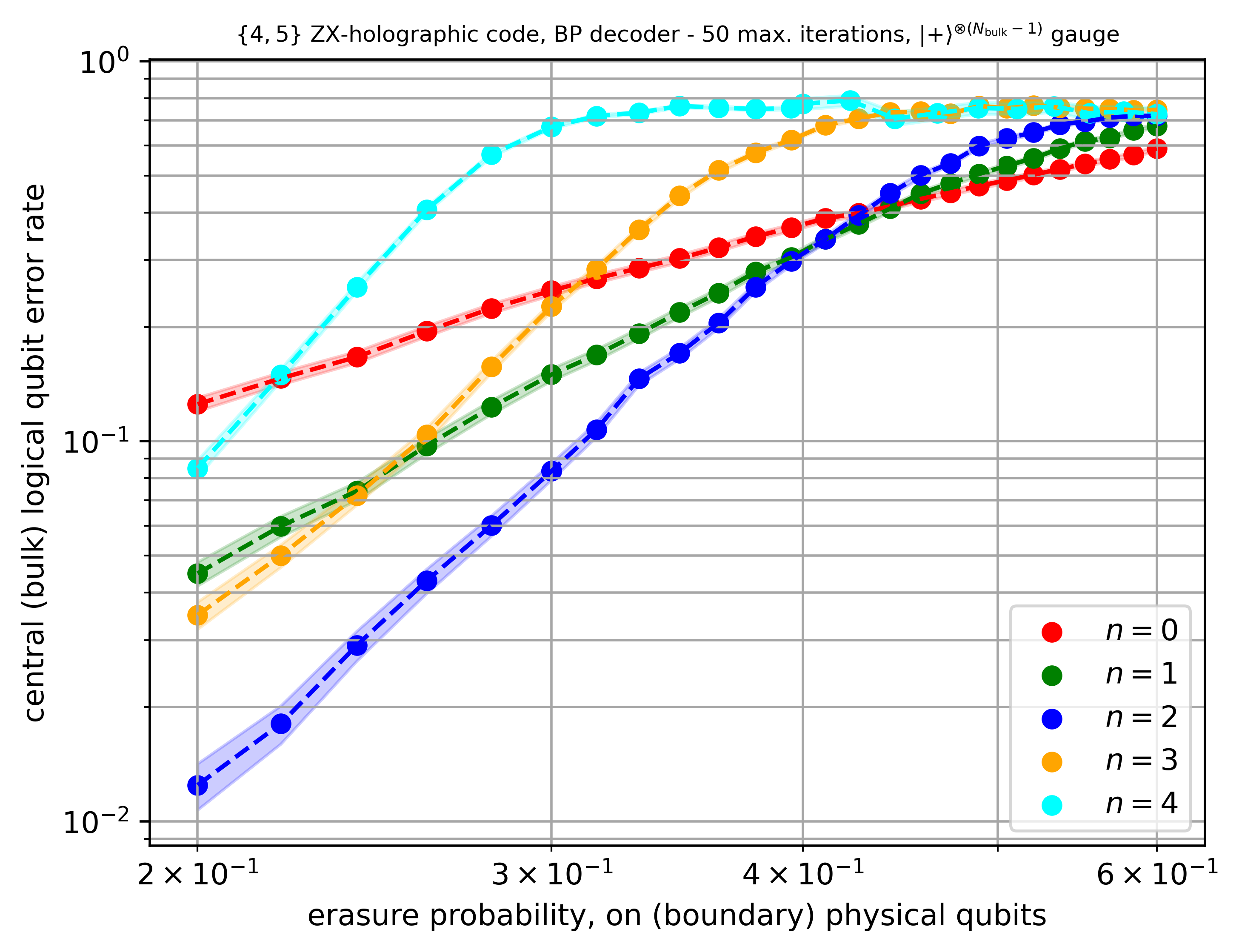}
    \caption{Logical error rate of the central bulk qubit for $\ket{+}^{\otimes (N_{\mathrm{bulk}}-1)}$ gauge fixing, decoded using BP.}
    \label{fig:54zx_holo_plus_gauge_BP}
\end{figure}

As $n$ increases, the intersection point (``crossing'') of the logical-error curves shifts to lower erasure rates. For example, the $n=3$ curve crosses the others at a smaller erasure rate in figure \ref{fig:54zx_holo_plus_gauge_BP}. This suggests that BP alone is not well matched to this decoding problem. Using BP decoding together with post-processing such as ordered-statistics decoding (OSD) \cite{Roffe_LDPC_Python_tools_2022,roffe_decoding_2020} may be beneficial.

\begin{figure}[!h]
    \centering
    \includegraphics[width=0.95\linewidth]{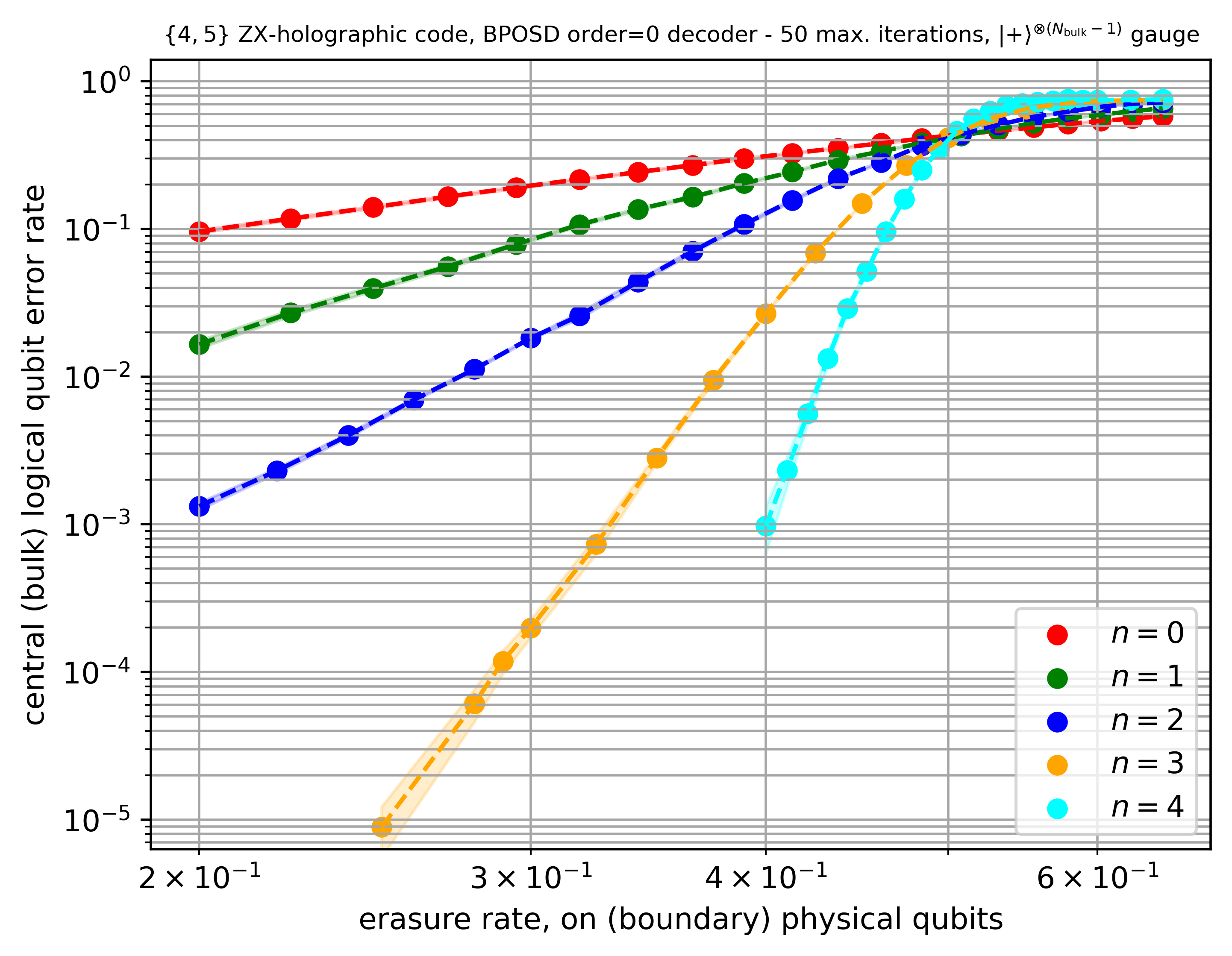}
    \caption{Logical error rate of the central bulk qubit for $\ket{+}^{\otimes (N_{\mathrm{bulk}}-1)}$ gauge fixing, decoded using BP+OSD$-0$, under erasure errors at the boundary.}
    \label{fig:54zx_holo_plus_gauge_BPosd_order0}
\end{figure}

An even stronger suppression of logical errors below threshold is obtained by decoding the $\ket{+}^{\otimes (N_{\mathrm{bulk}}-1)}$ gauged code using belief propagation with (order $=0$) ordered statistics decoding (BP+OSD-$0$), as shown in figure \ref{fig:54zx_holo_plus_gauge_BPosd_order0}. Surprisingly, despite its simplicity, BP+OSD-$0$ performs remarkably well, achieving near-optimal threshold at $\approx\frac{1}{2}$ erasure rate in the boundary qubits.

For optimal erasure decoding in the low-erasure regime, we expect the logical error rate ($p_\text{L}$) to scale as:
\begin{equation}
    p_\text{L} \propto p_e^{d} \ ,
\end{equation}
under the erasure error model with erasure rate $p_e$ and code distance $d$ \cite{Connolly_2024}. With this in mind, we can estimate an effective code distance for increasing $n$ by fitting the leading power-law dependence of the numerically obtained values of logical error rates from figure \ref{fig:54zx_holo_plus_gauge_BPosd_order0}, resulting in table \ref{tab:dist}.
\begin{table}[h!]
\centering
\resizebox{.95\columnwidth}{!}{%
\begin{tabular}{|c|c|c|c|}
\hline
\multicolumn{4}{|c|}{distance scaling} \\
\hline
layers ($n$) & fit ($\approx p_\text{L}$) & distance ($\approx d$) & $N_{\mathrm{boundary}}$\\
\hline
$0$ & $2\times p_e^2$     & $2$  & $4$ \\
$1$ & $10\times p_e^4$     & $4$  & $20$ \\
$2$ & $20\times p_e^6$     & $6$  & $76$ \\
$3$ & $50000\times p_e^{16}$  & $16$ & $284$ \\
$4$ & $10^{9}\times p_e^{30}$  & $30$ & $1060$ \\
\hline
\end{tabular}%
}
\caption{\label{tab:dist}Approximate distance scaling and qubit count for the $\{4,5\}$ ZX-holographic code with $X$-gauge, keeping only the central bulk qubit.}
\end{table}
We emphasise that the extracted distances should be interpreted as effective (approximate) distances inferred from finite error range fits, rather than exact minimum weight logical operators. The prefactors vary significantly with $n$ and are not expected to be universal.

Next, we will look at the code's error tolerance to random Pauli errors at the boundary qubits.

\subsection{Pauli error decoding}
\begin{figure}[!h]
    \centering
    \includegraphics[width=0.95\linewidth]{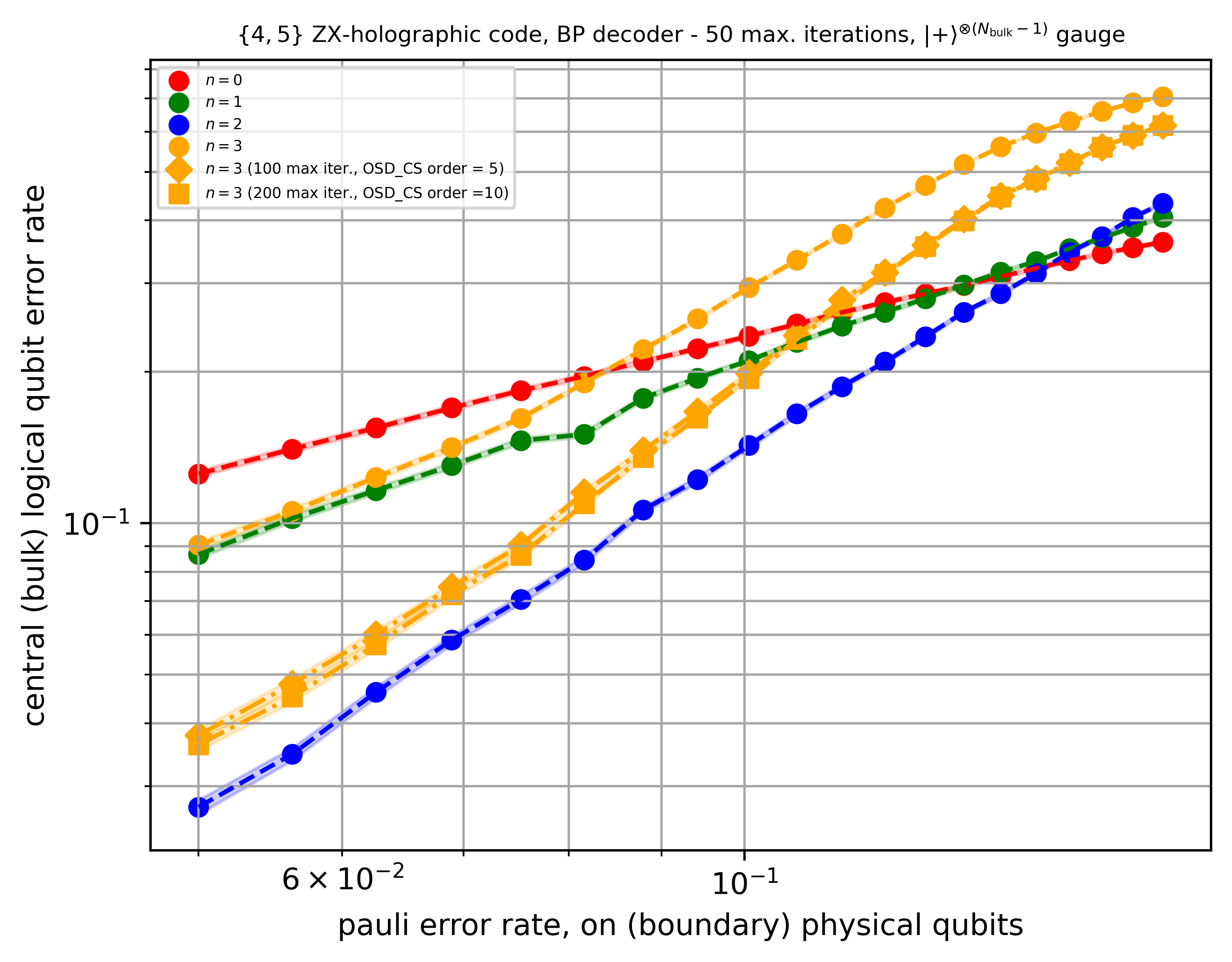}
    \caption{Logical error rate of the central bulk qubit for $\ket{+}^{\otimes (N_{\mathrm{bulk}}-1)}$ gauge fixing, decoded using BP and BP+OSD, under random Pauli depolarising errors at the boundary.}
    \label{fig:45_pauli}
\end{figure}

In the same $\ket{+}^{\otimes (N_{\mathrm{bulk}}-1)}$ gauge fixed ZX-diagram, we subjected each boundary qubits to an independent Pauli-$X$, $Y$ and $Z$ flip error, each with a probability $p/3$. This is the depolarising channel with the following map on every boundary qubit sub-system:
\begin{equation}
    \rho \rightarrow (1-p)\rho + \frac{p}{3}\big(X\rho X+ Y\rho Y + Z\rho Z\big) \ .
\end{equation}
We perform BP decoding under this depolarising Pauli noise model and observe that, while small instances show error suppression, the $n=3$ curve fails to improve over $n=2$, the logical error rate does not decrease with increasing code size, indicating that BP is not effectively exploiting the expected growth in distance. Augmenting BP with ordered statistics decoding provides only marginal improvement, even when increasing the maximum number of BP iterations to $200$ and the OSD order up to $10$ (see figure \ref{fig:45_pauli}). This suggests that decoding under Pauli depolarising noise is decoder-limited in our setting. 

Inspecting the stabiliser parity-check matrix $H\in\mathbb{F}_2^{m\times 2N_\text{boundary}}$ reveals that this may not primarily be a property of the code, but of the chosen generating set presented to BP. In particular, for larger $n$ the generator basis exhibits a heavy tail of very high-weight checks and large overlaps\footnote{Here ``large overlaps'' refers to pairs of stabiliser generators whose supports intersect on many Pauli variables (columns): for rows $H_i,H_j$ of the check matrix $H$, the overlap is
\(
|\mathrm{supp}(H_i)\cap \mathrm{supp}(H_j)|
=\sum_k (H_{i,k}\wedge H_{j,k}) \ ,
\)
and large values indicate highly correlated checks.}, well known to degrade BP \cite{Roffe_LDPC_Python_tools_2022}. To mitigate this, we smooth the stabiliser generating set by multiplying stabilisers together, which corresponds to elementary row operations over $\mathbb{F}_2$ on $H$. We represent Pauli operators on $N_{\text{boundary}}$ qubits (up to an overall phase) by the symplectic binary vector
\begin{equation}
h(g) = (x_1,z_1,\dots,x_{N_\text{boundary}},z_{N_\text{boundary}})\in\mathbb{F}_2^{2N_\text{boundary}} \ ,
\end{equation}
so that a stabiliser generator $g_i$ is encoded by the $i$-th row $H_i$. Since the product of two stabilisers is again a stabiliser, replacing $g_i$ by $g_i g_j$ yields an equivalent generating set. In the binary symplectic representation this multiplication is simply component-wise addition modulo $2$,
\begin{equation}
h(g_i g_j)=h(g_i)\oplus h(g_j) \ ,
\end{equation}
and hence corresponds to the row update
\begin{equation}
H_i \leftarrow H_i \oplus H_j \ .
\end{equation}

Such row operations preserve the row space of $H$ (and thus the stabiliser group and the code), but change the Tanner graph seen by BP. Our smoothing heuristic repeatedly selects a currently high-weight row $i$ and searches for another row $j$ whose addition decreases its Hamming weight, updating $H_i\leftarrow H_i\oplus H_j$ whenever this strictly reduces the weight. Iterating this procedure substantially reduces high check weights and overlaps.

\begin{figure}[!h]
    \centering
    \includegraphics[width=0.95\linewidth]{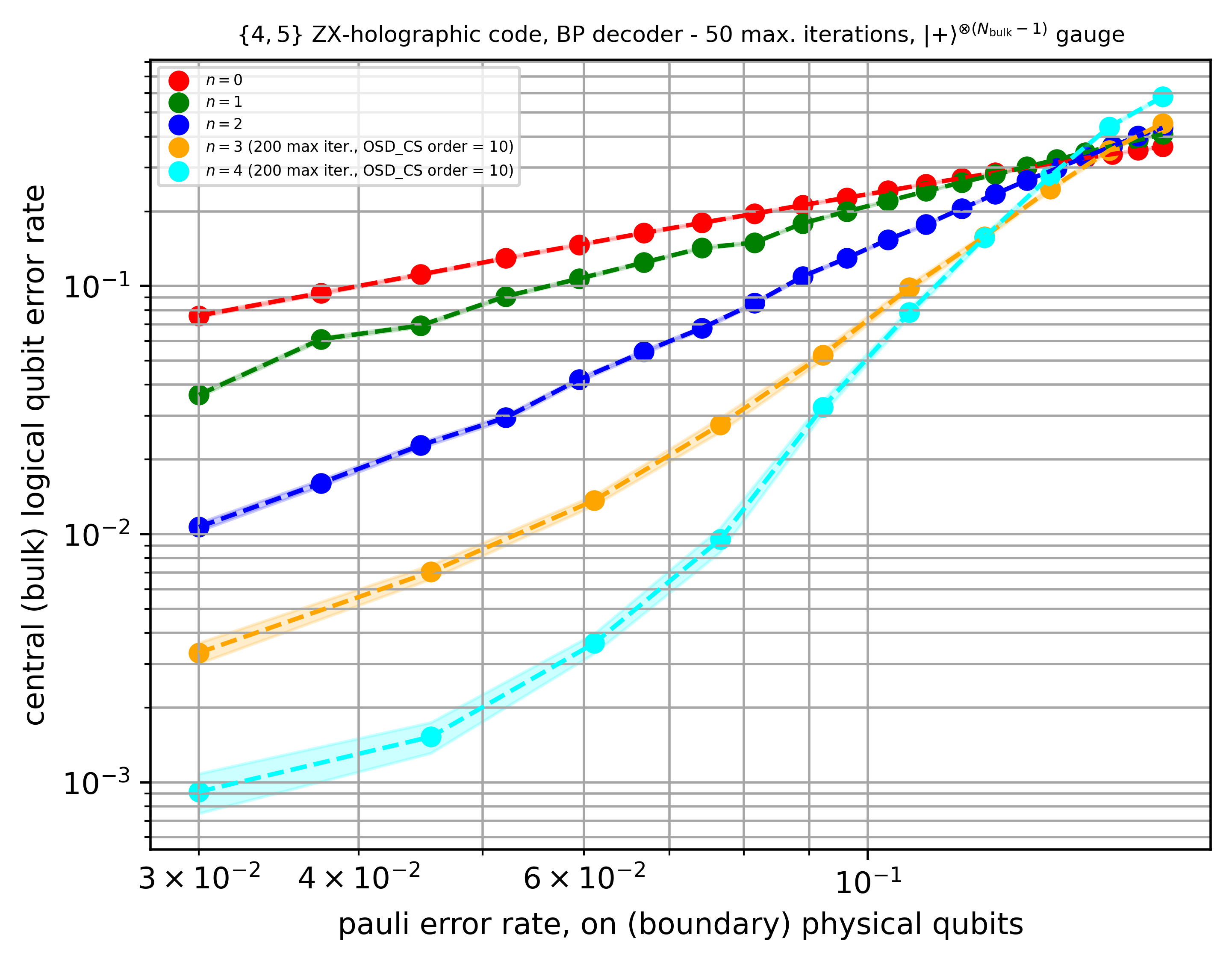}
    \caption{Logical error rate of the central bulk qubit for $\ket{+}^{\otimes (N_{\mathrm{bulk}}-1)}$ gauge fixing, decoded using BP (for $n=0,1,2$) BP+OSD (for $n=3,4$), under random Pauli depolarising errors at the boundary and heuristic row operations that reduce stabiliser weight.}
    \label{fig:45_pauli_better}
\end{figure}

In the simulations, we applied a fixed number of such row-combination attempts, using a random subsampling strategy. We ran up to $8000$ iterations in which the currently heaviest check row is selected and XOR-combined with one of $1200$ randomly chosen candidate rows whenever this strictly reduces its Hamming weight. The procedure terminates early once all check rows have weight at most $10$. This yields improved logical error suppression in figure \ref{fig:45_pauli_better}, where decoding is performed using BP+OSD (order $=10$ and maximum iterations $=200$) for $n=3$ and $n=4$. This further supports the conclusion that the observed performance is largely decoder limited. The weight/overlap profile of the chosen stabiliser generators, together with the limitations of BP (even when augmented with OSD), damages the achievable logical error rate.

\subsection{Erasure vs Pauli noise error-suppression region}
\label{sec:2d_error_suppression}
Until now we have looked at two extremes: (i) pure erasures on the boundary, and (ii) pure depolarising Pauli noise on the boundary. A simple way to combine these is to let the boundary experience both types of noise: some qubits are erased (with locations known), and the remaining qubits may also suffer random Pauli flips. This gives two parameters: an erasure rate $p_e$ and a Pauli error rate $p_r$. Each boundary qubit is erased with probability $p_e$; conditioned on not being erased, it undergoes a depolarising Pauli noise channel with probability $p_r$.
\begin{figure}[!h]
    \centering
    \includegraphics[width=0.95\linewidth]{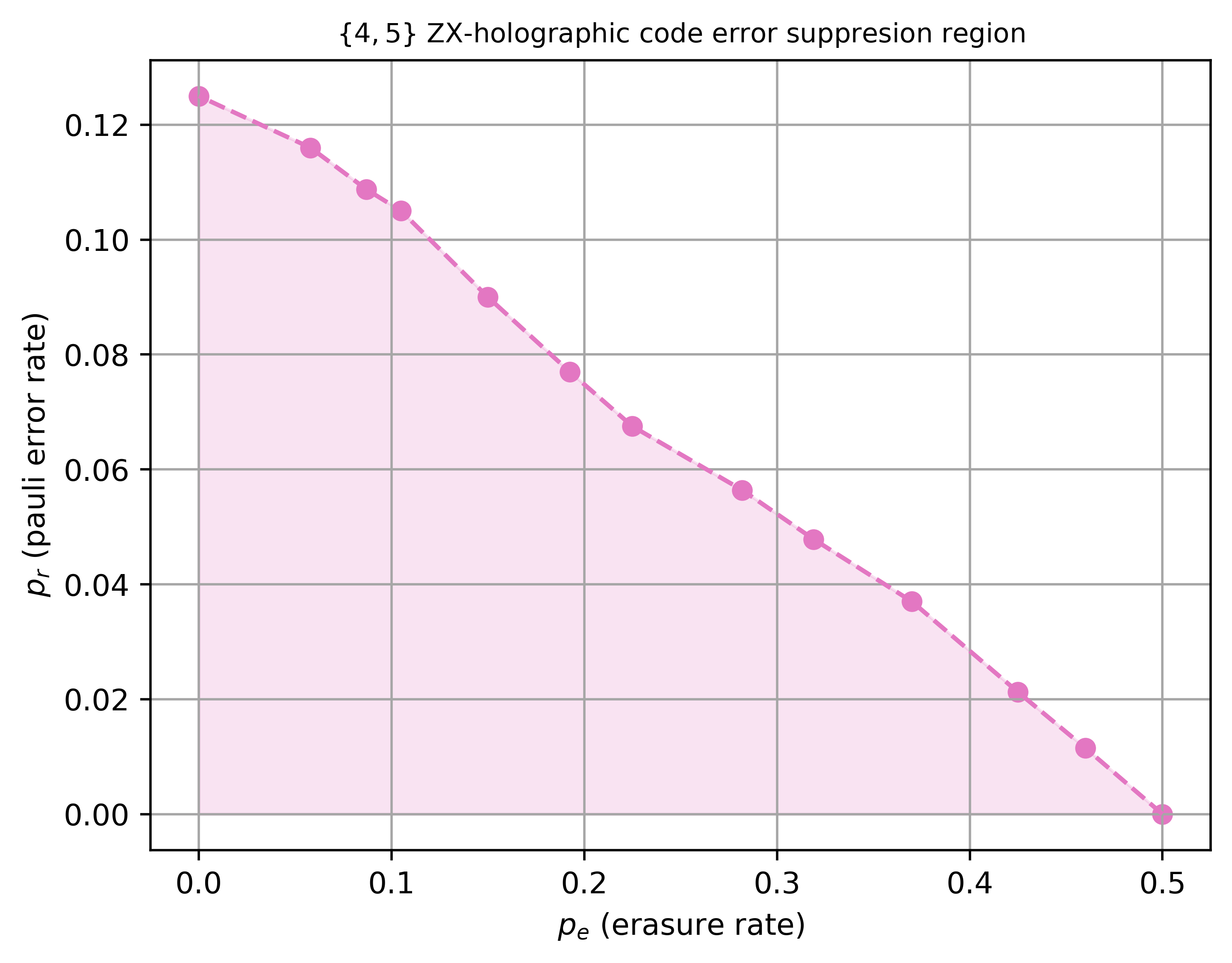}
    \caption{Error suppression region of the central bulk qubit in the $(p_e,p_r)$ plane. $\ket{+}^{\otimes (N_{\mathrm{bulk}}-1)}$ gauge fixing, decoded using BP+OSD with order $= 10$ and $200$ maximum iterations. The heuristic row operations that reduce stabiliser weight was also used.}
    \label{fig:es_region}
\end{figure}

We want a single figure that shows, in the $(p_e,p_r)$ plane, where the code looks like it is suppressing logical errors as we increase the number of layers $n$. In other words, we are recording the crossing points of all logical error rates curves for increasing layers. For each point $(p_e,p_r)$ we run the same decoding procedure as before (same gauge choice, same decoder, same stabiliser representation), and we compare the logical error rate of the central bulk qubit for two (or more) code sizes. If the larger code performs better than the smaller code at that noise point, we mark it as ``error suppression'', if it does not, we mark it as ``no suppression''. Plotting this over a grid of $(p_e,p_r)$ produces a two-dimensional phase diagram. 

We refer to the shaded region in figure \ref{fig:es_region}, where larger codes outperform smaller ones as an error-suppression region. This is intentionally a modest term, it is a finite-size, decoder-dependent region, and it should not be over interpreted as a fault tolerant region \cite{Paesani_2023}. Nonetheless, it is a convenient summary of how the apparent crossing behaviour shifts when we interpolate between the erasure dominated to Pauli error dominated regimes.

\section{Spacetime ZX-diagrams}
\label{sec:spacetimeZX}\label{app:spacetimeZX}
The constructions in this section are loosely inspired by the hyperbolic placement of tensors/ZX-diagrams in the holographic codes of the previous sections. The $\{4,5\}$ network supplies the tiles, bonds and boundary of the block, but the resulting spacetime code is a dynamical, repeatedly measured object in its own right rather than a static encoder.
An encoding map $V$ and its inverse $V^{\dagger}$ can be represented as the following quantum circuits:
\begin{equation}
\begin{array}{c}
\begin{tikzpicture}
\begin{yquant*}
init {$N_{\text{bulk}}$} (q[0-2]);
init {} (q[3, 4]);

discard q[1];
text {$\rvdots$} (q[1]);
discard q[3-];

hspace {.4cm} -;
box {\hspace{0.5cm}$V$\hspace{0.5cm}} (q);

settype {qubit} q[1,4];
discard q[3];
text {$\rvdots$} (q[3]);

output {$N_{\text{boundary}}$} (q[0-4]);
\end{yquant*}
\end{tikzpicture}
\\[2ex]
\begin{tikzpicture}
\begin{yquant*}
init {$N_{\text{boundary}}$} (q[0-4]);

discard q[3];
text {$\rvdots$} (q[3]);

hspace {.4cm} -;
box {\hspace{0.5cm}$V^{\dagger}$\hspace{0.5cm}} (q);

discard q[1];
text {$\rvdots$} (q[1]);

output {$N_{\text{bulk}}$} (q[0-2]);
discard q[3-];
\end{yquant*}
\end{tikzpicture}
\end{array} \ .
\end{equation}
Inspired by the wormhole construction in which the bulk legs of two copies of $V$ are fused, one way to transmit quantum information across time (left to right here) is to join $V$ and $V^{\dagger}$ together. This represents an `un-encode then encode' map:
\begin{equation}
\scalebox{0.85}{\begin{tikzpicture}
\begin{yquant*}
init {} q[0-4];
discard q[3];
text {$\rvdots$} (q[3]);
hspace {.5cm} -;
box {\hspace{0.5cm}$V^{\dagger}$\hspace{0.5cm}} (q);
discard q[1];
discard q[4];
hspace {.5cm} -;
text {$\rvdots$} (q[1]);
discard q[3-];
hspace {.5cm} -;
box {\hspace{0.6cm}$V$\hspace{0.6cm}} (q);
settype {qubit} q[1,4];
hspace {.5cm} -;
discard q[3];
text {$\rvdots$} (q[3]);
\end{yquant*}
\end{tikzpicture}} \ .
\end{equation}
For the $\{4,5\}$ ZX-holographic code, we can `foliate' the ZX-diagrams representing the encoding (and un-encoding) maps by joining the structures together in the following way:
\begin{equation}
\scalebox{0.58}{\begin{tikzpicture}
\begin{yquant*}
init {} q[0-4];
discard q[3];
text {$\rvdots$} (q[3]);
hspace {.5cm} -;
box {\hspace{0.5cm}$V^{\dagger}$\hspace{0.5cm}} (q);
discard q[1];
discard q[4];
hspace {.5cm} -;
text {$\rvdots$} (q[1]);
discard q[3-];
hspace {.5cm} -;
box {\hspace{0.6cm}$V$\hspace{0.6cm}} (q);
settype {qubit} q[1,4];
hspace {.5cm} -;
discard q[3];
text {$\rvdots$} (q[3]);
hspace {.5cm} -;
box {\hspace{0.5cm}$V^{\dagger}$\hspace{0.5cm}} (q);
discard q[1];
discard q[4];
hspace {.5cm} -;
text {$\rvdots$} (q[1]);
discard q[3-];
hspace {.5cm} -;
box {\hspace{0.6cm}$V$\hspace{0.6cm}} (q);
settype {qubit} q[1,4];
hspace {.5cm} -;
discard q[3];
text {$\rvdots$} (q[3]);
\end{yquant*}
\end{tikzpicture}} \ ,
\end{equation}
then we perform spider fusion on any bulk or boundary legs shared between adjacent $V$ and $V^{\dagger}$ blocks. The resulting object can be represented as a spacetime ZX-diagram; we depict the $n=1$ instance for the $\{4,5\}$ ZX-holographic code (with time running bottom to top) in figure \ref{fig:45foliated}a). Checks (closed Pauli webs) and logical correlators spanning bottom to top boundary legs exist (figures \ref{fig:45foliated}b) and \ref{fig:45foliated}c)). Decoding erasures (erasure permissible on every edge in this ZX-diagram) directly on this fused diagram shows no threshold-like behaviour, since fusion leaves only a handful of detectors ($7$ at $n=1$, $15$ at $n=2$) protecting a many-qubit logical channel, and larger diagrams are worse at every physical erasure rate. This motivates the bond-extended form below.

\begin{figure*}[!t]
\centering
\begin{tabular}{ccc}
\input{figures/spacetime45_foliated.tikz} & \input{figures/spacetime45_checks.tikz} & \input{figures/spacetime45_logical.tikz} \\
{\small (a)} & {\small (b)} & {\small (c)} \\
\end{tabular}
\caption{a) A spacetime ZX-diagram via joining together encoding and un-encoding maps of the $\{4,5\}$ ZX-holographic code ($n=1$, two $VV^{\dagger}$ units; time runs bottom to top). b) an example of a closed Pauli web representing a check, and c) an example of a logical correlator spanning time, both drawn on the same diagram.}
\label{fig:45foliated}
\end{figure*}

The resulting spacetime ZX-diagram should be understood as more than a fixed tensor-network contraction: it is a proposal for a \emph{spacetime code}, in which every edge of the diagram is a location where things can go wrong. Any wire may be erased or may experience a Pauli error, be it a worldline segment of a leg qubit, a join between adjacent $V$ and $V^{\dagger}$ blocks, or a measurement event. From this point of view, the diagram specifies simultaneously the code, the measurement schedule that protects it, and the complete set of fault locations.

To make this concrete, we recast the network in a \emph{bond-extended} form: every tensor leg-end hosts a qubit ($52$ qubits for the $n=1$ $\{4,5\}$ network), and each contracted bond between two tiles is replaced by a pair of two-qubit checks ($X\otimes Z$ and $Z\otimes X$ across the Hadamard-twisted bond) which are measured repeatedly rather than contracted once. Gauge fixing the non-central bulk legs to $|+\rangle$ turns each non-central tile into a fixed stabiliser state carried by its four leg qubits, while the central tile retains the single logical qubit, giving a $[[52,1]]$ code. Foliating the resulting group of $51$ checks over $K$ rounds produces exactly the object of figure \ref{fig:spacetime_conventions}: the tile tensors enter once at the initial time slice, their leg qubits continue as bare worldlines, and each check contributes one measurement event per round. Closed Pauli webs are the detectors of this code: an erased or faulty edge is flagged by the measurement events its web touches. Open webs carry the logical correlator across time. Examples of both, drawn on the block itself, are shown in figure~\ref{fig:newwebs}. Figure \ref{fig:spacetime_recipe} summarises this pipeline for the $n=1$ instance. Figure~\ref{fig:block_anatomy} overlays this dictionary directly on the spacetime block: with the rest of the diagram faded, one tile's four leg-qubit worldlines, one boundary-qubit worldline and one bond's measured stabiliser pair are singled out in the colours of figure~\ref{fig:spacetime_recipe}. Diagrammatically, a measurement event is nothing but a phase-free Z-spider hanging off the data worldlines by Hadamard edges (figure~\ref{fig:spacetime_conventions}), and where each edge lands determines which Pauli it measures on that qubit. Attaching \emph{after} the worldline's inter-round Hadamard measures $X$, and attaching \emph{before} it measures $Z$. The two events of a bond attach on opposite sides of their two strands and share a Hadamard edge with each other, realising exactly the measured pair $X\otimes Z$, $Z\otimes X$.

\begin{figure*}[!t]
\centering
\begin{tabular}{cc}
\begin{tikzpicture}[scale=0.8]
	\begin{pgfonlayer}{edgelayer}
		\draw[style=hadamard edge] (0.000, 2.125) to (-1.062, 1.840);
		\draw[style=hadamard edge] (0.000, 0.000) to (-1.062, 1.840);
		\draw[style=hadamard edge] (1.840, -1.063) to (2.125, -0.000);
		\draw[style=hadamard edge] (2.125, -0.000) to (1.840, 1.062);
		\draw[style=hadamard edge] (-1.840, -1.062) to (-1.062, -1.840);
		\draw[style=hadamard edge] (-0.000, -2.125) to (1.062, -1.840);
		\draw[style=hadamard edge] (-2.125, 0.000) to (-1.840, -1.062);
		\draw[style=hadamard edge] (0.000, 2.125) to (1.063, 1.840);
		\draw[style=hadamard edge] (-1.840, 1.063) to (-2.125, 0.000);
		\draw[style=hadamard edge] (0.000, 0.000) to (1.840, 1.062);
		\draw[style=hadamard edge] (-1.062, 1.840) to (-1.840, 1.063);
		\draw[style=hadamard edge] (0.000, 0.000) to (-1.840, -1.062);
		\draw[style=hadamard edge] (0.000, 0.000) to (1.062, -1.840);
		\draw[style=hadamard edge] (1.840, 1.062) to (1.063, 1.840);
		\draw[style=hadamard edge] (1.062, -1.840) to (1.840, -1.063);
		\draw[style=hadamard edge] (-1.062, -1.840) to (-0.000, -2.125);
		\draw (0.000, 2.125) to (-0.337, 2.823);
		\draw (0.000, 2.125) to (0.337, 2.823);
		\draw (-1.062, 1.840) to (-1.450, 2.511);
		\draw (-1.840, 1.063) to (-2.613, 1.119);
		\draw (-1.840, 1.063) to (-2.276, 1.703);
		\draw (-2.125, 0.000) to (-2.823, -0.337);
		\draw (-2.125, 0.000) to (-2.823, 0.337);
		\draw (-1.840, -1.062) to (-2.511, -1.450);
		\draw (-1.062, -1.840) to (-1.119, -2.613);
		\draw (-1.062, -1.840) to (-1.703, -2.276);
		\draw (-0.000, -2.125) to (0.337, -2.823);
		\draw (-0.000, -2.125) to (-0.337, -2.823);
		\draw (1.062, -1.840) to (1.450, -2.511);
		\draw (1.840, -1.063) to (2.613, -1.119);
		\draw (1.840, -1.063) to (2.276, -1.703);
		\draw (2.125, -0.000) to (2.823, 0.337);
		\draw (2.125, -0.000) to (2.823, -0.337);
		\draw (1.840, 1.062) to (2.511, 1.450);
		\draw (1.063, 1.840) to (1.119, 2.613);
		\draw (1.063, 1.840) to (1.703, 2.276);
		\draw (0.000, 2.125) to (0.000, 2.485);
		\draw (-1.062, 1.840) to (-1.062, 2.200);
		\draw (-1.840, 1.063) to (-1.840, 1.423);
		\draw (-2.125, 0.000) to (-2.125, 0.360);
		\draw (-1.840, -1.062) to (-1.840, -0.702);
		\draw (-1.062, -1.840) to (-1.062, -1.480);
		\draw (-0.000, -2.125) to (-0.000, -1.765);
		\draw (1.062, -1.840) to (1.062, -1.480);
		\draw (1.840, -1.063) to (1.840, -0.703);
		\draw (2.125, -0.000) to (2.125, 0.360);
		\draw (1.840, 1.062) to (1.840, 1.422);
		\draw (1.063, 1.840) to (1.063, 2.200);
		\draw (0.000, 0.000) to (0.000, 0.620);
	\end{pgfonlayer}
	\begin{pgfonlayer}{nodelayer}
		\node [style=stub dot] (s4) at (-0.337, 2.823) {};
		\node [style=stub dot] (s7) at (0.337, 2.823) {};
		\node [style=stub dot] (s8) at (-1.450, 2.511) {};
		\node [style=stub dot] (s12) at (-2.613, 1.119) {};
		\node [style=stub dot] (s13) at (-2.276, 1.703) {};
		\node [style=stub dot] (s17) at (-2.823, -0.337) {};
		\node [style=stub dot] (s18) at (-2.823, 0.337) {};
		\node [style=stub dot] (s21) at (-2.511, -1.450) {};
		\node [style=stub dot] (s25) at (-1.119, -2.613) {};
		\node [style=stub dot] (s26) at (-1.703, -2.276) {};
		\node [style=stub dot] (s29) at (0.337, -2.823) {};
		\node [style=stub dot] (s30) at (-0.337, -2.823) {};
		\node [style=stub dot] (s34) at (1.450, -2.511) {};
		\node [style=stub dot] (s38) at (2.613, -1.119) {};
		\node [style=stub dot] (s39) at (2.276, -1.703) {};
		\node [style=stub dot] (s42) at (2.823, 0.337) {};
		\node [style=stub dot] (s43) at (2.823, -0.337) {};
		\node [style=stub dot] (s47) at (2.511, 1.450) {};
		\node [style=stub dot] (s48) at (1.119, 2.613) {};
		\node [style=stub dot] (s51) at (1.703, 2.276) {};
		\node [style=tile box] (t0) at (0.000, 0.000) {};
		\node [style=stub dot] (bc) at (0.000, 0.620) {};
		\node [style=tile box] (t1) at (0.000, 2.125) {};
		\node [style=plus dot] (p1) at (0.000, 2.485) {};
		\node [style=tile box] (t2) at (-1.062, 1.840) {};
		\node [style=plus dot] (p2) at (-1.062, 2.200) {};
		\node [style=tile box] (t3) at (-1.840, 1.063) {};
		\node [style=plus dot] (p3) at (-1.840, 1.423) {};
		\node [style=tile box] (t4) at (-2.125, 0.000) {};
		\node [style=plus dot] (p4) at (-2.125, 0.360) {};
		\node [style=tile box] (t5) at (-1.840, -1.062) {};
		\node [style=plus dot] (p5) at (-1.840, -0.702) {};
		\node [style=tile box] (t6) at (-1.062, -1.840) {};
		\node [style=plus dot] (p6) at (-1.062, -1.480) {};
		\node [style=tile box] (t7) at (-0.000, -2.125) {};
		\node [style=plus dot] (p7) at (-0.000, -1.765) {};
		\node [style=tile box] (t8) at (1.062, -1.840) {};
		\node [style=plus dot] (p8) at (1.062, -1.480) {};
		\node [style=tile box] (t9) at (1.840, -1.063) {};
		\node [style=plus dot] (p9) at (1.840, -0.703) {};
		\node [style=tile box] (t10) at (2.125, -0.000) {};
		\node [style=plus dot] (p10) at (2.125, 0.360) {};
		\node [style=tile box] (t11) at (1.840, 1.062) {};
		\node [style=plus dot] (p11) at (1.840, 1.422) {};
		\node [style=tile box] (t12) at (1.063, 1.840) {};
		\node [style=plus dot] (p12) at (1.063, 2.200) {};
	\end{pgfonlayer}
\end{tikzpicture} & \begin{tikzpicture}[scale=0.8]
	\begin{pgfonlayer}{edgelayer}
		\draw[draw=gray] (-0.532, 2.611) to (-0.845, 2.528);
		\draw[draw=gray] (-0.275, 0.477) to (-1.102, 1.908);
		\draw[draw=gray] (2.528, -0.845) to (2.611, -0.532);
		\draw[draw=gray] (2.611, 0.532) to (2.528, 0.845);
		\draw[draw=gray] (-1.996, -1.766) to (-1.766, -1.996);
		\draw[draw=gray] (0.532, -2.611) to (0.845, -2.528);
		\draw[draw=gray] (-2.611, -0.532) to (-2.528, -0.845);
		\draw[draw=gray] (0.532, 2.611) to (0.845, 2.528);
		\draw[draw=gray] (-2.528, 0.845) to (-2.611, 0.532);
		\draw[draw=gray] (0.477, 0.275) to (1.908, 1.102);
		\draw[draw=gray] (-1.766, 1.996) to (-1.996, 1.766);
		\draw[draw=gray] (-0.477, -0.275) to (-1.908, -1.102);
		\draw[draw=gray] (0.275, -0.477) to (1.102, -1.908);
		\draw[draw=gray] (1.996, 1.766) to (1.766, 1.996);
		\draw[draw=gray] (1.766, -1.996) to (1.996, -1.766);
		\draw[draw=gray] (-0.845, -2.528) to (-0.532, -2.611);
	\end{pgfonlayer}
	\begin{pgfonlayer}{nodelayer}
		\node [style=event box] (e0) at (-0.660, 2.463) {};
		\node [style=event box] (e1) at (-0.717, 2.676) {};
		\node [style=none] (chk) at (0.000, 3.950) {\scriptsize measured stabiliser pair $X{\otimes}Z,\ Z{\otimes}X$ (orange)};
		\node [style=event box] (e2) at (-0.784, 1.138) {};
		\node [style=event box] (e3) at (-0.593, 1.248) {};
		\node [style=event box] (e4) at (2.463, -0.660) {};
		\node [style=event box] (e5) at (2.676, -0.717) {};
		\node [style=event box] (e6) at (2.463, 0.660) {};
		\node [style=event box] (e7) at (2.676, 0.717) {};
		\node [style=event box] (e8) at (-1.803, -1.803) {};
		\node [style=event box] (e9) at (-1.959, -1.959) {};
		\node [style=event box] (e10) at (0.660, -2.463) {};
		\node [style=event box] (e11) at (0.717, -2.676) {};
		\node [style=event box] (e12) at (-2.463, -0.660) {};
		\node [style=event box] (e13) at (-2.676, -0.717) {};
		\node [style=event box] (e14) at (0.717, 2.676) {};
		\node [style=event box] (e15) at (0.660, 2.463) {};
		\node [style=event box] (e16) at (-2.463, 0.660) {};
		\node [style=event box] (e17) at (-2.676, 0.717) {};
		\node [style=event box] (e18) at (1.138, 0.784) {};
		\node [style=event box] (e19) at (1.248, 0.593) {};
		\node [style=event box] (e20) at (-1.803, 1.803) {};
		\node [style=event box] (e21) at (-1.959, 1.959) {};
		\node [style=event box] (e22) at (-1.138, -0.784) {};
		\node [style=event box] (e23) at (-1.248, -0.593) {};
		\node [style=event box] (e24) at (0.784, -1.138) {};
		\node [style=event box] (e25) at (0.593, -1.248) {};
		\node [style=event box] (e26) at (1.803, 1.803) {};
		\node [style=event box] (e27) at (1.959, 1.959) {};
		\node [style=event box] (e28) at (1.803, -1.803) {};
		\node [style=event box] (e29) at (1.959, -1.959) {};
		\node [style=event box] (e30) at (-0.660, -2.463) {};
		\node [style=event box] (e31) at (-0.717, -2.676) {};
		\node [style=ghost box] (g0) at (0.000, 0.000) {};
		\node [style=ghost box] (g1) at (0.000, 2.754) {};
		\node [style=ghost box] (g2) at (-1.377, 2.385) {};
		\node [style=ghost box] (g3) at (-2.385, 1.377) {};
		\node [style=ghost box] (g4) at (-2.754, 0.000) {};
		\node [style=ghost box] (g5) at (-2.385, -1.377) {};
		\node [style=ghost box] (g6) at (-1.377, -2.385) {};
		\node [style=ghost box] (g7) at (-0.000, -2.754) {};
		\node [style=ghost box] (g8) at (1.377, -2.385) {};
		\node [style=ghost box] (g9) at (2.385, -1.377) {};
		\node [style=ghost box] (g10) at (2.754, -0.000) {};
		\node [style=ghost box] (g11) at (2.385, 1.377) {};
		\node [style=ghost box] (g12) at (1.377, 2.385) {};
		\node [style=qubit dot] (q0) at (-0.275, 0.477) {};
		\node [style=qubit dot] (q1) at (-0.477, -0.275) {};
		\node [style=qubit dot] (q2) at (0.275, -0.477) {};
		\node [style=qubit dot] (q3) at (0.477, 0.275) {};
		\node [style=rim dot] (r4) at (-0.240, 3.250) {};
		\node [style=qubit dot] (q5) at (-0.532, 2.611) {};
		\node [style=qubit dot] (q6) at (0.532, 2.611) {};
		\node [style=rim dot] (r7) at (0.240, 3.250) {};
		\node [style=rim dot] (r8) at (-1.652, 2.862) {};
		\node [style=qubit dot] (q9) at (-1.766, 1.996) {};
		\node [style=qubit dot] (q10) at (-1.102, 1.908) {};
		\node [style=qubit dot] (q11) at (-0.845, 2.528) {};
		\node [style=rim dot] (r12) at (-2.934, 1.418) {};
		\node [style=rim dot] (r13) at (-2.695, 1.832) {};
		\node [style=qubit dot] (q14) at (-2.528, 0.845) {};
		\node [style=qubit dot] (q15) at (-1.996, 1.766) {};
		\node [style=qubit dot] (q16) at (-2.611, 0.532) {};
		\node [style=rim dot] (r17) at (-3.250, -0.240) {};
		\node [style=rim dot] (r18) at (-3.250, 0.240) {};
		\node [style=qubit dot] (q19) at (-2.611, -0.532) {};
		\node [style=qubit dot] (q20) at (-2.528, -0.845) {};
		\node [style=rim dot] (r21) at (-2.862, -1.652) {};
		\node [style=qubit dot] (q22) at (-1.996, -1.766) {};
		\node [style=qubit dot] (q23) at (-1.908, -1.102) {};
		\node [style=qubit dot] (q24) at (-1.766, -1.996) {};
		\node [style=rim dot] (r25) at (-1.418, -2.934) {};
		\node [style=rim dot] (r26) at (-1.832, -2.695) {};
		\node [style=qubit dot] (q27) at (-0.845, -2.528) {};
		\node [style=qubit dot] (q28) at (-0.532, -2.611) {};
		\node [style=rim dot] (r29) at (0.240, -3.250) {};
		\node [style=rim dot] (r30) at (-0.240, -3.250) {};
		\node [style=qubit dot] (q31) at (0.532, -2.611) {};
		\node [style=qubit dot] (q32) at (1.102, -1.908) {};
		\node [style=qubit dot] (q33) at (0.845, -2.528) {};
		\node [style=rim dot] (r34) at (1.652, -2.862) {};
		\node [style=qubit dot] (q35) at (1.766, -1.996) {};
		\node [style=qubit dot] (q36) at (2.528, -0.845) {};
		\node [style=qubit dot] (q37) at (1.996, -1.766) {};
		\node [style=rim dot] (r38) at (2.934, -1.418) {};
		\node [style=rim dot] (r39) at (2.695, -1.832) {};
		\node [style=qubit dot] (q40) at (2.611, 0.532) {};
		\node [style=qubit dot] (q41) at (2.611, -0.532) {};
		\node [style=rim dot] (r42) at (3.250, 0.240) {};
		\node [style=rim dot] (r43) at (3.250, -0.240) {};
		\node [style=qubit dot] (q44) at (1.996, 1.766) {};
		\node [style=qubit dot] (q45) at (1.908, 1.102) {};
		\node [style=qubit dot] (q46) at (2.528, 0.845) {};
		\node [style=rim dot] (r47) at (2.862, 1.652) {};
		\node [style=rim dot] (r48) at (1.418, 2.934) {};
		\node [style=qubit dot] (q49) at (0.845, 2.528) {};
		\node [style=qubit dot] (q50) at (1.766, 1.996) {};
		\node [style=rim dot] (r51) at (1.832, 2.695) {};
	\end{pgfonlayer}
\end{tikzpicture} \\
{\small (a)} & {\small (b)} \\[6pt]
\begin{tikzpicture}
	\begin{pgfonlayer}{edgelayer}
		\draw[style=hadamard edge] (0.000, 0.000) to (0.552, 0.552);
		\draw[style=hadamard edge] (0.552, 0.552) to (-0.552, 0.552);
		\draw (0.552, 0.552) to (1.025, 1.025);
		\draw[style=hadamard edge] (0.000, 0.000) to (-0.552, 0.552);
		\draw[style=hadamard edge] (-0.552, 0.552) to (-0.552, -0.552);
		\draw (-0.552, 0.552) to (-1.025, 1.025);
		\draw[style=hadamard edge] (0.000, 0.000) to (-0.552, -0.552);
		\draw[style=hadamard edge] (-0.552, -0.552) to (0.552, -0.552);
		\draw (-0.552, -0.552) to (-1.025, -1.025);
		\draw[style=hadamard edge] (0.000, 0.000) to (0.552, -0.552);
		\draw[style=hadamard edge] (0.552, -0.552) to (0.552, 0.552);
		\draw (0.552, -0.552) to (1.025, -1.025);
		\draw (0.000, 0.000) to (0.000, 0.620);
		\draw[style=hadamard edge] (4.900, 0.000) to (5.452, 0.552);
		\draw[style=hadamard edge] (5.452, 0.552) to (4.348, 0.552);
		\draw (5.452, 0.552) to (5.925, 1.025);
		\draw[style=hadamard edge] (4.900, 0.000) to (4.348, 0.552);
		\draw[style=hadamard edge] (4.348, 0.552) to (4.348, -0.552);
		\draw (4.348, 0.552) to (3.875, 1.025);
		\draw[style=hadamard edge] (4.900, 0.000) to (4.348, -0.552);
		\draw[style=hadamard edge] (4.348, -0.552) to (5.452, -0.552);
		\draw (4.348, -0.552) to (3.875, -1.025);
		\draw[style=hadamard edge] (4.900, 0.000) to (5.452, -0.552);
		\draw[style=hadamard edge] (5.452, -0.552) to (5.452, 0.552);
		\draw (5.452, -0.552) to (5.925, -1.025);
		\draw (4.900, 0.000) to (4.900, 0.620);
	\end{pgfonlayer}
	\begin{pgfonlayer}{nodelayer}
		\node [style=Z dot] (gAr0) at (0.552, 0.552) {};
		\node [style=Z dot] (gAr1) at (-0.552, 0.552) {};
		\node [style=Z dot] (gAr2) at (-0.552, -0.552) {};
		\node [style=Z dot] (gAr3) at (0.552, -0.552) {};
		\node [style=Z dot] (gAh) at (0.000, 0.000) {};
		\node [style=plus dot] (gAb) at (0.000, 0.720) {};
		\node [style=none] (cA) at (0.000, 1.280) {\scriptsize gauge-fixed tile};
		\node [style=none] (sA) at (0.000, -2.000) {\scriptsize $S=\langle IXIX,\, XIXI,\, YYZZ,\, ZZZZ\rangle$};
		\node [style=Z dot] (gCr0) at (5.452, 0.552) {};
		\node [style=Z dot] (gCr1) at (4.348, 0.552) {};
		\node [style=Z dot] (gCr2) at (4.348, -0.552) {};
		\node [style=Z dot] (gCr3) at (5.452, -0.552) {};
		\node [style=Z dot] (gCh) at (4.900, 0.000) {};
		\node [style=stub dot] (gCb) at (4.900, 0.720) {};
		\node [style=none] (cC) at (4.900, 1.280) {\scriptsize central tile};
		\node [style=none] (sC) at (4.900, -2.000) {\scriptsize $S=\langle IXIX,\, XIXI,\, YYZZ\rangle$};
		\node [style=none] (sC2) at (4.900, -2.500) {\scriptsize $\bar{X}=ZZZZ,\ \bar{Z}=XZIZ$};
	\end{pgfonlayer}
\end{tikzpicture} & \begin{tikzpicture}
	\begin{pgfonlayer}{edgelayer}
		\draw[style=hadamard edge] (-0.300, -0.620) to (0.300, -0.620);
	\end{pgfonlayer}
	\begin{pgfonlayer}{nodelayer}
		\node [style=tile box] (L0) at (0.000, 0.000) {};
		\node [style=none, anchor=west] (T0) at (0.35, 0.000) {\scriptsize tile tensor ($V$ block)};
		\node [style=none, anchor=west] (T1) at (0.35, -0.620) {\scriptsize Hadamard edge (bond)};
		\node [style=qubit dot] (L2) at (0.000, -1.240) {};
		\node [style=none, anchor=west] (T2) at (0.35, -1.240) {\scriptsize leg-end qubit};
		\node [style=rim dot] (L3) at (0.000, -1.860) {};
		\node [style=none, anchor=west] (T3) at (0.35, -1.860) {\scriptsize boundary qubit};
		\node [style=plus dot] (L4) at (0.000, -2.480) {};
		\node [style=none, anchor=west] (T4) at (0.35, -2.480) {\scriptsize $|+\rangle$ gauge fix (Z-spider)};
		\node [style=stub dot] (L5) at (0.000, -3.100) {};
		\node [style=none, anchor=west] (T5) at (0.35, -3.100) {\scriptsize free leg};
		\node [style=event box] (L6) at (0.000, -3.720) {};
		\node [style=none, anchor=west] (T6) at (0.35, -3.720) {\scriptsize measured bond stabiliser ($X{\otimes}Z$ or $Z{\otimes}X$)};
	\end{pgfonlayer}
\end{tikzpicture} \\
{\small (c)} & {\small conventions} \\
\end{tabular}
\caption{From the holographic code to the spacetime code ($n=1$ $\{4,5\}$). a) the gauged network: non-central bulk legs fixed to $|+\rangle$, the central free leg carries the logical qubit. b) the bond-extended $[[52,1]]$ code: a qubit on every tensor leg-end, each bond replaced by the measured stabiliser pair $X\otimes Z$, $Z\otimes X$ (orange). c) the tile stabiliser groups and central-tile logical representatives after gauge fixing. The resulting spacetime ZX-diagram is shown in figure~\ref{fig:spacetime_conventions}.}
\label{fig:spacetime_recipe}
\end{figure*}

\begin{figure}
\centering
\begin{tikzpicture}[baseline=(current bounding box.north)]
	\begin{pgfonlayer}{edgelayer}
		\fill [yellow!30, rounded corners] (-0.60, -1.40) rectangle (3.60, 3.15);
		\draw (0.000, -0.650) to (0.000, 0.000);
		\draw[style=hadamard edge] (0.000, 0.000) to (0.000, 2.200);
		\draw (0.000, 2.200) to (0.000, 2.850);
		\draw (2.600, -0.650) to (2.600, 0.000);
		\draw[style=hadamard edge] (2.600, 0.000) to (2.600, 2.200);
		\draw (2.600, 2.200) to (2.600, 2.850);
		\draw[style=hadamard edge] (1.300, 1.650) to (0.000, 2.200);
		\draw[style=hadamard edge] (1.300, 1.650) to (2.600, 0.000);
		\draw[style=hadamard edge] (1.300, 0.550) to (0.000, 0.000);
		\draw[style=hadamard edge] (1.300, 0.550) to (2.600, 2.200);
		\draw[style=hadamard edge] (1.300, 1.650) to (1.300, 0.550);
	\end{pgfonlayer}
	\begin{pgfonlayer}{nodelayer}
		\node [style=Z dot] (a0) at (0.000, 0.000) {};
		\node [style=Z dot] (a1) at (0.000, 2.200) {};
		\node [style=none] (wa) at (0.000, -0.95) {\scriptsize wire $a$};
		\node [style=Z dot] (b0) at (2.600, 0.000) {};
		\node [style=Z dot] (b1) at (2.600, 2.200) {};
		\node [style=none] (wb) at (2.600, -0.95) {\scriptsize wire $b$};
		\node [style=Z dot] (anc1) at (1.300, 1.650) {};
		\node [style=none, anchor=west] (lb1) at (1.800, 1.85) {\scriptsize $X_a\otimes Z_b$};
		\node [style=Z dot] (anc2) at (1.300, 0.550) {};
		\node [style=none, anchor=west] (lb2) at (1.800, 0.32) {\scriptsize $Z_a\otimes X_b$};
	\end{pgfonlayer}
\end{tikzpicture}\\[6pt]
\input{figures/recipe_block_k1.tikz}
\caption{The spacetime ZX-diagram of the bond-extended $n=1$ $\{4,5\}$ code (one round shown, $K=1$; the simulations in figure~\ref{fig:erasure_family} use $K=2$). Each measurement event is a single phase-free Z-spider, one per stabiliser generator per round. Top, shaded yellow and highlighted in the block, are the two events of one bond, realising $X\otimes Z$ and $Z\otimes X$. Every wire and every event is a potential fault location.}
\label{fig:spacetime_conventions}
\end{figure}
\begin{figure*}
\centering
\begin{tabular}{ccc}
\input{figures/recipe_anatomy_tile.tikz} & \input{figures/recipe_anatomy_rim.tikz} & \input{figures/recipe_anatomy_bond.tikz} \\
{\small (a)} & {\small (b)} & {\small (c)} \\
\end{tabular}
\caption{Where figure~\ref{fig:spacetime_recipe} lives inside the spacetime block of figure~\ref{fig:spacetime_conventions} (faded copies of the same diagram). a) violet: the four leg-qubit worldlines of one gauged tile (the tile of fig.~\ref{fig:spacetime_recipe}c). b) red: the worldline of one boundary qubit (red in fig.~\ref{fig:spacetime_recipe}b). c) yellow: the two measurement events of one bond's stabiliser pair $X\otimes Z$, $Z\otimes X$, with their partner edge (orange in fig.~\ref{fig:spacetime_recipe}b).}
\label{fig:block_anatomy}
\end{figure*}
\begin{figure*}
\centering
\begin{tabular}{cc}
\input{figures/webs_new_closed.tikz} & \input{figures/webs_new_logical.tikz} \\
{\small (a)} & {\small (b)} \\
\end{tabular}
\caption{Pauli webs of the bond-extended spacetime code, computed at depth $K=2$ on the $n=1$ block of figure~\ref{fig:spacetime_conventions} (diagram faded; the block carries $51$ independent closed webs per pair of rounds; red/green/blue $=X/Z/Y$). a) a closed web, whose product of event outcomes has deterministic parity, so that a violated parity flags a fault on any wire it crosses. b) an open web spanning the bottom and top boundaries, which carries the logical correlator across time.}
\label{fig:newwebs}
\end{figure*}

\subsection{Erasure noise}
\begin{figure}
    \centering
    \includegraphics[width=\columnwidth]{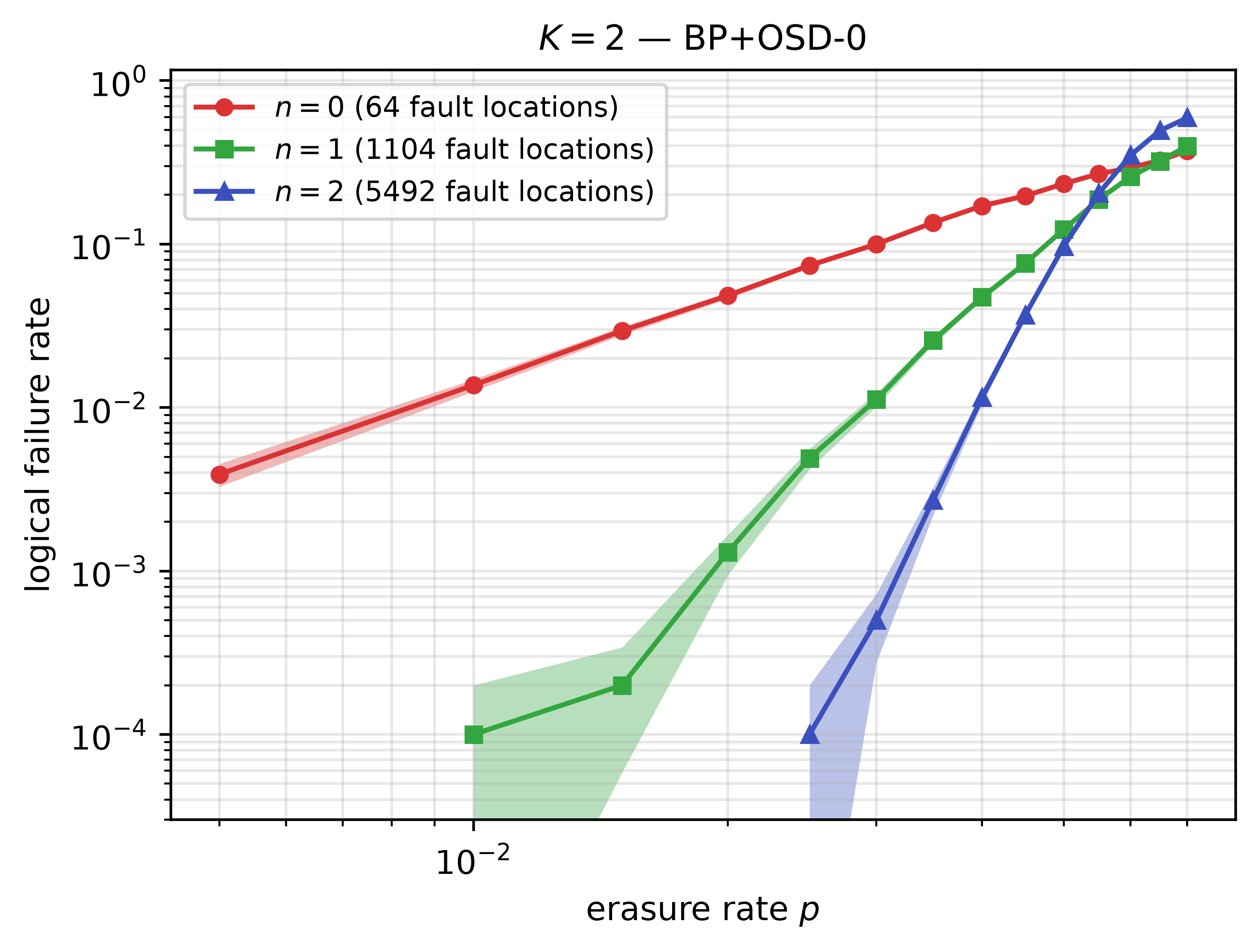}
    \caption{Erasure decoding of the spacetime $\{4,5\}$ family at matched depth $K=2$ under BP+OSD-0 ($64$, $1104$ and $5492$ fault locations for $n=0,1,2$). Every wire and measurement event is erased independently with probability $p$, an erased edge suffers a uniformly random Pauli error at a known location.}
    \label{fig:erasure_family}
\end{figure}

\begin{figure}
    \centering
    \includegraphics[width=\columnwidth]{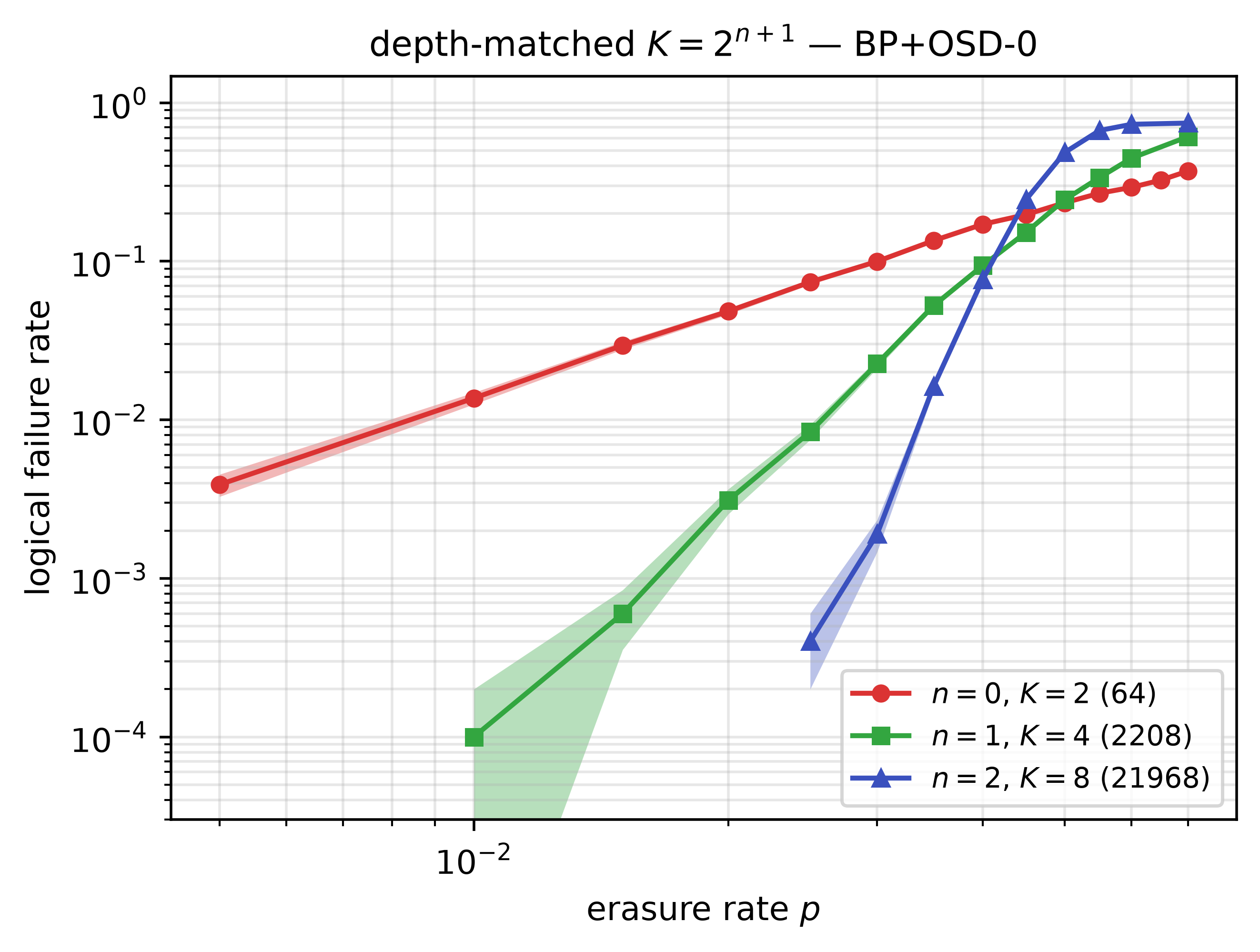}
    \caption{Depth-matched family, $K=2^{n+1}$, under the same BP+OSD-0 erasure model as figure~\ref{fig:erasure_family}: $n=0,1,2$ at $K=2,4,8$ ($64$, $2208$ and $21968$ fault locations). The family ordering inverts below a narrow crossing window at $4$--$5\%$. At $p=3\%$ the failure rate falls from $10\%$ to $2.3\%$ to $0.19\%$ with size, while above ${\approx}5\%$ larger blocks are worse. Extra rounds add fault locations without adding spatial distance, so depth scaling lowers the crossing relative to the fixed-depth comparison while preserving the ordering below it.}
    \label{fig:erasure_depthmatched}
\end{figure}

As a first quantitative probe of this proposal, we decode erasures on the spacetime code family. Unlike the error model of the previous sections, where noise acts only on the boundary qubits of the static code, here nothing is protected by fiat, and every wire and measurement event of the block is erased independently with probability $p$ ($64$, $1104$ and $5492$ fault locations for $n=0,1,2$ at matched depth $K=2$), each erased edge suffers a uniformly random Pauli, and the syndrome is decoded with BP+OSD-0 (product-sum belief propagation, $50$ iterations, combination-sweep OSD of order $0$, the same decoder family as in the previous sections, with the erased locations supplied as priors); a logical failure is recorded whenever the residual error flips the logical correlator. Our first attempt, measuring the generating set inherited directly from the tile stabilisers, shows no threshold-like behaviour at all. At fixed erasure rate the larger codes perform worse at every erasure error rate we simulated. The behaviour reverses under a \emph{rim-adapted basis}: keeping the bond stabilisers and re-choosing the remaining generators as low-weight operators supported on the boundary ($19$ rim generators of weight $3$--$8$ at $n=1$; the stabiliser group itself is unchanged). We stress that this is a change of the \emph{diagram}, not of the decoder. Each generator is measured by its own event, so re-generating the group replaces the events themselves; figure~\ref{fig:basis_change} contrasts a weight-$5$ tile-inherited generator's event with a weight-$3$ rim generator's event (the inherited non-bond generators run up to weight $32$). With the events, the spacetime diagram and its fault locations change ($1104$ wires in the rim basis at $K=2$), larger codes outperform smaller ones at low erasure error rates, with the pairwise crossings at $5.5$--$6.5\%$ (figure~\ref{fig:erasure_family}). We emphasise that this is a fixed-depth, finite-size comparison: a size-independent threshold is not yet claimed. As a first depth-scaled probe, matching the rounds to the geometric boundary growth via $K=2^{n+1}$ preserves the family ordering at low rates, with the pairwise crossings confined to a narrow $4$--$5\%$ window (figure~\ref{fig:erasure_depthmatched}). Larger $n$ is left to future work. 
\subsection{Depolarising noise}
\begin{figure}
    \centering
    \includegraphics[width=\columnwidth]{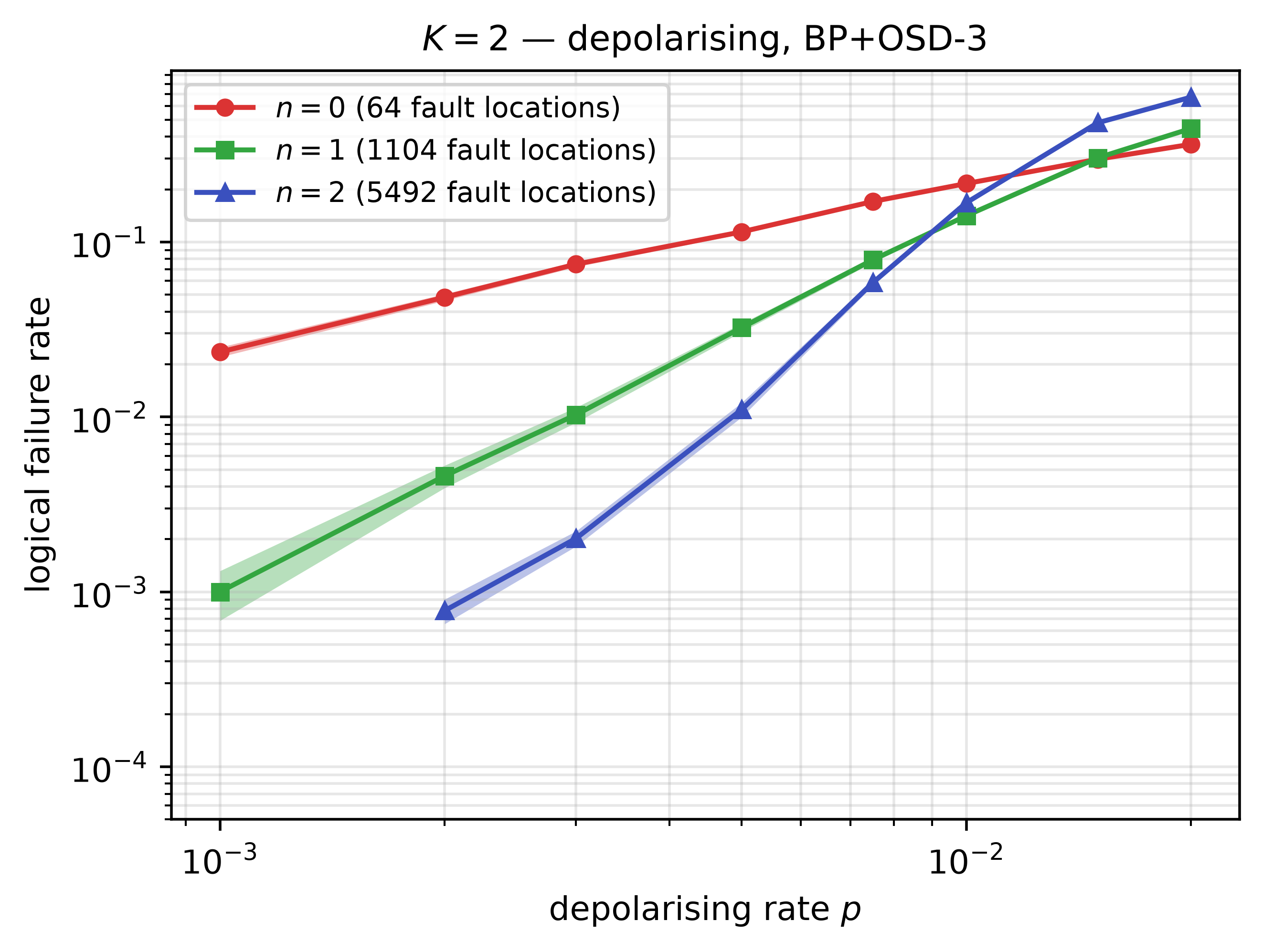}
    \caption{Depolarising noise on the spacetime $\{4,5\}$ family at fixed depth $K=2$ under BP+OSD-3. Every wire suffers $X$, $Y$ or $Z$ with probability $p/3$ each. Larger codes win at low rates, with pairwise crossings between ${\approx}0.9\%$ and $1.5\%$.}
    \label{fig:pauli_family}
\end{figure}

Under depolarising noise, where every wire suffers $X$, $Y$ or $Z$ with probability $p/3$ each and the syndrome is decoded with BP+OSD-3 (product-sum belief propagation, $50$ iterations, combination-sweep OSD of order $3$), the fixed-depth $K=2$ family shows the same qualitative behaviour as under erasures. Larger codes win at low rates, with an order-of-magnitude suppression per size step at $p=0.1\%$ and pairwise crossings between ${\approx}0.9\%$ and $1.5\%$, roughly a fifth of the erasure crossings (figure~\ref{fig:pauli_family}). In general, an arbitrary choice of web basis yields a poorly structured Tanner graph for belief-propagation decoding, and we did not search for a Pauli-adapted basis. The rim-adapted generators produce this threshold-like behaviour.

A practical remark on how the deep blocks were built. Extracting the detectors and logical correlator of a spacetime block by Pauli-web search (we use the web-finding tools of PyZX \cite{kissinger2020Pyzx}) becomes expensive with depth, and for the $n=2$, $K=8$ block ($21968$ wires) the direct computation ran for hours. Instead we exploit the round-periodicity of the block. The wire set, the detector matrix (block-banded, with two boundary bands and a repeating interior band, figure~\ref{fig:banded}) and a suitably chosen logical representative are all invariant under translation by one round, so the full decoding data for any depth $K$ can be assembled in seconds from the labelled $K=2$ block by time translation, after re-choosing the logical representative (by Gaussian elimination modulo the detectors) so that its supports on consecutive rounds match. We validated the assembly exactly at $n=1$, where the stacked $K=4$ matrices agree with the directly computed ones (identical wire sets, identical detector row spans, and logical representatives equal modulo detectors).

\begin{figure}
    \centering
    \begin{tikzpicture}[scale=0.82]
	\begin{pgfonlayer}{edgelayer}
		\draw (0.000, -0.55) to (0.000, 0.000);
		\draw[style=hadamard edge] (0.000, 0.000) to (0.000, 2.200);
		\draw (0.000, 2.200) to (0.000, 2.750);
		\draw[style=hadamard edge] (2.880, 1.050) to (0.000, 0.000);
		\draw (1.150, -0.55) to (1.150, 0.000);
		\draw[style=hadamard edge] (1.150, 0.000) to (1.150, 2.200);
		\draw (1.150, 2.200) to (1.150, 2.750);
		\draw[style=hadamard edge] (2.880, 1.050) to (1.150, 0.000);
		\draw (2.300, -0.55) to (2.300, 0.000);
		\draw[style=hadamard edge] (2.300, 0.000) to (2.300, 2.200);
		\draw (2.300, 2.200) to (2.300, 2.750);
		\draw[style=hadamard edge] (2.880, 1.050) to (2.300, 2.200);
		\draw (3.450, -0.55) to (3.450, 0.000);
		\draw[style=hadamard edge] (3.450, 0.000) to (3.450, 2.200);
		\draw (3.450, 2.200) to (3.450, 2.750);
		\draw[style=hadamard edge] (2.880, 1.050) to (3.450, 2.200);
		\draw (4.600, -0.55) to (4.600, 0.000);
		\draw[style=hadamard edge] (4.600, 0.000) to (4.600, 2.200);
		\draw (4.600, 2.200) to (4.600, 2.750);
		\draw[style=hadamard edge] (2.880, 1.050) to (4.600, 2.200);
		\draw (6.600, -0.55) to (6.600, 0.000);
		\draw[style=hadamard edge] (6.600, 0.000) to (6.600, 2.200);
		\draw (6.600, 2.200) to (6.600, 2.750);
		\draw[style=hadamard edge] (8.330, 1.050) to (6.600, 2.200);
		\draw (7.750, -0.55) to (7.750, 0.000);
		\draw[style=hadamard edge] (7.750, 0.000) to (7.750, 2.200);
		\draw (7.750, 2.200) to (7.750, 2.750);
		\draw[style=hadamard edge] (8.330, 1.050) to (7.750, 2.200);
		\draw (8.900, -0.55) to (8.900, 0.000);
		\draw[style=hadamard edge] (8.900, 0.000) to (8.900, 2.200);
		\draw (8.900, 2.200) to (8.900, 2.750);
		\draw[style=hadamard edge] (8.330, 1.050) to (8.900, 2.200);
	\end{pgfonlayer}
	\begin{pgfonlayer}{nodelayer}
		\node [style=Z dot] (Aa4) at (0.000, 0.000) {};
		\node [style=Z dot] (Ab4) at (0.000, 2.200) {};
		\node [style=none] (Al4) at (0.000, -0.85) {\scriptsize $q_{4}$};
		\node [style=none] (Ap4) at (0.000, 3.05) {\scriptsize $Z$};
		\node [style=Z dot] (Aa6) at (1.150, 0.000) {};
		\node [style=Z dot] (Ab6) at (1.150, 2.200) {};
		\node [style=none] (Al6) at (1.150, -0.85) {\scriptsize $q_{6}$};
		\node [style=none] (Ap6) at (1.150, 3.05) {\scriptsize $Z$};
		\node [style=Z dot] (Aa7) at (2.300, 0.000) {};
		\node [style=Z dot] (Ab7) at (2.300, 2.200) {};
		\node [style=none] (Al7) at (2.300, -0.85) {\scriptsize $q_{7}$};
		\node [style=none] (Ap7) at (2.300, 3.05) {\scriptsize $X$};
		\node [style=Z dot] (Aa49) at (3.450, 0.000) {};
		\node [style=Z dot] (Ab49) at (3.450, 2.200) {};
		\node [style=none] (Al49) at (3.450, -0.85) {\scriptsize $q_{49}$};
		\node [style=none] (Ap49) at (3.450, 3.05) {\scriptsize $X$};
		\node [style=Z dot] (Aa51) at (4.600, 0.000) {};
		\node [style=Z dot] (Ab51) at (4.600, 2.200) {};
		\node [style=none] (Al51) at (4.600, -0.85) {\scriptsize $q_{51}$};
		\node [style=none] (Ap51) at (4.600, 3.05) {\scriptsize $X$};
		\node [style=Z dot] (Aev) at (2.880, 1.050) {};
		\node [style=Z dot] (Ba7) at (6.600, 0.000) {};
		\node [style=Z dot] (Bb7) at (6.600, 2.200) {};
		\node [style=none] (Bl7) at (6.600, -0.85) {\scriptsize $q_{7}$};
		\node [style=none] (Bp7) at (6.600, 3.05) {\scriptsize $X$};
		\node [style=Z dot] (Ba8) at (7.750, 0.000) {};
		\node [style=Z dot] (Bb8) at (7.750, 2.200) {};
		\node [style=none] (Bl8) at (7.750, -0.85) {\scriptsize $q_{8}$};
		\node [style=none] (Bp8) at (7.750, 3.05) {\scriptsize $X$};
		\node [style=Z dot] (Ba13) at (8.900, 0.000) {};
		\node [style=Z dot] (Bb13) at (8.900, 2.200) {};
		\node [style=none] (Bl13) at (8.900, -0.85) {\scriptsize $q_{13}$};
		\node [style=none] (Bp13) at (8.900, 3.05) {\scriptsize $X$};
		\node [style=Z dot] (Bev) at (8.330, 1.050) {};
		\node [style=none] (la) at (2.300, -1.55) {\small (a)};
		\node [style=none] (lb) at (7.750, -1.55) {\small (b)};
	\end{pgfonlayer}
\end{tikzpicture}
    \caption{The rim-adapted basis amends the spacetime diagram itself. Two events measuring generators of the \emph{same} stabiliser group at $n=1$: a) a tile-inherited generator, $Z_{4}Z_{6}X_{7}X_{49}X_{51}$ (weight $5$); b) a rim-adapted generator, $X_{7}X_{8}X_{13}$ (weight $3$, boundary-supported). Attaching \emph{before} a worldline's inter-round Hadamard measures $Z$, \emph{after} it measures $X$; changing the generating set changes the events, wires and fault locations of the block, while the measured group, and hence the code, is unchanged.}
    \label{fig:basis_change}
\end{figure}
\begin{figure}
    \centering
    \begin{tikzpicture}[scale=0.95]
	\draw[thick] (0,0) rectangle (5.2,2.6);
	\foreach \x in {1.3,2.6,3.9} \draw[densely dotted, gray] (\x,0) to (\x,2.6);
	\fill[violet!30] (0.0,2.02) rectangle (1.95,2.52);
	\fill[gray!40] (1.30,1.28) rectangle (3.25,1.78);
	\fill[gray!40] (2.60,0.72) rectangle (4.55,1.22);
	\fill[violet!30] (3.25,0.08) rectangle (5.20,0.58);
	\node[style=none] at (0.97,2.27) {\scriptsize $T$};
	\node[style=none] at (2.27,1.53) {\scriptsize $M$};
	\node[style=none] at (3.57,0.97) {\scriptsize $M$};
	\node[style=none] at (4.22,0.33) {\scriptsize $B$};
	\node[style=none] at (0.65,-0.28) {\scriptsize round $1$};
	\node[style=none] at (1.95,-0.28) {\scriptsize round $2$};
	\node[style=none] at (3.25,-0.28) {\scriptsize $\cdots$};
	\node[style=none] at (4.55,-0.28) {\scriptsize round $K$};
	\node[style=none] at (-0.45,1.3) {\scriptsize $H$};
\end{tikzpicture}
    \caption{Block-banded structure of the spacetime detector matrix $H$. Columns are the wires of successive rounds. The boundary bands $T$ and $B$ act at the first and last rounds, while the interior band $M$ repeats along the diagonal, shifted by one round at a time. All three bands, and a periodic logical representative, can be read off the $K=2$ block, which is how deeper blocks are assembled.}
    \label{fig:banded}
\end{figure}

\section{Summary and outlook}
We provided various preliminary ideas on tensor network/Holographic codes, showcasing the usefulness of ZX-calculus and Pauli webs in the context of stabiliser codes. First, we re-visited the pentagon holographic code through the lens of ZX-calculus, using Pauli webs to visualise stabilisers, logical operators and parity-check matrices directly from ZX-diagrams. Then, we demonstrated how ZX-calculus related simplifications can be used to compute R\'enyi entropies. Next, motivated by the pentagon holographic code's construction, we introduced a family of ZX-diagram–inspired codes on the dual $\{4,5\}$ tessellation and studied its logical error rates under BP and BP+OSD decoding, including the impact of simple gauge-fixing prescriptions. These results suggest that decoder, gauge, or generator choices can be important in these holographic code constructions. Finally, we foliated these networks into spacetime ZX-diagrams, with numerical evidence of threshold-like erasure suppression when every internal edge is a fault location. In summary, ZX-diagrammatically motivated tensor networks may be a useful design lever towards interesting holographic and tensor network codes.

\clearpage
\section{Acknowledgments}
We thank  Luca Cocconi for a thorough review of an early draft of this manuscript. We acknowledge discussions with Matthew Steinberg, Steve Kolthammer, Amihay Hanany, Arshia Momeni, Kwok Chung Matthew Cheung and Evita Gamber. KHW was funded by the Imperial College London President's PhD scholarship The three-dimensional visualisations of ZX-diagrams and Pauli webs in this work were produced with the open-source \texttt{pyzx\_3d\_viewer} package \cite{pyzx_3d_viewer}. The companion code, cached decoding matrices and simulation data for this work are openly available \cite{zx_spacetime_code}. The wording in certain sections of this manuscript had been refined using LLMs.

\bibliography{main.bib}

\end{document}